\documentclass[11pt,a4paper]{article}
\usepackage{jheppub}
\usepackage[utf8]{inputenc}
\usepackage{amsfonts,amssymb}
\usepackage{acronym}
\usepackage{xspace}
\usepackage{cleveref}
\usepackage{siunitx}
\usepackage{booktabs}
\usepackage{pifont}
\usepackage{slashed}
\usepackage[noxspace]{listofsymbols}
\usepackage{slashed}

\DeclareSIUnit[number-unit-product = \;]{\permil}{\textperthousand}
\DeclareSIUnit[number-unit-product = \;]{\M}{M}
\DeclareBinaryPrefix\kibi{Ki}{10}
\DeclareBinaryPrefix\mibi{Mi}{20}
\DeclareBinaryPrefix\gibi{Gi}{20}

\crefformat{equation}{Eq.~(#2#1#3)}
\crefformat{section}{Section~#2#1#3}
\crefformat{figure}{Figure~#2#1#3}
\crefformat{table}{Table~#2#1#3}
\crefformat{chapter}{Chapter~#2#1#3}
\crefformat{appendix}{Appendix~#2#1#3}

\DeclareRobustCommand{\eq}[1]{Eq.~\eqref{eq:#1}}
\DeclareRobustCommand{\eqs}[2]{Eqs.~\eqref{eq:#1} and \eqref{eq:#2}}

\newcommand\standardwidth{.75\textwidth}
\newcommand\halfwidth{.49\textwidth}
\DeclareMathAlphabet{\mathcal}{OMS}{cmsy}{m}{n}

\newcommand\prog[1]{\textsc{#1}}

\newcommand\order[1]{\ensuremath{\mathcal{O}\!\left(#1\right)}}
\newcommand{\abs}[1]{\ems{\left\lvert#1\right\rvert}}

\newcommand{\ems}[1]{\ensuremath{#1}}
\newcommand{\T}[1]{\mathrm{#1}}
\newcommand{\MT}[1]{\quad\T{#1}\quad}

\newcommand\no{\notag{}\\}
\newcommand\po{\;.}
\newcommand\co{\;,}

\newcommand\Rcite[1]{Ref.~\cite{#1}}

\newcommand\Al{\bigl}
\newcommand\Ar{\bigr}
\newcommand\Bl{\Bigl}
\newcommand\Br{\Bigr}
\newcommand\Cl{\biggl}
\newcommand\Cr{\biggr}

\newcommand{\del}{\partial}
\newcommand{\E}{\ems{\T{e}}}
\newcommand{\I}{\ems{\T{i}}}
\newcommand{\D}{\ems{\T{d}}}

\newcommand{\of}[1]{\left[{#1}\right]}
\newcommand{\ofIT}[1]{\Al[{#1}\Ar]}
\newcommand{\absIT}[1]{\Al|{#1}\Ar|}
\renewcommand{\Re}{\operatorname{Re}}

\newcommand\Int[2][]{\ems{\int#1#2\:}}

\newcommand{\ket}[1]{\ems{\left|{#1}\right>}}
\newcommand{\bra}[1]{\ems{\left<{#1}\right|}}

\newcommand{\braketop}[3]
  {\ems{\bra{#1\vphantom{#2#3}\,}#2\ket{#3\vphantom{#1#2}}}}

\newcommand{\Pjet}{\ensuremath{j}}
\newcommand\jb{\ensuremath{\Pjet_{b}}\xspace}
\newcommand\jbbar{\ensuremath{\Pjet_{\bar b}}\xspace}
\newcommand{\projected}[1]{\hat{#1}}
\newcommand{\set}[1]{\{#1\}}
\newcommand{\lT}[1]{\losstring\T{#1}}
\opensymdef{}
\newsym[]{ee}{\ensuremath{e^+\,e^-}}
\newsym[]{WbWb}{\ensuremath{W^+bW^-\losstring\bar{b}}}
\newsym[]{eeWbWb}{\ensuremath{\ee\to\losstring\WbWb}}
\newsym[]{tT}{\ensuremath{t\losstring\bar{t}}}
\newsym[]{eetT}{\ensuremath{\ee\to\losstring\tT}}
\newsym[]{XSLO}{\sigma_\lT{LO}}
\newsym[]{XSLL}{\sigma_\lT{LL}}
\newsym[]{XSNLO}{\sigma_\lT{NLO}}
\newsym[]{XSLOLL}{\sigma_\lT{LO+LL}}
\newsym[]{XSNLOLL}{\sigma_\lT{NLO+LL}}
\newsym[]{XSNLONLL}{\sigma_\lT{NLO+NLL}}
\newsym[]{XSqcd}{\sigma_\lT{FO}}
\newsym[]{XSmatched}{\sigma_\lT{matched}}
\newsym[]{XSnrqcd}{\sigma_\lT{NRQCD}}
\newsym[]{XSresummed}{\sigma_\lT{NRQCD}^\lT{full}}
\newsym[]{XSexpanded}{\sigma_{\lT{NRQCD}}^{\lT{expanded}}}
\newsym[]{ME}{\mathcal{M}}
\newsym[]{MEfact}{\mathcal{M}_\lT{fact}}
\newsym[]{MEdec}{\mathcal{M}_\lT{dec}}
\newsym[]{MEdecT}{\mathcal{M}_{\lT{dec},t}}
\newsym[]{MEdecTbar}{\mathcal{M}_{\lT{dec},\losstring\bar{t}}}
\newsym[]{MEprod}{\mathcal{M}_\lT{prod}}
\newsym[is this a matrix element or the square of it?]{amp}{\mathcal{A}}
\newsym[]{ampDec}{\amp_\lT{dec}}
\newsym[]{ampProd}{\amp_\lT{prod}}
\newsym[]{FFLL}{F_\lT{LL}}
\newsym[]{FFNLL}{F_\lT{NLL}}
\newsym[]{FFLLmO}{\tilde{F}_\lT{LL}}
\newsym[]{FFNLLmO}{\tilde{F}_\lT{NLL}}
\newsym[]{FFexpLL}{F^\lT{exp}_\lT{LL}}
\newsym[]{FFexpNLL}{F^\lT{exp}_\lT{NLL}}
\newsym[]{switch}{f_s}
\newsym[]{ALweak}{\alpha_\lT{em}}
\newsym[]{ALstrong}{\alpha_s}
\newsym[]{AShard}{\alpha_{\lT{H}}} 
\newsym[]{ASfirm}{\alpha_{\lT{F}}}
\newsym[]{ASsoft}{\alpha_{\lT{S}}}
\newsym[]{ASusoft}{\alpha_{\lT{US}}}
\newsym[]{SW}{\sin^2{(\theta_{W})}}
\newsym[]{CW}{\cos^2{(\theta_{W})}}
\newsym[]{GF}{G_\mu}
\newsym[]{DeltaMM}{\Delta{M}}
\newsym[]{CF}{C_{F}}
\newsym[]{sqrts}{\sqrt{s}}
\newsym[]{rescaleH}{h}
\newsym[]{rescaleF}{f}
\newsym[]{pb}{p_{b}}
\newsym[]{pB}{p_{\losstring\bar{b}}}
\newsym[]{pWp}{p_{W^+}}
\newsym[]{pWm}{p_{W^-}}
\newsym[]{BornP}{p_{B}}
\newsym[]{RealP}{p_{R}}
\newsym[]{projBornP}{\projected{p}_{B}}
\newsym[]{projRealP}{\projected{p}_{R}}
\newsym[]{mpole}{m_t}
\newsym[]{mOneS}{M_t^{\lT{1S}}}
\newsym[]{DeltaM}{\Delta_{\losstring\mpole}}
\newsym[]{nustar}{\nu_{*}}
\newsym[]{propagators}{\mathcal{P}_{t\losstring\bar{t}}}
\newsym[]{NthreeLO}{\lT{N}^3\lT{LO}}
\newsym[]{hypergeo}{\losstring{}_2F_1}
\newsym[]{signal}{S}
\newsym[]{background}{B}
\closesymdef{}

\newcommand\lessspace{\hspace{-1.3em}}
\newcommand\halflessspace{\hspace{-0.6em}}
\newcommand\extraspace{\hspace{+1.3em}}
\newcommand\FFOne{\textcolor{blue}{1}\halflessspace}
\newcommand\FFOnePlus{\textcolor{blue}{(1+\FFLLmO{})}\halflessspace}
\newcommand\FFLLmOne{\textcolor{blue}{\FFLLmO{}}\lessspace}
\newcommand\FFNLLmOne{\textcolor{blue}{\FFNLLmO{}\lessspace}}

\newcommand{\diagram}[1]{%
\begin{array}{l}
  \includegraphics[height=.13\textwidth]{diagrams/#1-crop}
\end{array}
}
\newcommand{\diagramInText}[1]{\includegraphics[height=1em]{diagrams/#1-crop}}
\newcommand{\realOf}[2]{\left(#1\right.\left\lmoustache#2\right)}
\newcommand{\realOfDiagrams}[3][]{
  \left(#1\diagram{#2}\right.\left\lmoustache\diagram{#3}\right)
}

\newcommand\fastjet{\prog{FastJet}\xspace}

\newcommand\whz{\prog{Whizard}\xspace}
\newcommand\whizard\whz
\newcommand\OMega{\prog{O'Mega}\xspace}

\newcommand\powheg{\prog{Powheg}\xspace}

\newcommand\gosam{\prog{GoSam}\xspace}
\newcommand\openloops{\prog{OpenLoops}\xspace}
\newcommand\recola{\prog{Recola}\xspace}

\newcommand\toppik{\prog{Toppik}\xspace}

\newcommand\lrarrow{\ems{\leftrightarrow}}

\newacro{CMF}{center of mass frame}
\newacro{CLIC}{Compact Linear Collider}
\newacro{DPA}{double-pole approximation}
\newacro{EHA}{extra-helicity approximation}
\newacro{EW}{electroweak}
\newacro{HA}{helicity approximation}
\newacro{ILC}{International Linear Collider}
\newacro{ISR}{initial-state radiation}
\newacro{LL}{leading log}
\newacro{LO}{leading order}
\newacro{MC}{Monte Carlo}
\newacro{NLL}{next-to-leading log}
\newacro{NLO}{next-to-leading order}
\newacro{NNLO}{next-to-next-to-leading order}
\newacro{NRQCD}{nonrelativistic QCD}
\newacro{NWA}{narrow-width approximation}
\newacro{OLP}{One-Loop Provider}
\newacro{PA}{pole approximation}
\newacro{PDG}{Particle Data Group}
\newacro{QCD}{quantum chromodynamics}
\newacro{RG}{renormalization group}
\newacro{SCET}{soft-collinear effective theory}
\newacro{SD}{signal diagram}
\newacro{SM}{Standard Model}
\newacro{vNRQCD}{velocity NRQCD}

\DeclareRobustCommand{\Rcites}[1]{Refs.~\cite{#1}}
\newcommand{\bO}{\mbox{\boldmath $O$}}
\newcommand{\bq}{\mathbf q}
\newcommand{\bk}{\mathbf k}
\newcommand{\bp}{\mathbf p}
\newcommand{\be}{\mathbf e}
\newcommand{\bpp}{{\bp^\prime}}
\newcommand{\bsig}{\boldsymbol{\sigma}}
\def\O#1#2{\mbox{\boldmath $O$}_{\mbox{\scriptsize $\mathbf #1$},#2}}
\newcommand{\ord}[1]{\mathcal{O} (#1)}

\begin{document}
\title{Fully-differential Top-Pair Production at a Lepton Collider: From 
Threshold to Continuum
}

\author[a]{Fabian Bach,}
\author[b]{Bijan Chokouf\'e Nejad,}
\author[c,d]{Andr\'e H. Hoang,}
\author[e]{Wolfgang Kilian,}
\author[b]{J\"urgen Reuter,}
\author[f]{Maximilian Stahlhofen,}
\author[g]{Thomas Teubner,}
\author[b,e]{and Christian Weiss}

\affiliation[a]{European Commission, Eurostat, 2920 Luxembourg,
  Luxembourg~\footnote{Responsibility for the information and views
    set out in this article lies entirely with the author.}}
\affiliation[b]{DESY, Theory Group, Notkestr. 85, D-22607 Hamburg, Germany}
\affiliation[c]{University of Vienna, Faculty of Physics, Bolzmanngasse 5,
  A-1090 Wien, Austria}
\affiliation[d]{Erwin Schr\"odinger International Institute for Mathematical Physics, University of Vienna,
  Boltzmanngasse 9, A-1090 Vienna, Austria}
\affiliation[e]{University of Siegen, Department of Physics,
  Walter-Flex-Str. 3, D-57068 Siegen, Germany}
\affiliation[f]{PRISMA Cluster of Excellence, Institute of Physics, Johannes Gutenberg University,
  Staudingerweg 7, D-55128 Mainz, Germany}
\affiliation[g]{University of Liverpool, Department of Mathematical Sciences,
  Liverpool L69 3BX, United Kingdom}

\emailAdd{fabian.bach@t-online.de}
\emailAdd{bijan.chokoufe@desy.de}
\emailAdd{andre.hoang@univie.ac.at}
\emailAdd{kilian@physik.uni-siegen.de}
\emailAdd{juergen.reuter@desy.de}
\emailAdd{mastahlh@uni-mainz.de}
\emailAdd{thomas.teubner@liverpool.ac.uk}
\emailAdd{christian.weiss@desy.de}

\abstract{%
  We present an approach to predict exclusive $\WbWb{}$ production at lepton 
  colliders that correctly describes the top--anti-top threshold as well as the 
  continuum region.
  We incorporate $\tT{}$ form factors for the NLL threshold resummation derived 
in NRQCD into a factorized relativistic cross section using an extended 
double-pole approximation, which accounts for fixed-order QCD corrections to the 
top decays at NLO.
  This is combined with the full fixed-order QCD result at NLO
  for $\WbWb{}$ production to obtain predictions that are not only
  valid at threshold but smoothly transition to the continuum region.
  Our implementation is based on the Monte Carlo event generator \whz
  and the code \toppik and allows to compute fully-differential
  threshold-resummed cross sections including the interference
  with non-resonant background processes. For the first time it is now 
possible to systematically study general differential observables at future lepton colliders involving the decay products of the top quarks at energies close to the pair production threshold and beyond.
}

\keywords{top threshold, resummation, matching, event generators}

\arxivnumber{1712.02220}


\begin{flushright}
  \normalsize{} DESY 17--158, LTH 1143, MITP/17-077, SI-HEP-2017-20, UWThPh2017-35
  \vspace{+1em}
\end{flushright}
\maketitle{}
\section{Introduction}
\label{s:intro}
%
The top quark is the heaviest elementary particle in the Standard Model and thus
 provides a unique window to new physics models in which couplings are related to mass.
Top quark physics is therefore one of the corner stones in the physics program of all
future lepton colliders. One of the most important measurements is the
scan of the  top--anti-top resonance lineshape which will 
enable top mass and width measurements with unprecedented precision. The top
pair production cross section near threshold and in the transition region to the
continuum is also sensitive to the couplings of the top quark, like $\alpha_s$
or the top Yukawa coupling.


Reliable theory predictions in the threshold region crucially require the
resummation of Coulomb singular $(\alpha_s/v)^n$ terms to all orders in
perturbation theory, where here and throughout this paper $v$ denotes the (effective)
velocity of the top quarks in the center-of-mass (c.m.) frame. In the
threshold region we have $v \sim \alpha_s \sim 0.1$, which requires a
simultaneous expansion in both parameters. This indicates that
bound-state effects become important despite the fact that the top
quarks decay before they can form a would-be toponium
state. Furthermore, in the threshold region the concept of on-shell
top quark production loses its meaning and all kinematic
configurations in the resonance region governed by the top width are
equally important. We refer to this kinematic configuration as 
``resonant $\tT{}$ production''.\footnote{
Concerning terminology, we write ``on-shell'' when we refer to kinematic top 
quark configurations that exactly satisfy the on-shell relation $p_{t,\bar 
t}^2=\mpole^2$ and ``off-shell'' when we have
$p_{t,\bar t}^2\neq \mpole^2$ in general, where \mpole{} is the top quark pole 
mass and $p_{t,\bar t}$ denotes the (anti)top four-momentum.
We write ``resonant'' when we refer to kinematic top quark configurations where 
$|p_{t,\bar t}^2-\mpole^2| \lesssim \mpole \Gamma_{t}$ and  ``non-resonant'' 
when we refer to
$|p_{t,\bar t}^2-\mpole^2| \gg \mpole \Gamma_{t}$.
} 
The resummation of the singular velocity terms is performed within the
effective theory nonrelativistic QCD
(NRQCD)~\cite{Caswell:1985ui,hep-ph/9407339} by solving a Schr\"odinger-type
equation for the propagation of the top--anti-top system.
Cross section calculations in the NRQCD framework are
therefore always understood to include the resummation of the Coulomb
singular terms to the stated order.
In this context the ``fixed-order'' label is used to distinguish from a 
renormalization-group improved (RGI) calculation.

To be explicit concerning the nonrelativistic power counting in the threshold
region, we write the normalized total cross section (R-Ratio) schematically as
\begin{align}
R = \frac{\sigma_{t \bar t}}{\sigma_{\mu^+\!\mu^-}} = v \sum\limits_k
\bigg(\!\frac{\ALstrong}{v}\!\bigg)^{\!\!k} \sum\limits_i (\ALstrong\,\ln\, v
)^i
\times \bigg\{\! 1 \, (\T{LL});\;\ALstrong, v \, (\T{
	NLL});\;\ALstrong^2,\,
\ALstrong v,\, v^2\,(\T{NNLL});\,\ldots \! \bigg\}\,.
\label{eq:Rstruc}
\end{align}
The overall factor of $v$ emerges from the nonrelativistic phase space
integration. In the respective fixed-order counting (i.e.\ N$^k$LO instead of N$^k$LL) powers of
$\ln v$ are considered of order unity, so in the nonrelativistic power counting convention 
predictions at N$^k$LL order fully include all N$^k$LO contributions. 
So far, theory predictions for the $\tT{}$ threshold have mainly focused on the
total production cross section. For the fixed-order NRQCD expansion the total 
cross section is known at NNLO for a long time~\cite{Hoang:1998xf,Hoang:2000yr}, and since 
recently also to \NthreeLO{}~\cite{1506.06864}.
Besides the Coulomb singularities, close to threshold also logarithms of the velocity become sizable
requiring the additional resummation of $(\alpha_s \ln v)^n$ terms to all orders as indicated in Eq.\ (\ref{eq:Rstruc}).
Modified versions of NRQCD, namely the potential NRQCD
(pNRQCD)~\cite{hep-ph/9707481,hep-ph/9907240} and the velocity NRQCD
(vNRQCD)~\cite{hep-ph/9910209,hep-ph/9912226,hep-ph/0209340} frameworks, allow the systematic
resummation of these logarithmic terms via renormalization group (RG)
equations, where both frameworks differ with respect to the separation of soft and ultrasoft modes.%
\footnote{We will often refer to the nonrelativistic theory used to perform the 
  threshold resummation as NRQCD in general. The
  results for the total cross section obtained in vNRQCD and pNRQCD agree at NLL.}
Regarding RG improved predictions of $\tT{}$ threshold production the result
of \Rcite{1309.6323} is currently state-of-the-art combining NNLO fixed-order corrections with resummation of velocity logarithms at NNLL order.%
\footnote{A certain class of soft NNLL (mixing) terms starting at N${}^4$LO are
still missing in this calculation. There is, however, good evidence that these are
negligible compared to the fully known ultrasoft contributions and the residual uncertainty of the result~\cite{Hoang:2012us,1309.6323}.}
For the threshold production of resonant top quarks the vector
(S-wave) and axial-vector (P-wave) form factors 
differential in the top three momentum and the c.m.\ energy are available
at NNLO and provided from the numerical code
\toppik~\cite{Jezabek:1992np,Harlander:1994ac,hep-ph/9904468}.  
In the present work we will use \toppik in order to implement
the Coulomb resummation in our calculation of off-shell top-pair
production, as described below.

Once electroweak effects are included, the natural power counting scheme
near threshold also accounts for the electromagnetic coupling
$\ALweak$ and reads $v\sim\ALstrong\sim\sqrt{\ALweak}\sim 0.1$. 
Here the dominant effect is related to the top decay itself, which for
the total inclusive cross section determined via the optical theorem
enters the NRQCD prediction at LO simply via the shift
$\sqrt{s} \to \sqrt{s} + \I \Gamma_{t}$ in 
the top propagators~\cite{Fadin:1987wz,Strassler:1990nw}. The width
receives sizable QCD
corrections, which must be taken into account at NLO.
In addition there are non-resonant electroweak contributions to
$\WbWb{}$ production, e.g. where at least one of the top propagators
is non-resonant or absent. As was pointed out already in 
\Rcite{Hoang:2004tg}, these start contributing at 
NLO because instead of the overall nonrelativistic phase space factor $v$ 
shown in \cref{eq:Rstruc} these contributions have
at least one power of the electroweak coupling constant $\ALweak$. 
In \Rcites{Hoang:2004tg,1002.3223} the associated NNLO interference effects and 
a renormalization group method to resum large NLL phase space logarithms were 
provided accounting also for cuts on the top/anti-top invariant masses.
The full set of non-resonant non-logarithmic corrections were given in the 
fixed-order expansion at NLO and NNLO in \Rcite{1004.2188} and 
\Rcite{Beneke:2017rdn}, respectively,  (see also \Rcites{1110.1970,1307.4337}) 
for the total $\WbWb{}$ cross section and including top/anti-top invariant mass 
cuts.
The dominant purely hard electroweak
corrections to resonant nonrelativistic top pair production are from
QED initial state radiation enhanced by logarithms $\log(s/m_e^2)$,
which are treated in this work in the structure function approach. All
remaining ones carry the overall nonrelativistic phase space factor
$v$ shown in \cref{eq:Rstruc} as well as at least one power of the
electroweak coupling constant $\ALweak$ and therefore contribute only
at NNLL level~\cite{Guth:1991ab,hep-ph/0604104}.
The impact of non-resonant
contributions for measurements of the top quark mass was analyzed in 
\Rcite{1411.2355} focusing on the effects of single top production which 
represents the most important non-resonant correction. 

Adopting the nonrelativistic power counting, the full Born-level relativistic
$\WbWb{}$ production cross section contains the complete non-resonant contributions starting at NLO. Its $\order{\ALstrong}$ full QCD corrections contain the complete 
non-resonant contributions starting at NNLO.
It should be noted that the above counting scheme accounting for electroweak effects applies in a
small $\sim\SI{10}{\GeV}$ 
window around threshold, where the relation $v\sim \alpha_s\ll 1$ holds.

Above threshold, fixed-order full QCD corrections to the vector and
axial-vector current contributions to (on-shell) $\tT{}$ production have
been computed inclusively to \NthreeLO~\cite{0807.4173,0907.2120} and
differentially to NNLO~\cite{1410.3165,1610.07897}.
The off-shell process (with $\WbWb{}$ as the final state), which is also
defined below threshold, has been studied at NLO in \Rcites{Lei:2008ii,Liebler:2015ipp,1609.03390}.
Pure QCD fixed-order results without Coulomb
resummation are, however, only reliable in the relativistic continuum
sufficiently away from the $\tT{}$ threshold. In order to distinguish
the relativistic QCD fixed-order power counting, which refers to
powers of the strong coupling only, from the nonrelativistic power
counting, from here on we always supplement the relativistic counting
with the prefix ``QCD''.%
\footnote{We note that the contributions labeled ``QCD-LO" in this paper 
actually do not involve any strong interaction corrections.}

Despite their sophistication the current results for the top pair production are not yet suitable to address a number of key issues relevant for a fully realistic assessments of the top quark related measurements 
at a future lepton collider.
First of all, up to now a quantitative analysis addressing down to which c.m.\ energy 
threshold-resummed calculations can be neglected and pure fixed-order
continuum results can be used reliably has been missing.
Vice versa, it has not been shown how far one can trust a threshold-resummed
prediction when increasing the c.m.\ energy away from the nonrelativistic limit.
To approach these questions one can construct a \emph{matched}
computation for $\eeWbWb{}$ that 
combines the threshold-resummed and the fixed-order continuum
computations that is valid for 
all $\sqrt{s}$ and study its theoretical uncertainties. 
This is especially important for the proposed \SI{380}{\GeV} stage of the
\ac{CLIC} proposal, where one can probe the threshold region due to photon radiation
off the initial state, but also for the 350 GeV staging of the
International Linear Collider (ILC) and
the newly devised 350 GeV staging of CLIC.
The convolution of QCD threshold predictions with initial-state
radiation and realistic beam spectra have been already studied in
\Rcites{1303.3758,1603.04764}.
However, these results are only reliable as long as the QCD threshold 
predictions in the convolutions are employed in a small $\sim\SI{10}{\GeV}$ 
window around threshold, where the assumption $v\sim \alpha_s\ll 1$ holds.
Depending on the shape of the beam spectrum, this assumption can be 
questioned, and a fully matched prediction is therefore also quite desirable. 

The second issue concerns the fact that theoretical predictions that
provide a full description of the $\WbWb{}$ final state and in addition
account for the top pair threshold dynamics have not been provided up
to now. Currently theoretical examinations are known for the total
inclusive cross
section~\cite{Hoang:1998xf,hep-ph/9904468,Hoang:2000yr,Pineda:2006ri,hep-ph/0209340,1309.6323,1506.06864},  
the inclusive top three-momentum distribution~\cite{Jezabek:1992np,hep-ph/9904468},  
the top quark polarization~\cite{Harlander:1996vg,Jezabek:2000gr}
and for the dependence of the total cross section on cuts on the invariant mass $m_{bW}$ of the $b$-$W$ system arising from the top quark decay~\cite{Hoang:2008ud,1002.3223,1004.2188,1307.4337}. 
However, despite the fact that the $\tT{}$ measurements at threshold can be fairly inclusive, the physical final states are still the leptonic, semi-leptonic and hadronic decay
products.
Thus, any inclusive measurement suffers from systematic
uncertainties that result from extrapolating the measured
cross section in the fiducial phase space to the full phase space. Here it is also important to
have a full theoretical control of other $\WbWb{}$
production mechanisms such as single top production, which are characterized by
kinematical configurations that differ substantially from resonant $\tT{}$ production~\cite{1411.2355}. 
These types of systematic uncertainties have not been explored systematically and are not accounted for in current
experimental 
\cite{1303.3758,1603.04764,Simon:2016pwp} analyses.
They can only be addressed coherently by 
theoretical predictions that at the same time describe the full $\WbWb{}$ final state and account for the top pair threshold dynamics. 

In this paper, we present an approach which allows to address the issues above by providing a
matched computation for $\eeWbWb{}$ correct to QCD-NLO and including
threshold resummation with NLL precision. 
For observables that are inclusive concerning the top--anti-top
double-resonant portion of the $\WbWb{}$ phase space we achieve
QCD-NLO as well as NLL precision with respect to QCD effects. This
applies to the total cross section, but also includes the total cross
section with moderate acceptance cuts, e.g.\ related to the
reconstructed invariant mass of the top quarks or the typical event
selection procedures.   
To this end, we devise a master formula that matches the
nonrelativistic computation in the threshold region with the
relativistic fixed-order result in the continuum carefully avoiding any double counting of terms at QCD-NLO and NLL order.
We also maintain all irreducible backgrounds of $\WbWb{}$ and the
interference of resonant (i.e.\ involving a $\tT{}$ pair) and non-resonant
(i.e.\ not involving a $\tT{}$ pair) $W$-$b$ production in the 
threshold region, hereby reaching NNLO precision concerning these electroweak corrections in the threshold region at the differential level.
We construct a manifestly gauge-invariant result by applying an extended double-pole  approximation for the threshold-resummed top pair production form factors, which is also defined below the kinematical threshold.%
\footnote{First preliminary results of threshold resummation in
  \whizard{} have been presented in \Rcite{1411.7318}, which were,
  however, still based on a construction using ``signal diagrams'',
  i.e. full theory diagrams with a virtual $\tT{}$ pair that decays into
  the $\WbWb{}$ final state,  and therefore not manifestly gauge
  invariant.}
To achieve a QCD-NLO description of the top quark decay for these
threshold-resummed contributions, we include the corresponding QCD-NLO
corrections in the on-shell approximation.  

Concerning fully differential predictions in the threshold region our
approach is limited concerning the treatment of real and virtual ultrasoft gluon
radiation in the threshold region and concerning virtual
potential-type longitudinal gluon exchange involving the final state $b$
quarks. Both effects cancel at NLL order for the total inclusive cross
section~\cite{Fadin:1993dz,Melnikov:1993np} and also when acceptance
cuts are applied that do not resolve the top--anti-top double-resonant
portion of the $\WbWb{}$ phase space~\cite{1002.3223,1004.2188}.  
For many other types of distributions including top quark spin
measurements, however, they give additional non-trivial NLL
corrections which are typically at the level of up to 10\%,
see~e.g.\ \Rcite{Peter:1997rk}. 
A coherent treatment of these NLL effects in the context of our
matched setup is postponed to future work. 
Thus, at the fully differential level
our current results are strictly valid at LL order in the threshold region.
In any case, the first
gluon emission at QCD-NLO order  
is by construction fully accounted for in our matched prediction for all 
kinematic regions, and we expect that a significant part of the NLL corrections 
for fully differential observables are therefore included. We also 
note that in the context of this work we do not account for the summation of 
phase space logarithms coming from top--anti-top unstable particle 
effects~\cite{Hoang:2004tg} and for
Coulomb potential corrections due to photon exchange. The latter constitute the 
dominant electromagnetic corrections for the top--anti-top threshold
dynamics contributing at NLL order and can be trivially
implemented. The former come from  
phase space divergences in the nonrelativistic calculation and constitute NLL 
effects as well. Their treatment is also postponed to future work.    

With a completely differential and matched description for the threshold
region at hand, many kinematic observables and distributions in the threshold region can be
studied coherently for the first time. This may also allow to systematically access alternative 
methods to determine the top mass in the threshold region besides the
paradigmatic totally inclusive energy scan. Furthermore, it is possible to study observables in the intermediate region between threshold and high-energy continuum. In this paper, we will
discuss a small number of kinematic distributions at the exclusive level for the
matched setup as a proof of principle, but defer specific phenomenological studies for top threshold measurements to future publications. Here, we concentrate on the presentation and validation of our method for matching fixed-order QCD and resummed threshold corrections.

We note that apart from subtracting terms related to double counting
with respect to the QCD fixed-order and threshold-resummed
computations, an essential ingredient for a fully matched computation
that smoothly covers threshold, intermediate and continuum regions is
that the terms resummed in the nonrelativistic part are 
switched off away from the threshold region. 
This is necessary since the resummed terms determined in the
nonrelativistic expansion do not naturally transition into
relativistic expressions. Since there is no unique way to implement
the switch-off, the intermediate region between nonrelativistic and
relativistic domains carries an additional theoretical uncertainty,
which has to be estimated carefully. This uncertainty, however,
decreases with the order of the nonrelativistic and relativistic
expansions as long as both converge~\cite{Widlmatching}.

We also note that the computations presented in this work obviously do not have the
highest precision currently available for the inclusive total cross section concerning the QCD corrections in the threshold region.
On the other hand, our results are novel for the
phenomenologically more realistic differential final states and especially in the intermediate region between threshold and continuum.
It would be straightforward to augment our computations with K-factors such as $K^\T{NNLL}=\sigma^\T{NNLL}/\sigma^\T{NLL}$ to increase the precision for the inclusive total cross section. 

We also remark that there have been previous analyses that have
implemented nonrelativistic resummations within relativistic
calculations. In \Rcites{hep-ph/0504220,Farrell:2006xe} NLL
nonrelativistic resummations were embedded into a relativistic
calculation of the Higgs energy spectrum for $e^+e^-\to t\bar t H$ at
NLO. These results were inclusive with respect to the top quark
decays. In \Rcite{1602.00684} nonrelativistic resummation form factors
have been embedded into a factorized relativistic
tree-level computation for $e^+e^-\to \tT{}H \to \ell\bar{\ell}+X$ 
with the aim of examining the Higgs CP properties from the leptonic
angular correlations. Finally, in \Rcite{1007.0075}, the authors
incorporated $\tT{}$ threshold effects in a fully differential Monte
Carlo (MC) simulation at hadron colliders which combined NLO threshold
resummations with all resonant and non-resonant diagrams for $\WbWb{}$
production at tree-level by multiplying signal diagrams with
nonrelativistic form factors. Their approach is similar in scope to
ours as it provides a description in all kinematic regions. However,
they accounted for QCD-NLO relativistic corrections only through
K-factors in the $\tT{}$ invariant mass distribution (determined for
on-shell top quarks) rather than by including the full set of QCD-NLO
corrections for $\WbWb{}$ production. From a systematic point of view
their work represents a consistent leading order treatment. 

The outline of this paper is as follows.
In \cref{s:resummation}, we review the NRQCD derivation of the S- and P-wave 
form factors that are the key ingredients in our implementation of the threshold resummation.
The gauge-invariant embedding of these results within the relativistic setting in \whz{}
using an extended double-pole approximation for signal diagrams 
is discussed in \cref{s:implementation_in_whizard}.
In \cref{s:validation}, we verify that this implementation works as
expected by comparing to known results.
In \cref{s:matching}  we  discuss the necessary ingredients for the matching 
of the NLL resummations at threshold and the full QCD-NLO relativistic results
for $\WbWb{}$ production,
and we study inclusive and differential results in
\cref{s:inclusive_results} and \cref{s:differential_results},
respectively.
We summarize our findings and give an outlook to further developments and opportunities
in \cref{s:summary}.


\section{Threshold resummation}
\label{s:resummation}

The basis of our approach to implement the threshold resummation in a
MC event generator are the nonrelativistic S- and P-wave form
factors. They describe resonant $\tT{}$ production through the
vector and axial-vector currents in the presence of the bound state
effects. The latter are related to the  resummation of Coulomb
singular $(\alpha_s/v)^n$ terms as well as the resummation of the
velocity logarithms $\propto(\alpha_s \ln v)^n$ determined within the
NRQCD framework. As the form factors are fully differential concerning
the bound state dynamics, are by themselves gauge invariant and do not
depend on issues related to the final state particles originating from
the decays, they represent suitable building blocks for a MC
implementation.

In this section we review some details on the derivation of these form 
factors at NLL order using the
vNRQCD~\cite{hep-ph/9910209,hep-ph/9912226,hep-ph/0209340} framework. 
This forms the basis for
the theoretical setup used for their implementation in \whz{}.
We also briefly review the standard calculation of the total inclusive
threshold cross section using the optical theorem for later reference.
For brevity, throughout this section we denote the top pole mass by $m
\equiv m_t$.  

\subsection{Form factors in vNRQCD}
\label{ss:vnrqcd}
The effective field theory (EFT) vNRQCD involves soft, ultrasoft and potential degrees of freedom
whose dynamical effects are correlated via the nonrelativistic heavy quark dispersion relation $E\sim \bp^2/m$ 
which relates the ultrasoft and soft energy and momentum scales. 
The typical four-momenta $(q_0, \bq)$ of the soft, ultrasoft and potential degrees of freedom
scale like $(mv, mv)$ , $(mv^2 ,mv^2)$ and $(mv^2, mv)$, respectively.
Hard fluctuations with momenta $(q_0, \bq) \sim (m, m)$ and non-resonant modes, e.g. 
with momenta $(q_0, \bq) \sim (m v, m v^2)$, are integrated out.
The vNRQCD Lagrangian~\cite{hep-ph/9910209} reads
\begin{align}
\mathcal{L}(x) \supset \sum_{\bp}  \bigg[ \psi^\dagger_{\bp} \, 
\bigg(\I \del^0 - \frac{\bp^2}{2m} \bigg)
\psi_{\bp}  + (\psi_\bp \!\to\! \chi_\bp) \bigg]
- \sum_{\bp,\, \bpp}
\tilde V_c(\bp,\bpp) \;
\psi^\dagger_{\bpp}\, \psi_{\bp} \;
\chi^\dagger_{-\bpp}  \,\chi_{-\bp}
+\mathcal{L}_\mathrm{soft}
\co
\label{eq:LvNRQCD}
\end{align}
where
\begin{align}
 \tilde V_c(\bp,\bpp) &= \frac{\mathcal{V}_c^{(s)}}{\big(\bp - \bpp \big)^2}
\end{align}
is the (color-singlet) Coulomb potential and we only display the terms 
relevant to describe the dynamics of a
heavy quark pair in the color singlet configuration at NLO. We have suppressed 
higher order terms in the $v$ expansion as well as color indices. The vNRQCD 
field operators $\psi_{\bp}(x)$ and
$\chi_{\bp}(x)$ annihilate the heavy quark and anti-quark,
respectively. The (discrete) label $\bp$ denotes their soft
three-momentum, while their position argument $x$ (which is suppressed in Eq.~(\ref{eq:LvNRQCD})) is the Fourier conjugate to
their ultrasoft four-momentum components. All momenta are defined in
the center-of-mass frame of the heavy quark pair. 

At NLL (and NLO) the effects of soft interactions encoded in 
$\mathcal{L}_\mathrm{soft}$ can  effectively be accounted for by adding a
term~\cite{Fischler:1977yf} to the coefficient of the Coulomb
potential operator in \cref{eq:LvNRQCD}: 
\begin{align}
\mathcal{V}_c^{(s)}(\mu_\T{S}) \to \mathcal{V}_c^{(s)}(\mu_\T{S})
- C_F \big[\alpha_s(\mu_\T{S}) \big]^2 
\Bigg( - \beta_0 \ln \Bigg[ \frac{\big(\bp - \bpp \big)^2}{ \mu_\T{S}^2} \Bigg]
+\frac{31}{9} C_A - \frac{20}{9} n_f T_F
\Bigg)
\co
\label{eq:coulomb-pot}
\end{align}
where $\mathcal{V}_c^{(s)}(\mu_\T{S}) = -4 \pi C_F \alpha_s(\mu_\T{S})$ through 
NLL. Here and in the following, $\mu_\T{S}$ denotes the \emph{soft
renormalization scale}. For unstable top quarks the inclusive effects
of the decay width are taken into account by adding  
\begin{align}
 \sum_{\bp} \psi_{\bp}^\dagger \frac \I 2 \Gamma_{t} \psi_{\bp} +
 \sum_{\bp} \chi_{\bp}^\dagger \frac \I 2 \Gamma_{t} \chi_{\bp}
\end{align}
to the Lagrangian in \cref{eq:LvNRQCD}, which is sufficient to determine the form factors at NLO. 
In the EFT, $\Gamma_t$ is formally an input variable and can
be set to a measured value or to the width computed in the \ac{SM} at the desired order.
In our MC implementation, the width terms contribute to the virtual
unstable top and anti-top quark lines contained in the resummed diagrams,
while we describe the top decay fully 
differentially as discussed in \cref{s:implementation_in_whizard}.
For consistency it is therefore important that
the approximation and the parameters used for the width $\Gamma_t$ 
are equivalent to those used for the differential computation of the decays.  
In other words, $\Gamma_t$ must agree with the total
width computed with our program for the process $t \to W^+b$ to obtain the correct 
normalization of the total cross section.

Besides the operators in the vNRQCD Lagrangian we require
nonrelativistic currents that produce the heavy quark pair. They
couple to the virtual photon or $Z$ boson in the s-channel $\tT{}$
production process and are considered external from the EFT point of
view. For the purpose of this paper we need the leading vector
(S-wave) and axial-vector (P-wave) currents in the $v$ expansion. 
They are defined by
\begin{align}
\O{p}{1} & =  \psi_{\bp}^\dagger \; \bsig(\I\sigma_2)\; {\chi_{-\bp}^*}
\co
\label{eq:CurrentOp1}  \\
\O{p}{3} & =  \frac{-\I}{2m}\, \psi_{\bp}^\dagger \;
[\,\bsig,\bsig\cdot{\bp}\,]\,(\I\sigma_2)\;
{\chi_{-\bp}^*} \co
\label{eq:CurrentOp3}
\end{align}
respectively.

\begin{figure}[t]
\centering
\includegraphics[width=\textwidth]{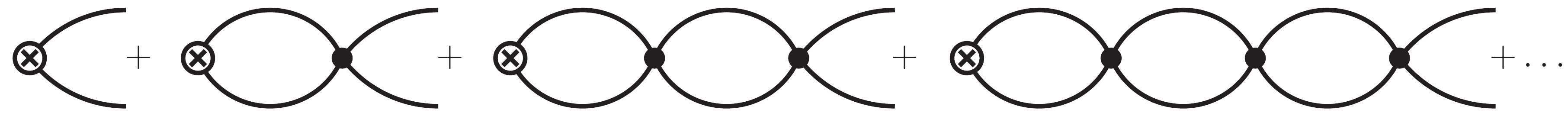}
\caption{Infinite sum of vNRQCD Feynman diagrams that contribute to the 
amputated S- or P-wave vertex functions in \eqs{Svertfct}{Pvertfct} depending 
on which of the production currents in \eqs{CurrentOp1}{CurrentOp3} is 
used for the crossed vertex. The black dots symbolize Coulomb potential 
interactions according to \cref{eq:LvNRQCD}. The first tree-level vertex 
diagram in the sum corresponds to the $1$ in \eqs{LSeqS}{LSeqP}.
For the form factors in \eqs{FF_vec}{FF_ax} all vertex diagrams are 
multiplied with the correponding current coefficient $c_1$ (S-wave) or $c_3$ 
(P-wave).
\label{fig:VertexFct}
}
\end{figure}

We can now form the operator matrix elements
\begin{align}
{\tilde G}_1 (E,\bp,\bpp) &= \frac{\I}{6 N_c} \int \! \mathrm{d}t \; e^{\I t 
E} 
\, \big\langle 0 \big| 
T\, \O{p}{1} (t) \, \O{p^\prime}{1}^\dagger (0) 
\big| 0 \big\rangle \co  \label{eq:G1}\\
{\tilde G}_3 (E,\bp,\bpp) &= \frac{\I (d-1)}{12 N_c} \frac{m^2}{\bp \cdot 
\bpp}\int \! \mathrm{d}t \; e^{\I t E} \, \big\langle 0 \big| 
T\, \O{p}{3} (t) \, \O{p^\prime}{3}^\dagger (0)
\big| 0 \big\rangle\co
\label{eq:G3}
\end{align}
where the first prefactor comes from the average over color and spin and 
where the soft momentum variables $\bp$ and $\bpp$ are now continuous
by including the residual (ultrasoft) three-momenta of the quark fields.
The functions ${\tilde G}_{1/3}$ in \eqs{G1}{G3} represent 
the 
S- and P-wave Green functions, respectively, of the Schr\"odinger 
equation for the quark--anti-quark system in 
momentum space~\cite{hep-ph/9904468,hep-ph/0107144}.
They describe the propagation of a heavy S/P-wave quark--anti-quark bound state 
with nonrelativistic energy 
\begin{align}
E \equiv \sqrts-2m \po
\end{align}
The relative three-momentum between the quarks is $2 \bpp$ in the initial and 
$2\bp$ in the final state.
Integrating over $\bpp$ is equivalent to producing the heavy quark and anti-quark 
at the same space-time point (at zero distance), i.e. from a local current as 
illustrated in \cref{fig:VertexFct}.
Accordingly we define the amputated S/P-wave vertex functions
\begin{align} 
\bar S(E,\bp) &=  \big[ G^f(E,\bp) \big]^{-1} \int\! \frac{\mathrm{d}^3\bpp}{(2 
\pi)^3} \,{\tilde G}_1 (E,\bp,\bpp)
\co
\label{eq:Svertfct}
\\
\bar P(E,\bp) &=  \big[ G^f(E,\bp) \big]^{-1} \int\! \frac{\mathrm{d}^3\bpp}{(2 
\pi)^3} \,{\tilde G}_3 (E,\bp,\bpp)
\co
\label{eq:Pvertfct}
\end{align}
with the free top--anti-top propagator (i.e.\ without potential
interactions, $V_c=0$) given by 
\begin{align}
G^f(E,\bk) &= \frac{- 1}{ E -\frac{\bk^2}{m} +\I \Gamma_t } \po
\end{align}
The inverse of $G^f(E,\bk)$  in the definition of the amputated vertex
functions removes the contribution of the resonant external legs of
the diagrams in \cref{fig:VertexFct}, such that $\bar{S}(E,\bp) = 1 +
\ord{\ALstrong}$ and $\bar{P}(E,\bp) = 1 + \ord{\ALstrong}$,
respectively.%
\footnote{Here we have introduced the notation $\bar S$ and $\bar P$
  for the amputated vertex functions to distinguish them from the
  non-amputated vertex functions $S$ and $P$ of \Rcite{hep-ph/9904468}
  that are defined without the factor of $\big[ G^f(E,\bk)
    \big]^{-1}$.} 
The vertex functions in \eqs{Svertfct}{Pvertfct} are key ingredients
to our vector and axial-vector form factors and we will need them to
NLO of their nonrelativistic expansion. Beyond that order, at NNLO and
beyond, one also has to take into account subleading currents,
e.g.\ describing D-wave or $v^2$-suppressed S-wave  effects.

The vNRQCD equations of motions for the amputated vertex functions
take the form of Lippmann-Schwinger equations~\cite{hep-ph/9904468}:
\begin{align}
 \bar S(E,\bk) &= 1 \;-\;  \int \frac{\mathrm{d}^3\bpp}{(2 \pi)^3}
 \;\tilde V_c(\bk,\bpp)\;G^f(E,\bpp)\; \bar S(E,\bpp)\co
\label{eq:LSeqS} 
\\
 \bar P(E,\bk) &= 1 \;-\;  \int \frac{\mathrm{d}^3\bpp}{(2 \pi)^3}
 \;\frac{\bk \cdot \bpp}{\bk^2} \;\tilde V_c(\bk,\bpp)\;G^f(E,\bpp)\; \bar 
P(E,\bpp)\po
\label{eq:LSeqP}
\end{align}
Solving these equations resums the contributions from the Coulomb 
interaction diagrams and the associated Coulomb singularities as illustrated in 
\cref{fig:VertexFct} to all orders.
The results at LO in the nonrelativistic counting according to \cref{eq:Rstruc} 
are given below in analytic form, see \eqs{FF_vec_LL}{FF_ax_LL}. 
For the NLO solutions of the Lippmann-Schwinger equation we have to account for the NLO corrections to the Coulomb potential shown in \cref{eq:coulomb-pot}. To the best of our knowledge no analytical results for the exact solutions are available in the literature at NLO and beyond.

At NLO we use the \toppik 
code~\cite{Jezabek:1992np,Harlander:1994ac,hep-ph/9904468} to obtain exact 
numerical solutions.
We note that the exact solutions also contain terms from beyond NLO from multiple iterations of 
the $\ord{\ALstrong^2}$ corrections in \cref{eq:coulomb-pot}. 
Due to the simple form of the Coulomb potential (which holds at all orders) the 
angular integrations in \eqs{LSeqS}{LSeqP} can be separated. 
The three-dimensional integral equations thus effectively become 
one-dimensional, with an additional non-trivial angular integration remaining 
in the kernel, which can be performed analytically. 
For the numerical solution, the one-dimensional integral equations are then
written in discretized form as sums over a suitably chosen grid of
momenta. This transforms the two integral equations into two systems of
linear equations which can be solved directly by matrix inversion
methods. The manifest (but integrable) singularities at momenta $\bpp=\bk$ in 
\eqs{LSeqS}{LSeqP} are avoided by subtracting and adding back $\bar 
S(E,\bk)$, and respectively  $\bar P(E,\bk)$, under the integral. The resulting 
additional
integrals on the RHS can be calculated analytically, while the
subtraction removes the manifest singularities in the integrand/summand. 

In \toppik, to improve efficiency and accuracy, momentum grids have been chosen 
based on Gauss-Legendre integration methods. To achieve the required accuracy 
($\lesssim 1\%$), a grid of typically 600 points is then
sufficient. It should be noted that this method is extremely robust in
the case of the S-wave, where, even in the case of the long-range
Coulomb interaction, the convergence properties of $\bar S(E,\bk)$ guarantee
a finite integral of the momentum distribution $|\bar S(E,\bk)|^2$ even for
unphysically large, nonrelativistic momenta up to infinity. In
contrast, the corresponding integral over the P-wave momentum distribution $|\bar P(E,\bk)|^2$ is ill-defined
without a cut-off. While the solution of \cref{eq:LSeqP} as described is 
still possible, the worse convergence behavior of the P-wave as compared to
the S-wave leads to numerical instabilities and hence a somewhat
limited accuracy of the method as implemented in TOPPIK.
The problems arise from the fact that in order to achieve a good accuracy for 
the un-regularized vertex function, very large momenta must be sampled in
the grid. These in turn lead to instabilities at very small
momenta. In practice, as the behavior at small momenta is known,
this problem can be fixed easily and the method remains powerful.
As a cross check of the numerical solutions by \toppik and to estimate their 
accuracy, we have inserted the solutions into the RHSs of the
Lippmann-Schwinger equations of \eqs{LSeqS}{LSeqP} and compared the
result to the original results. We found that for our numerical P-wave
vertex function, both sides of \cref{eq:LSeqP} agree within $\lesssim
1\%$ for all relevant energy values. This numerical precision is
sufficient for our purposes. For the S-wave the precision is better by
roughly an order of magnitude.

In the vNRQCD framework the nonrelativistic vector and axial-vector current 
operators in 
\eqs{CurrentOp1}{CurrentOp3} are multiplied with the Wilson coefficients 
$c_1(h,\nu)$ and $c_3(h,\nu)$, respectively, which account for the summation of the
velocity logarithms $\propto(\alpha_s \ln v)^n$.
They depend on the matching 
parameter $h$~\cite{1309.6323} and the vNRQCD renormalization parameter $\nu$, 
the so-called subtraction velocity~\cite{hep-ph/9910209}, see 
\cref{ss:resummation_and_scales}.
The matching of the EFT currents to the full theory vector and axial-vector 
currents is performed at the (hard) matching scale $\mu_\T{H} \equiv h m$, i.e. 
at $\nu=1$ and for a choice of $h$ close to unity. This gives at one loop
\begin{align}
c_1(h,1) &= 1 - 2 C_F\, \frac{\alpha_s(\mu_\T{H})}{\pi} + \ord{\alpha_s^2}\co
\label{eq:c1match}
\\
c_3(h,1) &= 1 - C_F\, \frac{\alpha_s(\mu_\T{H})}{\pi} + \ord{\alpha_s^2}\po
\label{eq:c3match}
\end{align}
These matching coefficients are by now in fact already known to 
$\ord{\alpha_s^3}$~\cite{Marquard:2014pea} and 
$\ord{\alpha_s^2}$~\cite{Beneke:2013kia}, respectively.
At LL order the anomalous dimensions of the currents vanish, such that their RG running at this order is trivial~\cite{hep-ph/9910209}: 
\begin{align}
c_1^\mathrm{LL}(h,\nu) 
= c_3^\mathrm{LL}(h,\nu) = 1 \po
\end{align}
The NLL evolution of $c_1$ has been computed in 
\Rcites{Pineda:2001et,hep-ph/0209340}. Also the (dominant) NNLL contributions 
are 
known~\cite{Hoang:2006ht,Hoang:2011gy}, see also \Rcite{Pineda:2011aw}.
The NLL result for $c_3$ can be found in \Rcite{hep-ph/0609151}.
For completeness we quote the NLL expressions for $c_{1,3}$ in 
\cref{app:NLLc1c3}.
Setting $\nu \sim |v|$, with
\begin{align}
  v = \sqrt{\frac{\sqrts{} - 2 \mpole + \I \Gamma_{t}}\mpole}
  \label{eq:velocity}
\end{align}
being the effective velocity of the heavy quarks, resums large velocity logarithms $\sim (\alpha_s\ln 
v)^n$ from the production process into the Wilson coefficients 
$c_{1,3}(h,\nu)$ and the soft coupling $\alpha_s(\mu_\T{S}\equiv h m \nu)$ in 
the vertex functions.%
\footnote{The precise definition of the mass parameter $m$ in the renormalization scales $\mu_\T{H} = hm$ and $\mu_\T{S}=hm\nu$ is not important physically since $h$ and $\nu$ are varied anyway. In 
\cref{ss:resummation_and_scales} we will use the 1S mass $\mOneS$ 
instead of the pole mass $m$ in the parameterization of $\mu_\T{H}$ and 
$\mu_\T{S}$ for convenience.}

With the ingredients discussed before we can now define the vector 
and axial-vector form factors that we will use to implement the 
threshold (Coulomb and log) resummations  
in \whz{} (see \cref{s:implementation_in_whizard}):
\begin{align}
 F_{V}(E,|\bp|,h,\nu) &= c_1(h,\nu) \; \bar S(E,\bp,h,\nu) \co
 \label{eq:FF_vec}
 \\
 F_{A}(E,|\bp|,h,\nu) &= c_3(h,\nu) \; \bar P(E,\bp,h,\nu)
 \po
 \label{eq:FF_ax}
\end{align}
The superscripts $V$, $A$ stand for the vector (S-wave) and axial-vector 
(P-wave) current, respectively, and we have listed the renormalization
scaling parameters $h$ and $\nu$ as arguments to be explicit.%
\footnote{
At higher orders (beyond N$^3$LO) there will also be 
contributions associated with other partial waves.}
We stress that the only kinematic variables the form factors depend on
are the modulus of the heavy quark three-momentum $|\bp|$ and the
total nonrelativistic energy of the quark-pair $E$.

The LL expressions read
\begin{align}
 F_{V}^\mathrm{LL}&(E,|\bp|,h,\nu) = \nonumber \\ & 
  \frac{ m^2 v^2 \!- \bp^2}{4 mv |\bp| (1\!-\!\lambda)}
 \Bigg[{}_2F_1\bigg(2,1;2\!-\!\lambda; \frac{ mv\!+\! |\bp|}{ 2 mv} \bigg)
-\, {}_2F_1\bigg(2,1;2\!-\!\lambda; \frac{ mv\!-\!|\bp|}{2 mv} \bigg)\Bigg]\co
\label{eq:FF_vec_LL}
\\[2 ex]
F_{A}^\mathrm{LL}&(E,|\bp|,h,\nu) = \nonumber \\ & 
\frac{m v (m^2 v^2 \!- \bp^2)}{2 |\bp|^3} \Bigg\{ \!
\Bigg[{}_2F_1^{(1,0,0,0)}\!\bigg(\!0,3;2\!-\!\lambda;\frac{ mv\!-\!|\bp|}{2 
mv}\bigg)- 
{}_2F_1^{(1,0,0,0)}\!\bigg(\!0,3; 2\!-\!\lambda;\frac{mv\!+\!|\bp|}{2 
mv}\bigg)\Bigg]
\nonumber\\
&
+(1\!-\!\lambda) 
\Bigg[ 
{}_2F_1^{(1,0,0,0)}\!\bigg(\! 0,2;1\!-\!\lambda;\frac{mv\!+\!|\bp|}{2 
mv}\bigg)-{}_2F_1^{
(1,0,0,0) }
\!\bigg(\!0,2;1\!-\!\lambda;\frac{mv\!-\!|\bp|}{ 2 mv}\bigg)\Bigg] 
-   \frac{2 |\bp|}{mv} \Bigg\}\co
\label{eq:FF_ax_LL}
\end{align}
where $\lambda\equiv \I C_F\, \alpha_s(\mu_\T{S})\, /(2 v)$, 
${}_2F_1(a,b;c;z)$ denotes the ordinary hypergeometric function and 
${}_2F_1^{(1,0,0,0)}\!(a,b;c;z) \equiv \frac{\mathrm{d}}{\mathrm{d} a}\, 
{}_2F_1(a,b;c;z)$.
For the NLL form factors we use the NLL current coefficients $c_{1,3}$ as given 
in \cref{app:NLLc1c3} and the NLO vertex functions 
from \toppik{} with NLL running for $\ALstrong(\mu_\T{S})$.

For later reference we also expand the NLL form factors to 
first order in $\alpha_s$:
\begin{align}
  F^\mathrm{exp}_{V,\mathrm{NLL}}\of{\ALstrong} &= 1 +
  \I\,  \ALstrong \CF{} \frac{m}{2 |\bp|} \log{\frac{mv + |\bp|}{mv - |\bp|}} 
  - 2\CF{} \frac{\ALstrong}{\pi}
\co
  \label{eq:FFexpNLL} 
\\
  F^\mathrm{exp}_{A,\mathrm{NLL}}\of{\ALstrong} &= 1 
  -\ALstrong C_F \frac{m }{2  |\bp|} \Bigg(  \frac{ \I m v}{|\bp|}
 +\frac{m^2 v^2+\bp^2}{4 \pi \bp^2}  \Big[\log ^2(-m v-|\bp|)-\log ^2(-m 
v+|\bp|)
  \nonumber\\
  &\qquad +\log ^2(m v-|\bp|)-\log ^2(m v+|\bp|)\Big] \Bigg)
  - \CF{} \frac{\ALstrong}{\pi} \po
  \label{eq:FFAexpNLL}
\end{align}
The first two terms in these expressions are the expansions of 
\eqs{FF_vec_LL}{FF_ax_LL} and the last terms come from the matching 
coefficients in \eqs{c1match}{c3match}, respectively.

\subsection{Inclusive cross sections}
\label{ss:sigmaincl}
For validation purposes we will often refer to the direct analytic computation 
of the $\tT{}$ threshold production cross section in NRQCD via the optical 
theorem and compare to its results.
These predictions are only valid in a small $\sqrts$ range of a few GeV around 
the threshold, but have already reached 
NNLL~\cite{1309.6323} (see also \Rcites{hep-ph/0107144,Pineda:2006ri}) and N$^3$LO~\cite{1506.06864} 
level.
Here we shall briefly review the vNRQCD NNLL calculation following 
\Rcites{hep-ph/0107144,1309.6323}.

The total inclusive cross section for $\tT{}$ production can be written as
\begin{align}
  {\sigma_\T{tot}(\ee\to\gamma^*/Z^*\to \tT{})}
    = \frac{4\pi\alpha_\mathrm{em}^2}{3s}\left(f_v R_v + f_a R_a\right) \co 
    \label{eq:sigmatot}
\end{align}
where the prefactors $f_v$ and $f_a$ account for tree-level $\gamma$
and $Z$ exchange and are given e.g. in ~\Rcite{hep-ph/0107144}.
In the full theory (SM), the  vector/axial-vector R-ratios can be computed  via
\begin{align}
  R^{v/a} = \frac{4\pi}{s} \T{Im}\left[-\I\Int{\D^4 x}\E^{\I q x}
    \braketop{0}{\T{T}j_\mu^{v/a}(x){j^{v/a}}^\mu(0)}{0}\right]\co
\label{eq:Rratios}
\end{align}
with $q=(\sqrt{s},0)$ and $j_\mu^{v/a}$ being the vector/axial-vector currents 
that produce a top--anti-top pair.
In the effective theory these currents are replaced by
their nonrelativistic counterparts and up to NNLL we  have
\begin{align}
  R^v &= \frac{4\pi}{s} \T{Im}
  \left[c_1^2 \mathcal{A}_1 + 2 c_1 c_2 \mathcal A_2 \right]\co\qquad
  R^a = \frac{4\pi}{s} \T{Im} \left[c_3^2 \mathcal{A}_3\right]\po
\label{eq:Rr}
\end{align}
Here the effective current correlators are defined by
\begin{align}
{\cal A}_i(v, m, \nu,h) &= \I\,
\sum\limits_{\bp,\bpp}
\int\! d^4x\: e^{\I E x_0}\:
\Big\langle\,0\,\Big|\, T\, \bO_{\bp,i}(x) \, \bO_{\bpp,i}^\dagger(0)
\Big|\,0\,\Big\rangle \co
\label{A1}
\end{align}
where the top-pair is produced and annihilated both at zero distance.
Taking the imaginary part in \cref{eq:Rr} corresponds to cutting through these 
zero-distance correlators in all possible ways.
The currents operators $\O{p}{1}$ and $\O{p}{3}$ are given in 
\eqs{CurrentOp1}{CurrentOp3} and $\O{p}{2}$ is the subleading S-wave current 
suppressed by two powers of $v$ as given in \Rcite{hep-ph/0107144}.
The calculation of the ${\cal A}_i$ correlators is detailed in 
\Rcites{hep-ph/0107144,Hoang:2003ns,hep-ph/0209340} and based on 
\toppik~\cite{hep-ph/9904468} as far as the (Coulomb) contributions to ${\cal 
A}_1$ at leading order in $v$ are concerned.
For NNLL precision of the total cross section the NNLL expression for the 
current coefficient $c_1$~\cite{Hoang:2011gy} and the LL expressions for $c_2$ 
and $c_3$~\cite{hep-ph/0107144} are required.

In \Rcite{1002.3223} this approach to compute the total cross section was 
extended to allow also for (moderate) cuts on the reconstructed top invariant 
masses. We will validate our threshold resummation implementation in \whz{} 
against such inclusive threshold predictions with invariant mass cuts in 
\cref{s:validation}.
Concretely, we compute ``analytic''%
\footnote{We use the term ``analytic'' here to distinguish the cross section 
from the MC generated one. Nevertheless the integrations in 
\cref{eq:sigmaCut} are in practice carried out numerically.}
inclusive cross 
sections with top invariant mass cuts at (N)LL using the formula
\begin{align}
\sigma_\Lambda^\mathrm{(N)LL} &= \frac{4\pi\alpha_\mathrm{em}^2}{3s} f_v
 \frac{6 N_c m^3 \Gamma_{t}^2}{\pi^2 s}
\int\limits_{\Delta(\Lambda)} \!\!\! \mathrm{d}t_1 \mathrm{d}t_2 \,
\frac{\sqrt{m E - \frac12 (t_1 + t_2)}}{\big(t_1^2 + m^2 \Gamma_{t}^2 \big)
\big(t_2^2 + m^2 \Gamma_{t}^2 \big)} 
\nonumber \\
&\quad \times
\Big|F_V^\mathrm{(N)LL}\Big( E, \sqrt{m E - {\textstyle \frac12} (t_1 + t_2)} 
,h,\nu \Big) 
\Big|^2
\po
\label{eq:sigmaCut}
\end{align}
As for the total inclusive cross section in \cref{eq:sigmatot} the axial-vector 
current (P-wave) first starts to contribute at NNLL. The form factor $F_A$ is 
therefore not required here. For our validation purposes, we only
consider the top decay at LO, i.e. $\Gamma_{t}= \Gamma_{t}^\mathrm{LO}$.
The integration in \cref{eq:sigmaCut} is over the 
\emph{nonrelativistic} invariant mass variables
\begin{align}
t_{1,2} = 2 m \bigg(E_{1,2} - \frac{\bp^2}{2 m} \bigg) \co
\end{align}
which represent the nonrelativistic expansion of the top and anti-top 
off-shellness variables $p^2_{1,2}-m^2$.
Here, $p_{1,2}$, $E_{1,2}$ and $\pm\bp$ are the 
four-momenta, the kinetic energies and three-momenta of the top and the 
anti-top, respectively. In general we have $E_{1,2}\neq \frac{\bp^2}{2 m}$ 
since the resonant top and anti-top quarks can be off-shell.
The integration region in $(t_1,t_2)$-space is given by
\begin{align}
\label{eq:t1t2limits}
\Delta (\Lambda) = \Big\{(t_1,t_2) \in \mathbb{R}^2:
\Big(|t_{1,2}| < \Lambda^2 \Big) \wedge 
\Big(m E - {\textstyle \frac12} (t_1+t_2) > 0 \Big) \Big\}\po
\end{align}
The first inequality in \cref{eq:t1t2limits} defines the phase space cut and 
the second one represents a kinematic constraint.
In order to relate the $\Lambda$-cut on $t_{1,2}$ to the (usual) cut
on the relativistic invariant masses,
\begin{align}
(\mOneS -\DeltaM) &  \,\le\, \sqrt{p^2_{1,2}} \,\le\, (\mOneS +\DeltaM) 
\co
\label{eq:lambdac}
\end{align}
as implemented for the validation of our \whz{} MC result in 
\cref{s:validation}, 
we use the approximation~\cite{1002.3223}
\begin{align}
 \Lambda^2 &= 2 \mOneS  \DeltaM - \frac{3}{4} \DeltaM^2 + \ldots
 \po
 \label{eq:LambdaApprox}
\end{align}
The ellipses stand for terms suppressed by additional powers of
$\DeltaM/\mOneS$ and $\mOneS v^2/ \DeltaM$, while the mass parameter
$\mOneS \sim m$ is defined in the next section. 


\subsection{Top mass parameter}
\label{ss:mass}
It is well known that the pole mass parameter $m_t$ in QCD is plagued by an 
intrinsic renormalon ambiguity of $\ord{\Lambda_\mathrm{QCD}}$, 
see~\Rcite{Beneke:2016cbu,Hoang:2017btd}.
In order to avoid an unnecessary theory error one should therefore use a 
suitable renormalon-free short-distance mass parameter as input in calculations 
that strongly depend on the heavy quark mass.
This is particularly important for $\tT$ threshold production, where the shape 
of the cross section (peak position), is very sensitive to the top mass.

In this work we employ the 1S top quark mass scheme~\cite{hep-ph/9904468}, 
where the top mass is defined as half of the mass of the fictitious $^3S_1$ 
toponium ground state for stable top quarks.
The pole mass can then be expressed order by order as a function of the 1S mass 
parameter $\mOneS$.
At (N)LL + QCD-(N)LO the relation we need is~\cite{hep-ph/0107144}
\begin{align}
\label{eq:mPoleRelation}
\mpole\of{\mOneS{}} &= \mOneS{} \left(1 + \DeltaMM_\mathrm{(N)LL} 
\big[\ALstrong(\mu_\T{S})\big] \right)\co
\end{align}
with
\begin{align}
\DeltaMM{}_\T{LL} [\ALstrong] &= \frac {C_F^2 \,\ALstrong^2}8 \co \\
  \DeltaMM{}_\T{NLL} [\ALstrong]  &= \DeltaMM{}_\T{LL} [\ALstrong]  +
      \frac{C_F^2 \, \ALstrong^3}{8\pi} \left\{
      \beta_0 \bigg[1 + \log \bigg( \frac{h \, \nu}{C_F \ALstrong} 
\bigg) \bigg] + \frac{a_1}2
    \right\}\co
\end{align}
and
\begin{align}
 \beta_0 = \frac{11}{3} C_A - \frac{4}{3} n_f T_F \co \qquad
 a_1 = \frac{31}{9}C_A - \frac{20}{9} n_f T_F\po
 \label{eq:b0a1}
\end{align}
For our (SM) process the color constants are $C_A=3$, $C_F=4/3$, $T_F=1/2$ and 
we use $n_f=5$ for the number of active fermion flavors throughout this work.

We note that there is a mismatch in the order counting for
short-distance masses suitable for the top--anti-top threshold and in
the high energy continuum. This can be seen in \cref{eq:mPoleRelation}
for the definition of the 1S mass, where $\Delta M$ starts with
$\ord{\ALstrong^2}$. This is in contrast to the definition of the
$\overline{\mathrm{MS}}$ mass which is suitable at high energies and starts
to differ from the pole mass at $\ord{\alpha_s}$. So, using the 1S
mass in the continuum at very high energies far above threshold is not
appropriate. This problem can be resolved~\cite{Widlmatching} by using
a subtraction-scale dependent short-distance mass that interpolates
between relativistic and nonrelativistic counting such as the MSR
mass~\cite{Hoang:2008yj,Hoang:2017suc}. A full implementation of this
approach in our MC framework is postponed to future work. 
\subsection{Renormalization and matching scales}
\label{ss:resummation_and_scales}
%

As outlined in \cref{ss:vnrqcd} the vNRQCD expressions for the form factors in 
\eqs{FF_vec}{FF_ax} depend on the hard matching scale $\mu_\T{H}$ and the soft 
and ultrasoft renormalization scales $\mu_\T{S}$ and $\mu_\T{US}$, 
respectively.%
\footnote{The ultrasoft scale $\mu_\T{US}$ first enters at NLL through the 
current coefficients $c_{1,3}$, see \cref{app:NLLc1c3}.}
In \cref{ss:vnrqcd} we have already introduced the matching parameter $h$ and 
the subtraction velocity $\nu$.
The three matching/renormalization scales can be  
parametrized by the terms $h$ and $\nu$~\cite{1309.6323}, respecting the 
natural correlation between soft and ultrasoft scales:
\begin{align}
  \mu_\T{H} = h \mOneS{} \co \quad
  \mu_\T{S} = h \nu \mOneS{}  \co \quad 
  \mu_\T{US} = h \nu^2 \mOneS{}  \po
\label{eq:NRQCDmus}
\end{align}
By setting $h\sim 1$ and $\nu \sim |v|$ we make sure that large logs 
$\sim (\ln v)^n$ are resummed in the (N)LL form factors.
Following~\cite{1309.6323}, we define the (energy-dependent) default 
value for $\nu$:
\begin{align}
\nustar\of{\sqrts} = 0.05 +
      \abs{\sqrt{\frac{\sqrts{} - 2 \mOneS{} + \I \Gamma_{t}^*}\mOneS{}}}\co
\label{eq:nustar}
\end{align}
with $\Gamma_{t}^* \equiv \Gamma_{t}^\T{(N)LO}\ofIT{\mpole=\mOneS}$.
In \Rcite{1309.6323}, a detailed study of uncertainties of the total cross 
section (with invariant mass cuts) at (N)NLL has been performed.
In the present work, we will adopt the same conventions for the combined scale 
variations. These will play a major role in assessing the theoretical 
uncertainties of our results, see \cref{ss:theoretical_uncertainties}. 
To this end, we define the renormalization parameter $f$ by 
\begin{align}
\nu = f \nustar \po
\end{align}
Scale variations are then performed in the $h$-$f$ plane around the 
default values $h=f=1$ within the boundaries set by
\begin{align}
1/2\le h f^2 \le 2\co \quad 1/2<h<2\po
\label{eq:hfvar}
\end{align}
The $h$ variation is equivalent to a correlated variation of all three scales, 
while $f$ variation only affects the soft and ultrasoft renormalization scales 
allowing the ultrasoft scale $\mu_\T{US}$ to take values between 1/2 and 2 
times its default value.
These constraints maintain the natural scale hierachy, and the small offset 
($0.05$) in \cref{eq:nustar} ensures that the
renormalization scales always remain in the perturbative regime: 
$\mu_\T{S} > \mu_\T{US} > 0.01\, \mOneS$.

Within the form factor expressions the strong coupling constant is evaluated at 
the three different scales $\mu_\T{H}$, $\mu_\T{S}$ and $\mu_\T{US}$.
For convenience we therefore define
\begin{align}
 \AShard=\ALstrong\Al[\mu_\T{H} \Ar] \co \quad
 \ASsoft=\ALstrong\Al[\mu_\T{S} \Ar] \co \quad
 \ASusoft=\ALstrong\Al[\mu_\T{US} \Ar] \,.
\label{eq:NRQCDalphas}
\end{align}
The hard coupling $\AShard$ is determined from $\alpha_s(M_Z)$, which is an 
input parameter in our calculation, using three-loop running ($n_f=5$).
The soft coupling $\ASsoft$ is then obtained by running down from $\mu_\T{H}$ 
to $\mu_\T{S}$ using one- and two-loop running for LL and NLL precision, 
respectively. Analogously, we use one-loop running for the LL 
ultrasoft coupling $\ASusoft$, which enters the NLL form factors via the NLL 
current coefficients $c_{1,3}^\mathrm{NLL}(h,\nu)\equiv 
c_{1,3}^\mathrm{NLL}(\AShard,\ASsoft^\mathrm{LL},\ASusoft^\mathrm{LL})$, see 
\cref{app:NLLc1c3}.

%
\section{Implementation in WHIZARD}
\label{s:implementation_in_whizard}

In this section we discuss how to implement the threshold summation encoded in
the (N)LL form factors in \eqs{FF_vec}{FF_ax} into the framework of an event
generator in order to combine it with a fully differential fixed-order
continuum calculation. For this goal we use the multi-purpose
event generator \whz{}~\cite{0708.4233}, which is a universal
MC generator for lepton and hadron colliders. It has its own
completely general matrix element generator for tree-level matrix
elements, \prog{O'Mega}~\cite{Moretti:2001zz,Nejad:2014sqa}. The usage
of an adaptive multi-channel phase space generation~\cite{Ohl:1998jn}
allows the integration and simulation of complex hard processes with
up to ten particles in the final state. Color quantum numbers
are handled via the color-flow algorithm~\cite{Kilian:2012pz}. Two
different parton shower algorithms ($k_T$-ordered and
analytic)~\cite{Kilian:2011ka} are available, while hadronization and
hadronic decays have to be simulated with interfaced external tools.

\whz{} is particularly well suited for the simulation of linear
collider events as it allows for completely arbitrary beam polarization, has
fully-inclusive soft photon corrections to all
orders~\Rcite{Gribov:1972ri,Gribov:1972rt} and
hard-collinear photon corrections up to third
order~\Rcite{Kuraev:1985hb,Skrzypek:1990qs} implemented in terms of an
initial-state (ISR) beam function, and allows for the simulation of
classical beamstrahlung. The latter -- originating from
ultra-collimated electron and positron bunches -- is simulated via the
dedicated \prog{CIRCE}~\cite{Ohl:1996fi} subpackages. For precision
simulations at QCD-NLO, \whz{} uses virtual amplitudes from external
one-loop providers (OLP) like \prog{OpenLoops}~\cite{Cascioli:2011va},
\prog{GoSam}~\cite{Cullen:2011ac,Cullen:2014yla}, or
\prog{Recola}~\cite{1211.6316,1605.01090}. \whz{} uses the FKS
subtraction formalism~\cite{hep-ph/9512328,0908.4272} which is completely
automatized, also allowing to preserve resonance masses in the
subtraction following the algorithm in~\Rcite{1509.09071}. The QCD-NLO
capabilities of \whz{} have recently been demonstrated in the fully
off-shell leptonic top decays with and without additional Higgs
radiation in $e^+e^-$ collisions~\cite{1609.03390}, building on earlier 
calculations for LHC
physics~\cite{Binoth:2009rv,Greiner:2011mp}. In fact, for QED
corrections matching between fixed-order and the structure function in
the initial state have been implemented in~\Rcite{Kilian:2006cj}.

Besides its extensive capabilities for simulating precision SM
processes, \whz{} supports many extensions beyond the SM (like
e.g. supersymmetry, composite Higgs, Little Higgs etc.) either by
direct implementations or via its interface to tools operating at the 
Lagrangian level like \prog{FeynRules}~\cite{Christensen:2010wz}. 


\subsection{General remarks on the implementation of the form factors}

We implement and study different variants of embedding the form factors of
the last section in the relativistic computation.
Let us first remark that we can in principle modify the vector
$\bar{t} \,\gamma_\mu t$ and axial-vector $\bar{t}\,\gamma_\mu\gamma_5\, t$ 
couplings to the $A$ and $Z$ fields directly in the tree-level matrix element 
by multiplying them with the corresponding nonrelativistic form factors.
This is a straightforward modification, and we refer to it as the
\emph{signal diagram} method. As explained in more detail in 
\cref{ss:gauge_invariance}, in \Rcite{1007.0075} an improved version of the
signal diagram method was adopted to implement the nonrelativistic S-wave
form factor.

However, this modification manifestly breaks gauge invariance for resonant (and off-shell) 
top production already
for the naive insertion in the tree-level matrix elements due to gauge
cancellations among the $t\bar{t}$ signal diagrams (resonant) and the rest of the
diagrams (non-resonant) describing $\WbWb{}$ production. 
The problem can also be seen at the Lagrangian level.
The term in the SM Lagrangian containing the covariant derivative $\bar 
t_{(L)}\,
\I\slashed{D}\,t_{(L)}$, which includes the $\bar{t} \,\gamma_\mu t$ (and the
$\bar{t}\,\gamma_\mu\gamma_5\, t$) vertex, is invariant under local 
gauge
variations $t \to e^{-\I Q_t g \theta(x)} t$.
This is no longer the case when we naively insert the nonrelativistic form 
factors for the vector and axial-vector currents and thus modify the covariant 
derivative (unless the tops are on-shell and the derivative cancels the mass 
term by construction).%

It is preferable to multiply the form factors to a gauge invariant
quantity. Let us consider a schematic factorized ansatz for 
the S-matrix element of double-resonant top pair production:
\begin{equation}
  \ME \simeq
    \underbrace{\braketop{e^+e^-}{\mathcal T_\T{NRQCD}}{{t \bar t}}}_{\equiv \MEprod}
    \braketop{{t \bar t}}{\mathcal T}{\WbWb}\po
\label{eq:factorized}
\end{equation}
Here the form factor only enters the on-shell production matrix element \MEprod{}
and QCD-NLO corrections to the decay can be computed separately.
We note that in \Rcite{1602.00684} a similar ansatz was employed to
study the CP properties 
of the Higgs boson from lepton angular correlations in 
$e^+e^-\to t\bar{t} H\to (\ell\bar{\nu} \bar{b})(\bar{\ell}\nu b)H$ for
$\sqrt{s}=500$~GeV. 
We discuss the exact realization of \cref{eq:factorized} in
\cref{ss:factorization}, after a more general discussion of the
possible violations of gauge invariance in the treatment of unstable
particles in \cref{ss:gauge_invariance}.
\subsection{Possible violations of gauge invariance}
\label{ss:gauge_invariance}
%
The treatment of unstable particles such as the top quark is technically challenging from a
perturbative point of view.
The Breit-Wigner distribution of their invariant mass in the resonance region is a result of
resumming absorptive self-energy corrections to the propagator, called Dyson summation.
This procedure mixes perturbative orders from the point of view of 
a strict Feynman diagrammatic expansion.
As gauge invariance can only be guaranteed order by order, the
associated Ward, Slavnov-Taylor and Nielsen identities may be violated if the summation is carried out naively as the off-shell particle self energies are in general not gauge-invariant.
While these violations are associated with higher orders, they can be
made arbitrarily large by applying extreme gauge
transformations~\cite{Stuart:1991xk}. 

Theoretically the problem is resolved by 
considering a simultaneous expansion in the couplings and the off-shellness $q^2-m^2$ in the resonance region 
set by the particle width and by Dyson summing only gauge-invariant portions of the self energy. 
As the approach requires an additional expansion, the predictions unavoidably acquire an additional scheme dependence that is, however, of higher order and therefore consistent with the perturbative expansion. 
A possible guiding principle to identify the gauge-invariant parts of the unstable particle self energy is  the fact that the complex pole
$p^2 = \mu^2$, where $\mu^2 = m^2 - i m \Gamma$, of the propagator of an
unstable particle is a gauge-invariant
quantity~\cite{hep-th/0005149,hep-ph/9907254}.
This property is the basis of two approaches that we will use in this work:
the \emph{complex-mass scheme} and the \emph{pole approximation}.

For calculations that involve complete matrix elements (i.e.\ not the
factorized parts constructed according to Eq.~(\ref{eq:factorized})), intermediate unstable particles are treated in the
\emph{complex-mass scheme}~\cite{hep-ph/0505042,hep-ph/0605312}.
The idea behind the complex-mass scheme is to add and subtract the width $\Gamma$ (defined through the complex pole)
in the bare Lagrangian.
While one of the terms is absorbed in the complex renormalized mass
definition, the other one adopts the role of a complex counterterm that is treated 
perturbatively order-by-order.\footnote{The approach can be
  generalized by adopting other gauge-invariant definitions of
  $\Gamma$ which differ from the complex pole definition by higher
  order terms.} 
This leads to a gauge invariant treatment of finite width effects, as
Ward or Slavnov-Taylor identities are exactly respected, while
maintaining perturbative unitarity~\cite{1406.6280}.
The \emph{pole approximation} is used for the parts of the calculation that are based on the factorization 
ansatz in Eq.~(\ref{eq:factorized}) and
will be discussed in \cref{ss:factorization}.

Aside from problems associated with the implementation of the width close to 
a particle resonance, it is essential to
treat the resonant signal and the non-resonant background diagrams in a coherent
fashion such that the respective gauge cancellations between them can take place. 
In fact, the smallest strictly gauge-invariant subset that contains the $\tT{}$ signal
diagrams for $\WbWb{}$ production consists of \emph{all} diagrams.
In the language of \Rcite{hep-ph/9903357}, such a subset is called a
\emph{grove} and can be systematically constructed.
Modifications of the signal diagrams such as the simple attachment of a form
factor will, therefore, in general spoil gauge invariance as we have already pointed out
above. 
As we show in \cref{ss:validation_of_factorization_approaches}, 
at tree-level these gauge cancellations become gigantic at high energies. 
In the threshold region, the dominant 
gauge cancellation concerning the signal diagram method is associated to contributions 
originating from the matrix elements and phase space integrations related to off-shell
top and anti-top quark decays. This can also been seen from the fact that the imaginary 
part of the off-shell top quark self energy is gauge-dependent.\footnote{
We note that the real part of the off-shell top quark self energy is gauge-dependent as
well. However, this contribution is part of the renormalized hard electroweak corrections
to $\tT{}$ production and constitutes NNLO matching corrections to the nonrelativistic
current Wilson coefficients  $c_1$ and $c_3$ in Eqs.~(\ref{eq:c1match}) and (\ref{eq:c3match}),
respectively.
}
In \Rcite{1007.0075} this imaginary part was used to construct a global
correction factor that allowed to define modified signal diagrams that
lead to gauge-invariant results at least for 
observables that are inclusive concerning the top and anti-top quark decays.
We note, however, that when QCD-NLO corrections 
are considered, the gauge cancellations are uncontrollable and large
at any energy as they also affect 
the structure of the QCD infrared divergences. At this level the signal diagram
method becomes practially meaningless. For this reason and in order to
also achieve manifest gauge-invariance 
for observables that are differential concerning the top and anti-top
quark decays we incorporated the nonrelativistic form factors within a
factorized approach as explained in more detail below.

To conclude this section, we remark that for the full SM tree-level
electroweak diagrams  used in this work we employ unitary gauge,
while for determination of the QCD-NLO corrections we adopt
t'Hooft-Feynman gauge for the gluon lines. 

\subsection{Factorization in the Double-Pole Approximation (DPA)}
\label{ss:factorization}
%
Different approaches to treating unstable particles close to and above
threshold have been compared in \Rcite{hep-ph/9303236} for $WW$
production.
Above threshold, the differences between the approaches in unitary gauge
have been found to be at the per cent level.
We will discuss here three of these methods and refer to them in the following 
as the \acf{NWA}, the \ac{SD}, and the \ac{PA}.
The \ac{NWA} is based on the on-shell $\tT$ production cross section
and results in the simple formula $\sigma=\sigma_\T{prod}\cdot
\T{BR}$, where BR is the corresponding on-shell particle branching
ratio. The \ac{NWA} is gauge-invariant, allows for factorizable
corrections but incorporates no off-shell behavior and defines no cross
section below threshold.
The latter property is inherited from the on-shell production cross
section $\sigma_\T{prod}$.
The \ac{SD} approach singles out the resonant signal diagrams and includes the 
decay width in the unstable particle propagators. In the  \ac{SD} approach the 
signal process can be evaluated with off-shell tops and is finite below 
threshold.
However, it is neither gauge-invariant nor suitable for computing QCD-NLO
corrections, as already pointed out in \cref{ss:gauge_invariance}.
In fact, we have encountered negative cross sections when taking QCD
corrections to the signal diagrams far above threshold into account. 
This is yet again related to the electroweak gauge-dependence of the signal 
diagrams. 

The \emph{pole expansion scheme}~\cite{Veltman:1963th,Stuart:1991xk} was one of the first
schemes to enable gauge-invariant computations for kinematics where
unstable particles are described close to resonance.
To this end, a Laurent expansion of the full scattering amplitude is performed 
in the unstable particle off-shellness around the unstable particle pole keeping the 
pole and a number of higher-order 
non-resonant terms that vanish on-shell.
The location of the pole in the complex plane, the pole residue and the 
non-resonant terms are by construction gauge invariant. 
In the \ac{PA}~\cite{hep-ph/9312212}, one drops the (higher-order) non-resonant
contributions as a first approximation for computational simplicity.
Thus, in the PA the gauge-invariant denominators of the unstable particle (Breit-Wigner) 
propagators are equivalent to the \ac{SD} approach, but the numerators are 
constructed to be gauge invariant as well.
This is achieved by projecting the (anti)particle momenta onto a suitable 
on-shell configuration in the calculation of the matrix elements, which allows to express the numerators as (fermion) 
spin sums projecting onto the corresponding on-shell (anti)particle state.
The property of gauge-invariance at the amplitude level then follows
directly from the
gauge-invariance of on-shell production and decay matrix elements.
The required on-shell projections cannot be uniquely defined and introduce an 
ambiguity in the results, which is of the order of the neglected non-resonant 
terms, i.e. $\order{\Gamma/m}$~\cite{hep-ph/0006307}.

\begin{figure}[htbp]
\centering
\includegraphics[width=0.5\textwidth]{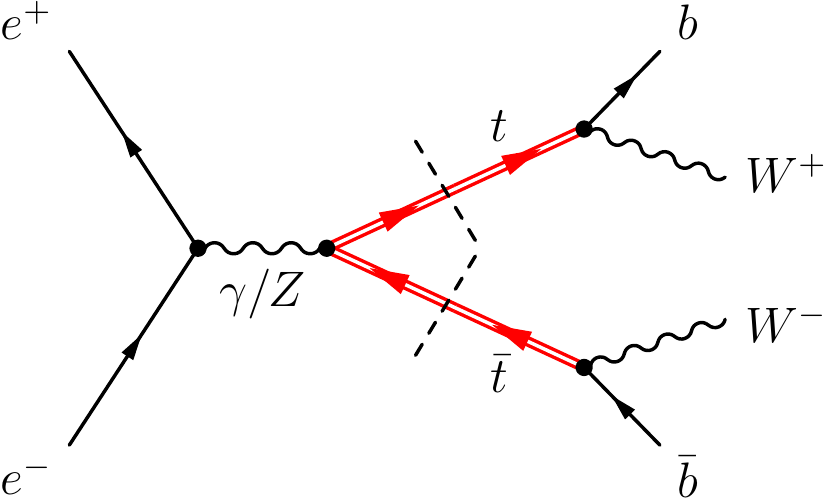}
\caption{Depiction of the factorized computation in the double-pole
  approximation (\ac{DPA}), exemplified at the Born level. The (red)
  double lines represent the top  
  propagators and a dashed line through them represents the on-shell
  projection.} 
\label{fig:factorized}
\end{figure}
In our case, we have to deal with two top-quark resonances and thus have
to use a \emph{double-pole approximation} (\ac{DPA})~\cite{hep-ph/9912261,hep-ph/9811481}.
Diagrammatically, we depict the factorized computation as shown in
\cref{fig:factorized}.
The factorized matrix element in this approximation can be written as
\begin{equation}
  \MEfact = \sum_{h_t,h_{\bar{t}}} \underbrace{\frac 1
  {(p_t^2-\mu_t^2)}\frac 1 {(p_{\bar{t}}^2-\mu_t^2)}}_{\equiv \propagators}
  \MEprod^{h_t,h_{\bar t}}[\set{\projected{p}}] \,
  \MEdecT^{h_t}[\set{\projected{p}}] \,
  \MEdecTbar^{h_{\bar{t}}}[\set{\projected{p}}]\co
  \label{eq:ME_factorized}
\end{equation}
where $h_t$, $h_{\bar t}$ are the polarizations of the top quark
resonances and $\mu_t^2 = \mpole^2 - \I \mpole\Gamma_{t}$ is the complex
top quark pole, respectively. In our work we choose the top polarization states 
to be helicity states. The term
$\set{\projected{p}}$ denotes a set of momenta that have been projected 
on-shell (concerning the top and anti-top quarks) 
such that $\projected{p_t}^2,\projected{p}_{\bar t}^2=\mpole^2$, while 
$\set{p}$ represents the set of physical momenta with $p_t^2,p_{\bar t}^2\neq 
\mpole^2$ in general.
So, in $\MEfact$ physical off-shell top momenta are still used 
for the top propagator denominators (and the event output).
The details of the projection procedure are discussed in
\cref{sss:on_shell_projection}.
We note that if $p_t$ and $p_{\bar t}$ are on-shell,
we can use the fermion spin sums
\begin{align}
  \sum_{h_t} u_{h_t}(\bp_t) \, {\bar u}_{h_t}(\bp_t) =
      \slashed{p}_t + \mpole \co \qquad
  \sum_{h_{\bar{t}}} v_{h_{\bar{t}}}(\bp_{\bar{t}}) \,
      \bar{v}_{h_{\bar{t}}}(\bp_{\bar{t}}) =
      \slashed{p}_{\bar{t}} - \mpole
\label{eq:fermion_helsum}
\end{align}
over top ($u$) and anti-top ($v$) spinors
to show that \cref{eq:ME_factorized} is identical to the signal diagram at 
tree-level.
We have used this property to verify the correct implementation of the 
factorized matrix elements:
Within the numerical precision of the (complex) amplitudes we found perfect 
agreement for a given on-shell phase-space point and given external helicity.
Considering the fully relativistic four-body phase space, which probes also all 
possible top and anti-top quark off-shell regions, we stress that \cref{eq:fermion_helsum} and 
therefore the equality of \MEfact{} and signal diagram does in general
not hold. Nevertheless, as we also show in
\cref{ss:validation_of_factorization_approaches} in the absence of
QCD-NLO corrections, computations with signal diagrams are numerically
quite close to the ones with \MEfact{} at least in the threshold region. When 
accounting for QCD-NLO corrections, however, using signal diagrams leads to 
results that are far off the correct physical predictions, as already mentioned 
above.

The crucial aspect of the \ac{DPA}/\ac{PA} is that factorizable and
nonfactorizable corrections are separately gauge-invariant.
The latter arise e.g from the crosstalk among the top, anti-top 
decay and production subprocesses via (ultrasoft) gluon exchange.
Depending on the (inclusiveness of the) observable the factorizable corrections 
to the production and decay matrix elements are often dominant.
We will neglect nonfactorizable corrections when implementing the threshold 
resummation for \whz{} predictions in this work, see 
\cref{sss:fixed_order_corrections}.
In passing, we note that the nonfactorizable one-loop corrections due to soft 
photon exchange are universal and have been given analytically for any number 
of unstable particles in \Rcite{1511.01698}.

The DPA expression in \Cref{eq:ME_factorized} is our preferred setup to single out the $\tT{}$
resonant production part for the $\WbWb{}$ cross section and will also be
referred to as \emph{factorized, on-shell evaluated}.
\emph{Off-shell evaluated} would correspond to replacing
$\set{\projected{p}}$ with $\set{p}$ and is only used as a test in
\cref{sss:high_energy_behavior,sss:around_threshold}.
To include the nonrelativistic S- and P-wave form factors in
this approach, we multiply them with the corresponding factorized
production matrix elements. 
Since the nonrelativistic form factors are gauge-invariant, the resulting 
$\WbWb{}$ amplitudes are still factorizable and gauge-invariant.
In this setup it is then straightforward to also include (hard)
QCD-NLO corrections to the on-shell decay, as described in 
\cref{sss:fixed_order_corrections}.
\subsubsection{Helicity correlations}
\label{sss:helicity_correlations}
Concerning helicity correlations, \cref{eq:ME_factorized} can be
considered as complete as possible.
This implies that in general there are also interference terms in the top and 
anti-top density matrices that arise in the squared factorized matrix element,
\begin{equation}
  \abs{\MEfact}^2 = \left(\sum_{h_t,h_{\bar{t}}} \propagators \,
  \MEprod^{h_t,h_{\bar t}} \MEdecT^{h_t} \MEdecTbar^{h_{\bar t}}\right)
  \left(\sum_{h_t',h_{\bar{t}}'} \propagators\,
  \MEprod^{h_t',h_{\bar t}'} \MEdecT^{h_t'} \MEdecTbar^{h_{\bar t}'}\right)^*\co
  \label{eq:ME2_factorized}
\end{equation}
where $h_t\neq h_{t}'$ or $h_{\bar t}\neq h_{\bar t}'$ in the helicity
basis. This expresses the fact that the top and anti-top quark
polarizations in general differ from the helicity eigenstates.  
These off-diagonal contributions in the spin density matrices are
known to be sizable at
threshold~\cite{Harlander:1994ac,Jezabek:2000gr}, but  
at the implementation level, they are currently not yet
available in the one-loop provider \openloops, which we use to obtain
the QCD-NLO virtual corrections for the top decay. For the time being
this is a limitation, which we intend to lift with future releases of
\openloops. From \whz{}, by default events at QCD-LO for the
factorized processes are generated using the full spin
correlations. But there is also the option to switch off the
off-diagonal entries in the helicity basis completely, in accordance to the
available terms from  \openloops at QCD-NLO. For spin-independent
observables the off-diagonal entries in the spin density matrices are
irrelevant and the current implementations (also described in the next
paragraph) provide the correct (and equivalent) results.  

For testing purposes in comparisons with previous spin-independent
results, we thus define the following \ac{HA}, which covers the
diagonal correlation entries (for $h_t=h_t'$ or $h_{\bar t}=h_{\bar
  t}^\prime$), but neglects all off-diagonal entries (for $h_t\neq h_t'$ or
$h_{\bar t}\neq h_{\bar t}^\prime$): 
\begin{align}
  \abs{\MEfact^\T{HA}}^2 &= \sum_{h_t,h_{\bar{t}}} \abs{\propagators}^2\left(
  \MEprod^{h_t,h_{\bar t}} \MEdecT^{h_t} \MEdecTbar^{h_{\bar t}}\right)
  \left(\MEprod^{h_t,h_{\bar t}} \MEdecT^{h_t} \MEdecTbar^{h_{\bar 
t}}\right)^*\no
  &= \sum_{h_t,h_{\bar{t}}} \abs{\propagators}^2 \abs{\MEprod^{h_t,h_{\bar 
t}}}^2
  \abs{\MEdecT^{h_t}}^2 \abs{\MEdecTbar^{h_{\bar t}}}^2\po
  \label{eq:ME2_factorized_approx}
\end{align}
The \ac{HA} allows for predictions where in addition to the measurement 
of interest the top and anti-top helicities are measured as well. 
We may neglect correlations even further concerning
comparisons with analytic results for the total cross section,
obtained as described in \cref{ss:sigmaincl}, where 
predictions are made for spin-averaged states. To this end we apply an
uncorrelated average to the decay matrix elements which we call the
\ac{EHA}:
\begin{equation}
  \abs{\MEfact^\T{EHA}}^2 = \sum_{h_t,h_{\bar{t}}} \abs{\propagators}^2
  \abs{\MEprod^{h_t,h_{\bar t}}}^2
  \left(\frac 12\sum_{h_{t}'}\abs{\MEdecT^{h_t'}}^2\right)
  \left(\frac 12\sum_{h_{\bar t}'}\abs{\MEdecTbar^{h_{\bar t}'}}^2\right)
\label{eq:ME2_factorized_EHA}
\end{equation}
So, in the EHA all spin correlations between production and decay are
removed entirely. We stress that for physical applications, in all the
above approximations, it is implied that the external helicities of
$\eeWbWb{}$ are summed over, as we only investigate unpolarized
configurations in this paper.

\subsubsection{On-shell projection}
\label{sss:on_shell_projection}
%
A generic algorithm and formulae to obtain on-shell projected momenta for any
number of resonances can be found in~\Rcite{1511.01698}.
In our case, the expressions simplify to
\begin{align}
  \projected{p}_t = \Bl(\frac{\sqrts{}}2, \frac{\sqrt{s-4\mpole^2}}2 
 \,\be_t\Br) \co \qquad
  \projected{p}_{\bar t} = \Bl(\projected{p}_t^0, - \projected{\bp}_t\Br)\co
\label{eq:projection}
\end{align}
where $\be_t=\bp_t/\absIT{\bp_t}$ is the top flight-direction in 
the collision system determined from the original final state momenta of the 
process.
Again, we denote on-shell projected momenta with a hat, and they fulfill
by definition the on-shell conditions 
\begin{align}
  \projected{p}_t^2 =\mpole^2\co \qquad \projected{p}_{\bar t}^2=\mpole^2 \po
\label{eq:onshello}
\end{align}
Our definition of $\be_t$ guarantees that the projection leaves
final state momentum configurations that correspond to on-shell top
quarks unchanged. It also maintains spatial correlations, which
are important for example for the forward-backward asymmetry.
Furthermore, it is crucial to retain the three-momentum directions for the
interference terms of the factorized with the full amplitude in 
\cref{ss:contributions}.

We note that the projection in \cref{eq:projection} cannot be applied literally in this form over the whole
physical kinematic range.
Below threshold, $\sqrts{}<2\mpole$, it is meaningless as it would yield complex momenta.
Thus, it is only defined for $\sqrt{s}>2\mpole$.
The resummed computation, however, reaches its peak at $\sqrts\approx 2\mOneS <
2\mpole$.
Therefore, we define the extended \ac{DPA} as follows:
For $\sqrts{}>2\mpole$, the extended \ac{DPA} is identical to the normal
\ac{DPA}.
For $\sqrts{}\leq2\mpole$, we project to a set of momenta that
correspond to $\sqrts{}=2\mpole+\epsilon$, where
$\epsilon$ is a very small number that is introduced to
avoid potential numerical instabilities in the matrix elements, that occur for 
$\projected{\bp}_t=0$.
The three-momentum direction $\be_t$ is defined in the same way as above 
threshold, i.e.\ it is unchanged.
This extended \ac{DPA} 
transitions smoothly into the normal \ac{DPA} at threshold
and, crucially, provides finite and gauge-invariant results below threshold.

Having defined $\projected p_t$ and $\projected p_{\bar t}$, we also
have to project the momenta of the decay products. Let us e.g. discuss the
case of the top decay. At the Born level, this is a simple $1\to2$
decay with the well-known 
kinematics~\cite{Byckling1973,1511.01698}
\begin{align}
  \abs{\projected{\bp}_W} = \abs{\projected{\bp}_b} &=
  \frac{\sqrt{\lambda(\mpole^2,m_{W}^2,m_{b}^2)}}{2 \mpole}\co\quad
  E_W = \frac{\mpole^2 + m_{W}^2 - m_{b}^2}{2 \mpole}\co\quad
  E_b = \frac{\mpole^2 - m_{W}^2 + m_{b}^2}{2 \mpole} \co \no
  \lambda(x,y,z)&=x^2+y^2+z^2-2xy-2yz-2xz\co
\label{eq:decay_projection}
\end{align}
in the frame where $\projected{p}_t = (\mpole, \mathbf{0})$, called the top 
rest 
frame in the following. Analogous to the top three-momentum, we choose
$\projected{\bp}_W =  
\absIT{\projected{\bp}_W}\,\be_W$,
where $\be_W$ is the actual direction of the $W^+$ in the top rest
frame for the given kinematics (generated by \whz{}).
Note that this also ensures that the flight direction of the $b$ quark is
unchanged in the top rest frame.
These momenta are then boosted back to the collision (lab) frame.
The analogous procedure is applied to the anti-top decay products.
The QCD-NLO case, involving the $1\to3$ decay with an additional gluon, is discussed in
\cref{ss:whizard_nlo}.
\subsection{Input parameters}
\label{ss:input_parameters}

Here we list the SM quantities that enter our 
calculation as input parameters together with their default values we 
use throughout this work.
The top 1S mass plays a special role in this work and we choose
\begin{align}
 \mOneS =\SI{172}{\GeV}\po
\end{align}
Unless stated otherwise the other default values are taken from 
\Rcite{Agashe:2014kda}.
Namely, we use 
\begin{align}
  m_{Z} &= \SI{91.1876}{\GeV}\co & m_{W} &= \SI{80.385}{\GeV}\co \\
  m_{b} &= \SI{4.2}{\GeV}\co     & m_{H} &= \SI{125}{\GeV}\co
\end{align}
for the vector boson, Higgs and bottom quark pole masses\footnote{As
  we work at LO in the electroweak couplings, we directly take the
  numerical values of the Breit-Wigner masses given in
  \Rcite{Agashe:2014kda} as (real) pole masses and transfer
  them to the complex-mass scheme. The numerical differences are
  negligible for our purposes.}, while the electron,  
the muon, as well the first two quark generations are treated massless.
Besides that, we have the $W$/$Z$ and Higgs boson widths
\begin{align}
\Gamma_{W}= \SI{2.049}{\GeV} \co  \qquad 
\Gamma_{Z} = \SI{2.443}{\GeV}\co  \qquad 
\Gamma_{H} &= \SI{4.143}{\MeV} \co
\end{align}
entering the particle propagators in the complex-mass scheme.
For the CKM matrix we assume the unit matrix, which for the most relevant 
element of our computation $V_{tb}$ is consistent with the measured 
value ($1.021\pm0.032$~\cite{Agashe:2014kda}).
Furthermore, we set the initial value for the strong (running) coupling 
constant to~\cite{Bethke:2015etp}
\begin{align}
\ALstrong\ofIT{\mu=m_{Z}} = 0.118\co
\end{align}
and fix the electromagnetic coupling to its value in the 
\GF-scheme:
\begin{align}
  \alpha_{\mathrm{em},G} \equiv \frac{\sqrt{2}}{\pi}\, m_{W}^2 
\bigg(1-\frac{m_W^2}{m_Z^2} \bigg) \GF\po
  \label{eq:ALweak}
\end{align}
Using the precisely measured value for the Fermi constant, 
$\GF=\SI{1.1663787e-5}{\per\GeV\squared}$, absorbs the dominant electroweak 
corrections to the top decay.
As in this work we treat electroweak effects at the LO level, it is 
advisable to use a scheme where the electroweak corrections to the top decay are 
small. 
This is the case for the \GF-scheme~\cite{Denner:1990ns}.
Of course, this choice is not fully capturing the production dynamics,
where a $\sqrts$-dependent electromagnetic coupling 
$\ALweak\ofIT{\mu=\sqrt{s}}$, which absorbs oblique vacuum polarization 
corrections, is more
appropriate~\cite{hep-ph/0604104,hep-ph/0302259}.
Note, however, that the relative difference of
$\alpha_{\mathrm{em},G}=1/\num{132.233}$ to
$\ALweak\ofIT{\mu=2\mOneS{}}=1/\num{125.924}$ (which is roughly 
\SI{5}{\percent}) is smaller
than to the Thomson limit $\ALweak\ofIT{\mu=0}=1/\num{137.036}$
(roughly \SI{9}{\percent}).
Either way, one can simply reweight our predictions with
$(\ALweak\ofIT{\mu=2\mOneS{}}/ \alpha_{\mathrm{em},G})^2$, which increases them
by about \SI{10}{\percent}, to account for the dominant effects from
summing s-channel photon and $Z$-vacuum polarization diagrams. 

As noted in \cref{ss:gauge_invariance}, we will use the complex-mass
scheme for the QCD-LO and QCD-NLO SM (and top decay) matrix elements.
That is, we introduce complex-valued renormalized masses
\begin{equation}
  \label{eq:complexmasses}
  \mu^2_i = m_i^2-\I\,\Gamma_i\,m_i \MT{for}\;i=W,Z,H,t\co
\end{equation}
which implies a complex-valued weak mixing angle
\begin{equation}
  \label{eq:defsintheta}
  \SW=1-\CW=1-\frac{\mu_{W}^2}{\mu_{Z}^2}\po
\end{equation}
We use \eq{defsintheta} for explicit values of the weak mixing angle e.g. in
charged and neutral current couplings. We will, however, refrain from
using the complex-valued mixing angle to define \ALweak{} and rather
use the (real-valued) definition of \cref{eq:ALweak}. This allows us
to use one consistent definition in all parts of our calculation:
It is consistent with all calculations in this work being at leading
order in the electroweak couplings, and retains the gauge-invariance
properties of the complex-mass scheme.

With the above input we can translate the 1S top mass to the respective 
pole mass at a given precision according to \cref{ss:mass} and
\cref{ss:resummation_and_scales}.
The relation between the two masses, \cref{eq:mPoleRelation}, depends (also via 
$\ASsoft$) on the renormalization parameters $h$ and $\nu = f \nustar$ with 
$\nustar$ as defined in \cref{eq:nustar}, which in turn depends on the c.m. 
energy $\sqrts{}$.
So, using our default scale choice $h=f=1$ we obtain e.g. at threshold (thr) for
$\sqrts{}=2 \mOneS=\SI{344}{\GeV}$:
\begin{align}
 \mpole^\T{LL}\Al[h\!=\!f\!=\!1\Ar] \Big|_\mathrm{thr} 
=\SI{172.802}{\GeV}\co \qquad
 \mpole^\T{NLL}\Al[h\!=\!f\!=\!1\Ar] 
\Big|_\mathrm{thr}
=\SI{173.128}{\GeV} \po
\end{align}
The (N)LL soft coupling \ASsoft{}, see \cref{eq:NRQCDalphas}, which enters
these numbers, is evaluated as described in \cref{ss:resummation_and_scales}.
Of course, the same $\sqrts{}$-dependent (N)LL top pole mass is used for the 
(N)LL form factors, the corresponding factorized on-shell production/decay 
(N)LO matrix elements, as well as the QCD-(N)LO fixed-order cross sections.

Also the QCD-(N)LO top decay width depends on the (N)LL top pole mass.
For fixed renormalization parameters $h$ and $f$ it therefore also varies 
slightly with $\sqrts{}$ and is (like the pole mass) automatically recomputed
by \whz{} whenever $\sqrts{}$ changes.
At NLO the top width also depends on the QCD coupling
$\alpha_\T{F} = \alpha_{s}\of{ \mu_\T{F}}$, where 
$\mu_\T{F}=h \mOneS \sqrt{\nustar}$  is the
same ($\sqrts{}$-dependent) renormalization scale we use for the fixed-order 
QCD-NLO cross section computation, see \cref{s:matching}.
Similar to the hard NRQCD coupling \AShard{} in 
\cref{ss:resummation_and_scales}, we determine \ASfirm{} by three-loop RG 
running from $\mu=m_Z$ to $\mu=\mu_\T{F}$ with $n_f=5$ active flavors.
At threshold ($\sqrts{}=2 \mOneS=\SI{344}{\GeV}$) we then end up with
\begin{align}
\Gamma_{t}^\T{LO}\ofIT{h\!=\!f\!=\!1}=\SI{1.4866}{\GeV} \Big|_\mathrm{thr}
\co \qquad
\Gamma_{t}^\T{NLO}\ofIT{h\!=\!f\!=\!1}=\SI{1.3491}{\GeV} \Big|_\mathrm{thr}
\po
\end{align}
By evaluating fixed-order QCD matrix elements (in particular also for the top 
decay in the factorized computations) and the top decay width at the same
perturbative order, we guarantee that
$t\to W^+b(+\mbox{gluon})$ branching ratios obtained from
the differential final state computations remain consistently equal to unity at 
QCD-(N)LO upon full integration over the decay phase
space, as recently demonstrated in \Rcite{1609.03390}.

\subsection{Validation of factorized approach}
\label{ss:validation_of_factorization_approaches}
In this section we are going to validate the factorized setup for the
threshold calculation. This includes checks concerning the high-energy
behavior (in connection with potential inconsistencies related to
gauge-dependent terms), the behavior of the factorized process around
the threshold, the consistency of the complex phase in the different
pieces of the interference terms, as well as the properties of the
factorized amplitudes under charge conjugation and parity
transformations. If not stated otherwise, all results shown in this 
subsection are obtained with a tree-level form factor of unity, LO top
decays, and setting $\mpole=\mOneS$.
\subsubsection{High-energy behavior}
\label{sss:high_energy_behavior}
\begin{figure}[htbp]
  \centering
  \includegraphics[width=\standardwidth]{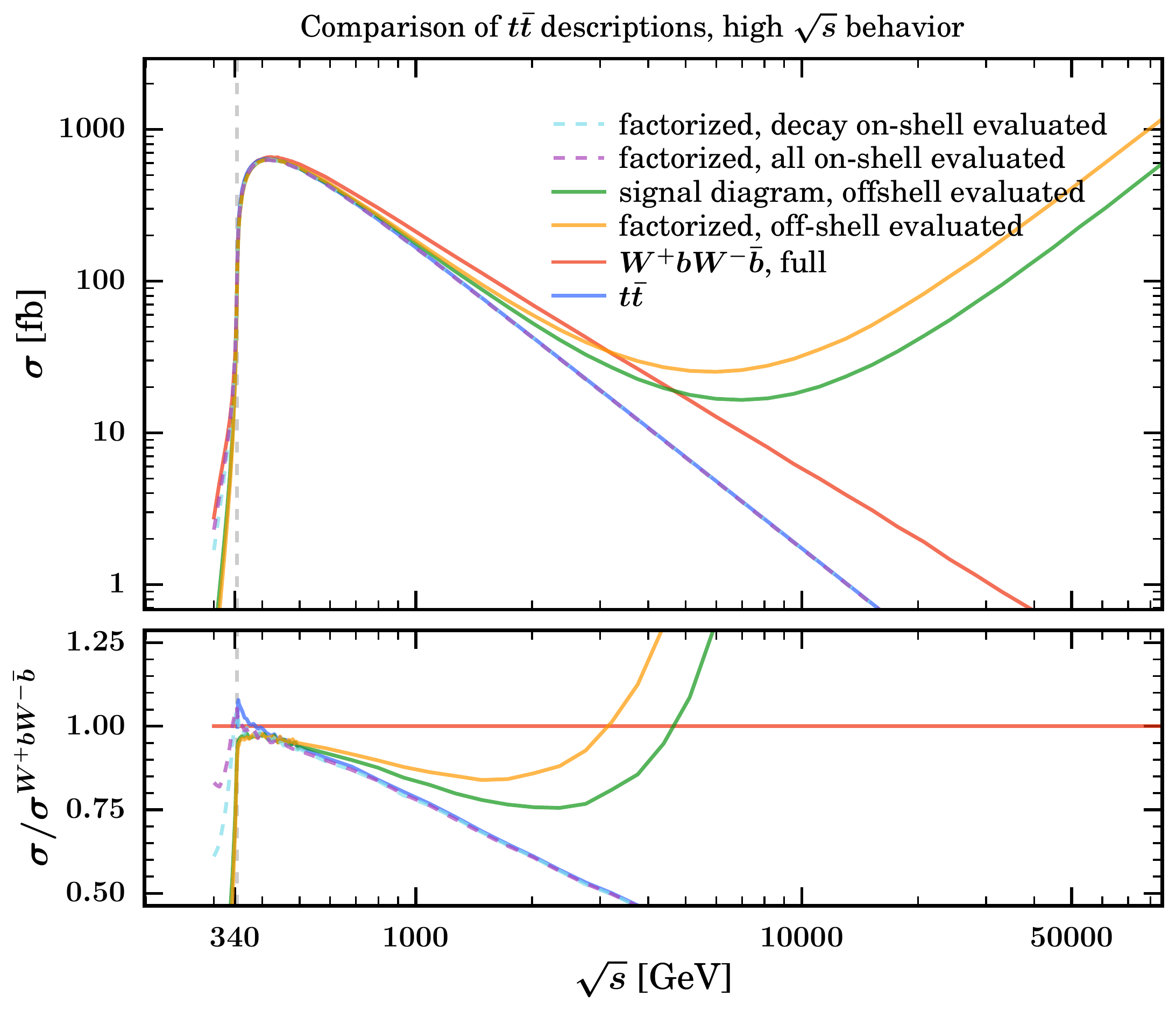}
  \caption{Total cross section at LO (without any QCD corrections) over $\sqrt{s}$ for different factorization approaches for high energies as described in
  more detail in \cref{sss:high_energy_behavior}. The dashed gray line
  indicates $\sqrt{s}=2\mOneS$.}
  \label{fig:high_sqrts_behavior}
\end{figure}
%
The high-energy behavior of total cross sections can serve as a test for
gauge-invariance. As known for example from $WW$
production~\cite{Beenakker:1996kt}, the total cross section is
expected to fall off with \sqrts{} in the high-energy limit due to
perturbative unitarity. This may not be the case, if it contains
unphysical gauge-dependent terms. In \cref{fig:high_sqrts_behavior},
we compare different factorization approaches implemented in \whz.
The full $\WbWb{}$ LO cross section (red) serves as reference in this case, as
it is gauge-invariant and valid below threshold.
One can see that while it agrees with the cross section for on-shell $\tT{}$  production (blue) around
threshold up to $\sim\SI{10}{\percent}$, the $\tT{}$ curve falls off faster with 
energy and constitutes only \SI{50}{\percent} of the full $\WbWb{}$
cross section at \SI{2}{\TeV}, in full agreement with the corresponding results shown in \Rcite{1411.2355}.
At high energies, the $\tT{}$ cross section is much smaller than the
total cross section for the final state 
$\WbWb{}$ because the latter gets sizable contributions from single-top and
non-$\tT{}$-resonant processes which are missing in the $\tT{}$ cross section.
We furthermore see that using the gauge-dependent $\tT{}$ signal diagram only (green) instead of the sum over
all diagrams for $\WbWb{}$ production  leads to an unphysical rise of the cross section at high
energies.
The same holds for the cross section based on the factorized matrix element,
\cref{eq:ME_factorized}, evaluated with off-shell momenta (orange),
i.e.\ without the on-shell projection of
\cref{sss:on_shell_projection}. 
The difference between the latter two descriptions, which both show
unphysical behavior and should not be used due to unitarity violation,
is related to the fact that \cref{eq:fermion_helsum} only holds on-shell.
Finally, we have two descriptions that closely follow the $\tT{}$ curve for high
energies: the factorized computation with on-shell momentum
projections in the decay and off-shell momenta in the production
matrix element (dashed cyan) and the one with on-shell momentum
projections for both production and decays (dashed purple), which is our
default approach. The similarity of the latter cross sections indicates
that the unphysical high-energy behavior arises from using off-shell
momenta for the decay matrix elements. This reconfirms the approach 
used in \Rcite{1007.0075} to construct their global correction factor to
eliminate the gauge-dependence of the signal diagrams. 

While it is generally known that evaluating only signal diagrams
leads to gauge-dependent results, our numerical analysis shows the impact 
of the associated numerical effects. 
Given the large (unphysical) differences between the SD curve (green) and 
the full cross section (red) observed in \cref{fig:high_sqrts_behavior}, we 
conclude that using signal diagrams, particular at TeV energies,
leads to unreliable results and should be avoided in general.
\subsubsection{Threshold region}
\label{sss:around_threshold}
\begin{figure}[htbp]
  \centering
  \includegraphics[width=\standardwidth]{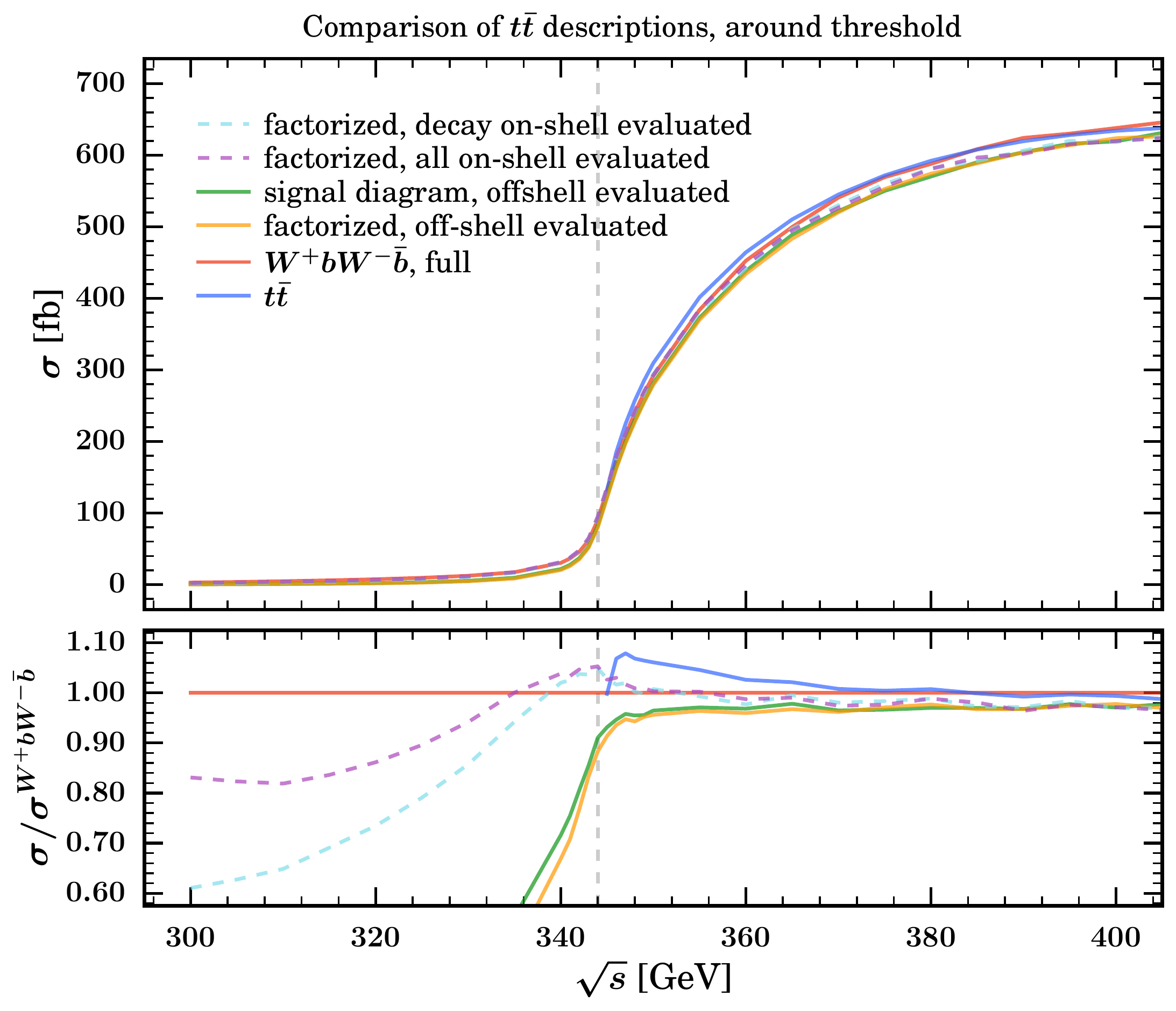}
  \caption{Total cross section at LO (without any QCD corrections) for different factorization approaches in the threshold region as described in
  	more detail in \cref{sss:around_threshold}. The dashed gray line
  	indicates $\sqrt{s}=2\mOneS$.
     Line colors as in \cref{fig:high_sqrts_behavior}.}
  \label{fig:high_sqrts_behavior_zoom}
\end{figure}
The considerations in the last paragraph were instructive to
understand the numerical impact of gauge dependence for the
high-energy behavior. However, in this work we actually only rely on the 
factorized computation within the threshold region for the construction of our 
matched cross sections, see \cref{s:matching}. 
In \cref{fig:high_sqrts_behavior_zoom} we therefore examine the different
prescriptions once again zooming in a \SI{100}{\GeV} window around
threshold. Obviously, the on-shell $\tT{}$ cross section (blue) is
only finite above $2\mpole$ and therefore not discussed further in the
following.  

We see that 10~GeV above threshold and higher all other curves are quite 
close to each other and moreover agree with the full $\WbWb{}$ cross section 
(red) to within \SI{4}{\percent} or better. This indicates that the relative 
effects from gauge-dependent terms are very small in this region and that the 
difference to the full $\WbWb{}$ cross section predominantly arises from 
non-signal diagrams, i.e. diagrams without the (virtual) $\tT{}$ pair. 
Below threshold, on the other hand, where all cross sections decrease strongly
and a double on-shell configuration is kinematically not allowed anymore, 
the other approaches deviate significantly from the full $\WbWb{}$ cross 
section.
Whereas the cross sections based on the signal diagram (green) and the 
off-shell evaluated factorized matrix elements (orange) fall off quickly far 
below \SI{60}{\percent} of the full $\WbWb{}$ cross section, the factorized
descriptions with on-shell projections for the decay (dashed cyan and
purple) are also smaller than the full $\WbWb{}$ cross section but
remain relatively close. We observe that the non-signal
 $\WbWb{}$ production diagrams in general yield positive
contributions. The behavior of the results also indicates that the
unphysical gauge-dependent off-shell effects are quite dramatic below
threshold and should not be used for predictions. In this context,
employing off-shell evaluation for the decay matrix elements
represents a much bigger (unphysical) effect than for the production
matrix element.

In any case, the cross section computed in the proper (extended) \ac{DPA} 
(dashed purple), which consistently applies on-shell projection for production 
and decay matrix elements, is closest to the full $\WbWb{}$ result and thus 
provides the best factorized approximation.
From now on we will stick to this method of factorization and simply refer to 
it as the factorized approach.

\subsubsection{Helicity correlations}
\label{sss:validation_helicity_correlations}
\begin{figure}[htbp]
\centering
\includegraphics[width=\standardwidth]{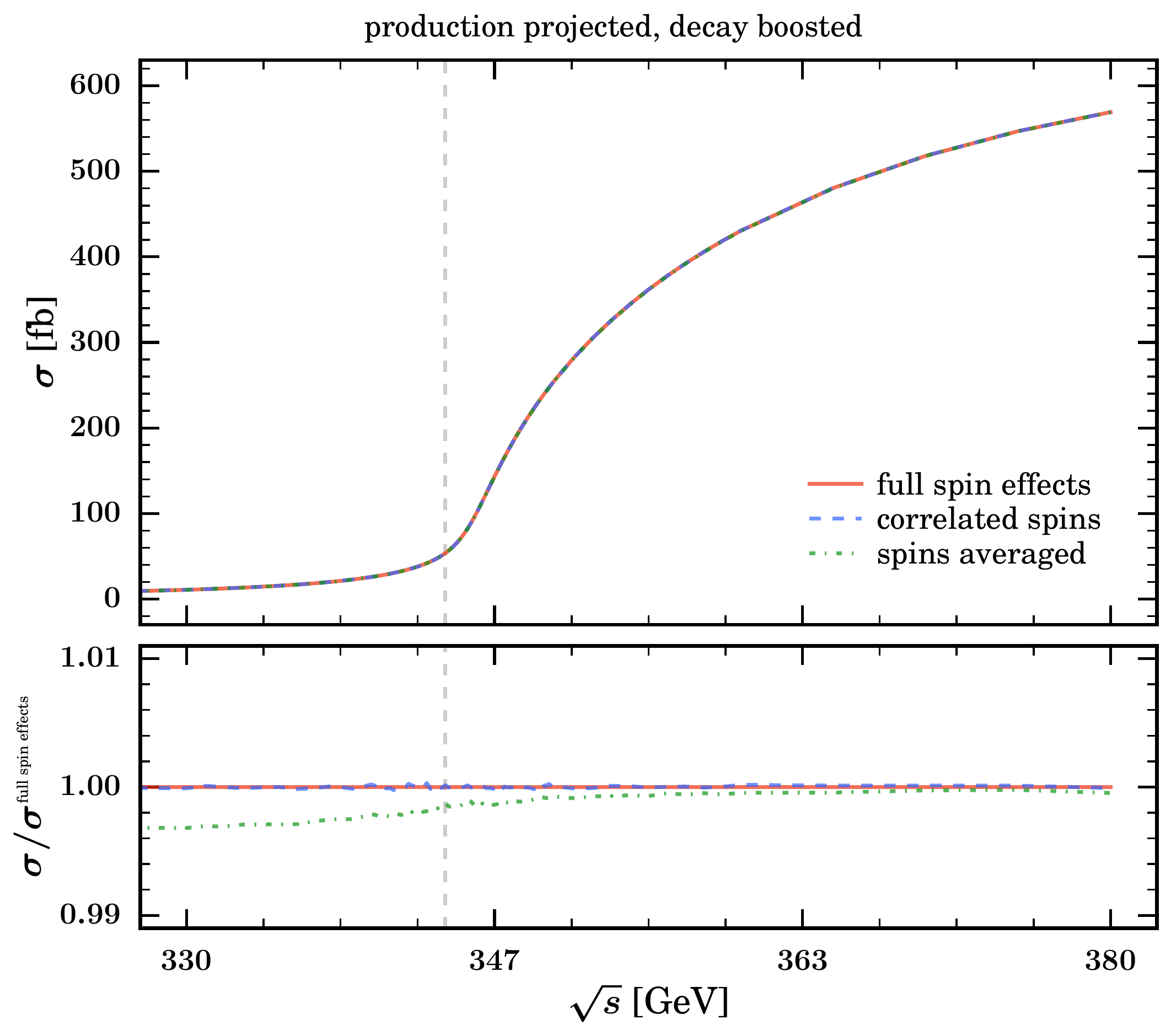}
  \caption{
    Total cross section at LO (without any QCD corrections) for
    our factorized calculation in the threshold region. For
    the top and anti-top quark wave function we use the complete
    spin-dependence (red), the diagonal entries in the helicity
    basis (HA, dashed blue) and the
    spin-averaged approach (EHA,
    dash-dotted green).
    The dashed gray line indicates $\sqrt{s}=2\mOneS$.}
  \label{fig:influence_of_OS_projection_for_production_zoom} 
\end{figure}

In \cref{fig:influence_of_OS_projection_for_production_zoom}, we show the 
effects of using different approximations concerning the top--anti-top spin 
density matrix for the total cross section in the threshold region calculated 
from our factorized approach, as discussed in \cref{sss:helicity_correlations}.
The total cross section is displayed using the full spin correlations
according to \cref{eq:ME2_factorized} (red), using only the diagonal
spin-density entries in the helicity basis for the top and anti-top
wave functions, called helicity approximation (HA), according to
\cref{eq:ME2_factorized_approx} (dashed blue), and using spin summation
and averaging for the production and decay, respectively, called
extra-helicity approximation (EHA) according to
\cref{eq:ME2_factorized_EHA}. Since all three approaches lead to
equivalent results for spin-independent observables we expect that the
results for the total cross section agree. We see that this is true 
at the level of the numerical precision of our analysis for the full
spin correlation (red) and the HA results (dashed blue). On the other
hand, for the EHA (dash dotted green) we find a notable relative
difference to the other two approaches at the level of up to a few
permille below threshold. The discrepancy is a numerical artefact of
the combination of spin-averaging and on-shell projection within the
current implementation. This is, however, irrelevant
phenomenologically since below threshold the contributions from
non-resonant $\WbWb{}$ production are much larger than this discrepancy,
cf.~\cref{fig:high_sqrts_behavior_zoom}. In any case, as our default, we
use full spin correlations in all parts of our factorized calculations
except for when QCD-NLO corrections to the top quark decays enter,
where we use the HA for the reason explained in \cref{sss:helicity_correlations}. 

\subsubsection{Interference terms}
\label{sss:interference_terms}
\begin{figure}[htbp]
\centering
\includegraphics[width=\halfwidth]{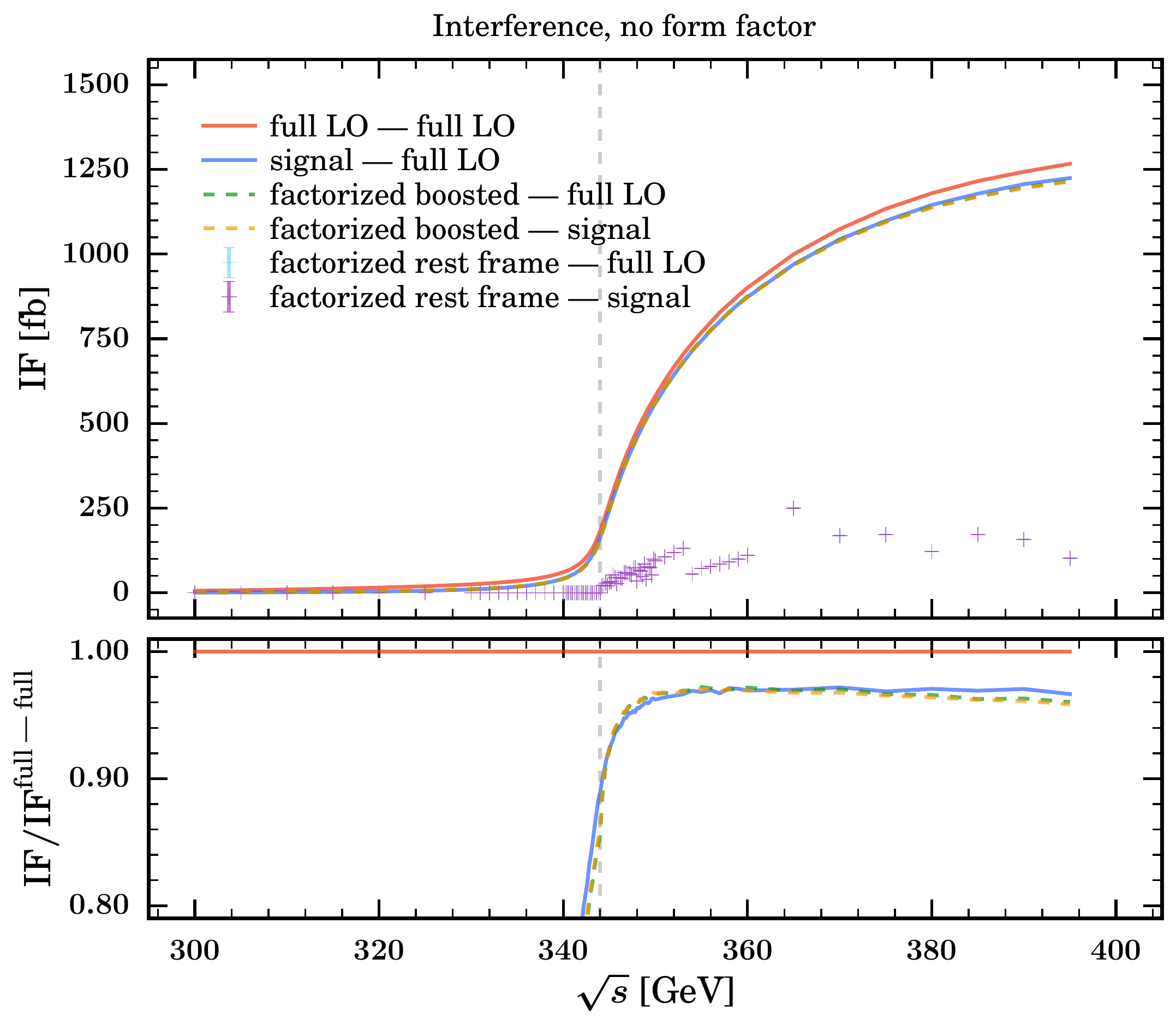}
\includegraphics[width=\halfwidth]{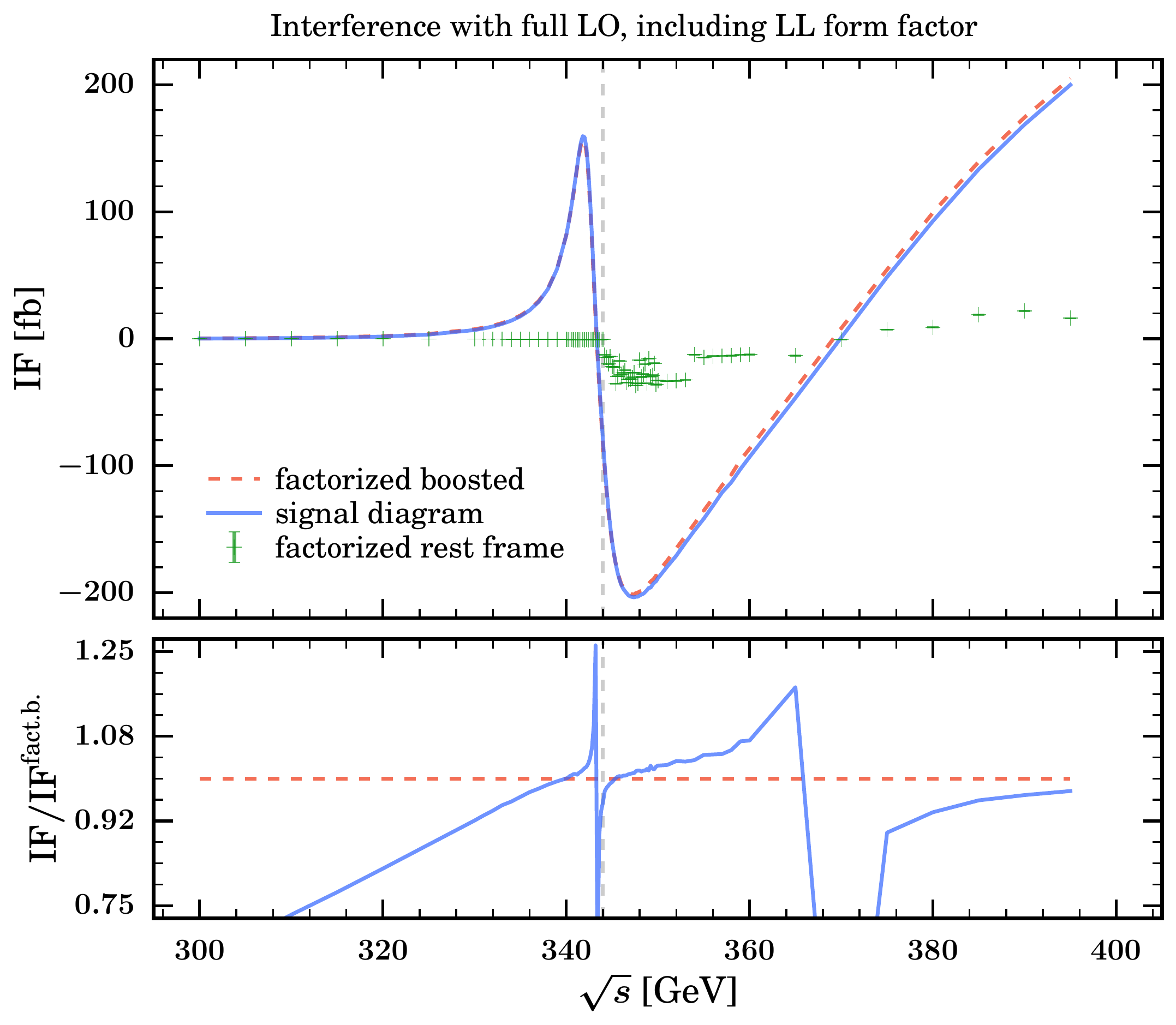}
\caption{Integrated inclusive interference (IF) terms: In the left panel
we show the full QCD-LO matrix element squared (times a factor two) as reference curve (red solid).
The solid blue curve corresponds to the IF of the signal diagram with the 
full QCD-LO matrix element. 
The IF terms involving the proper (boosted) factorized matrix element 
are represented by dashed lines.
For comparison we also show the IF terms involving a factorized 
matrix element, where the decay part is (inconsistently) evaluated with
momenta in a different Lorentz frame than the production part, namely the top 
rest frame. 
These are indicated by the $+$ symbols. MC integration errors are too 
small to be visible.
In the right panel, we show again the IFs with the full QCD-LO matrix element, where 
this time the vector-current signal diagram and factorized amplitude are multiplied with  $F^\mathrm{LL}_V-1$, where $F^\mathrm{LL}_V$ is the LL S-wave form factor.
The corresponding ratio is ill-defined at the two zero crossings, but is 
still shown to visualize quantitative differences, especially below 
threshold.
  }
\label{fig:interferences}
\end{figure}
In \cref{sss:around_threshold} we have verified that the factorized 
(total) cross  section in the DPA represents a good approximation for
our purposes in the threshold region. The matched cross sections we
construct in \cref{s:matching} will in addition  
include interference (IF) terms between factorized and full QCD-LO matrix 
elements. It is therefore necessary to also examine the relative
complex phases of the  involved matrix elements. In this subsection we
compare inclusive (i.e.\ fully phase-space integrated) IF contributions  
among QCD-LO matrix elements obtained in the factorized DPA and  
signal diagram (SD) approach as well as from the full Born-level SM calculation.
For example, the IF term between full and factorized matrix element reads
$2\Re[\MEfact \cdot \ME_\T{full}^*]$ and is integrated over the 
total $\WbWb{}$ PS. By also considering \MEfact{} multiplied in addition with the corresponding 
nonrelativistic form factor, which contains a non-trivial QCD phase, we can 
fully examine the relative complex phase of this inclusive IF contribution. 
Moreover, we illustrate the effects of Lorentz boosting the momenta in the 
individual parts of the factorized amplitude.
The factorized matrix elements studied here include full spin 
correlations between top production and decay, cf. \cref{eq:ME2_factorized}.

In \cref{fig:interferences} the inclusive IF terms are shown as a functions of \sqrts{}. 
In the left panel only QCD-LO matrix elements are evaluated,
i.e.\ without accounting for the effects  
of the nonrelativistic form factors. The solid red line represents the full 
matrix element squared (times a factor of two) and serves as a reference curve.
The signal-full IF term is represented by the solid blue line. 
This term is around 8\% below the red curve above threshold and
substantially smaller below threshold where the IF terms becomes small.
The difference comes from the non-signal diagrams contained in the full amplitude.
The factorized-full (dashed green) and factorized-signal (dashed orange) 
IF terms are practically equal and very close to the signal-full IF curve.
This is perfectly consistent with our findings for the factorized and signal 
cross sections in \cref{sss:around_threshold}, cf. 
\cref{fig:high_sqrts_behavior_zoom}, and shows that (the implementation of) 
our factorized approach based on the DPA does not lead to additional large 
relative complex phases.

The factorized total cross section can be cast into the EHA form in 
\cref{eq:ME2_factorized_EHA} without loss of generality, and thus factorizes 
into separately Lorentz-invariant and
spin-independent parts associated with top production and decay.
For the factorized total cross section 
we can therefore independently boost the momenta in the different parts even 
without transforming the spin-vectors (spinors) accordingly.
This is, however, not the case on the amplitude level and in particular when 
spin correlations between the production and decay matrix elements are taken 
into account. This can be illustrated considering the resulting IF terms with the 
full (and unmodified) matrix element.  
For example, the effect of boosting all momenta and wave functions in the decay matrix elements to 
the top rest frame, while all momenta and wave functions in the production matrix 
elements remain in the lab frame, is demonstrated by the curves represented 
by the $+$ symbols and labeled ``factorized rest frame'' in 
\cref{fig:interferences}.
We find that the resulting inclusive IF contributions 
with the full (cyan crosses) and signal (purple crosses) matrix 
elements differ completely from the corresponding IF terms using the correctly
factorized (``factorized boosted'') matrix elements with all momenta and spins 
consistently in the same Lorentz frame. This is due to an additional
unphysical and inconsistent relative complex phase that arises in the
``factorized rest frame'' amplitude. 
We stress that here we have only performed the inconsistent boost on the decay 
matrix element in order to illustrate the dramatic effect of an
(admittedly extreme) incorrect implementation of the factorized matrix
elements. As was also demonstrated in~\Rcite{Kilian:2006cj},
maintaining the correct phase between production and decay matrix
elements is essential.

In our matching procedure explained in detail in \cref{s:matching} we eventually multiply 
the factorized matrix elements with the corresponding nonrelativistic S- and P-wave form 
factors. Because the nonrelativistic form factors are sizable and strongly varying 
concerning modulus and complex phase within the threshold region
this substantially modifies the behavior of the factorized amplitude close to 
threshold.
In the right panel of \cref{fig:interferences} we show the effect of multiplying 
the QCD-LO correctly factorized and signal matrix elements with the \tT{} vector current with $F^\mathrm{LL}_V-1$.
The `$-1$' subtraction is introduced in \cref{s:matching} to avoid double counting, and we
therefore account for it here as well.
We find that the correctly factorized-full (dashed red curve) and
signal-full (solid blue curve) inclusive IF terms are very similar,
because the respective QCD-LO amplitudes are very similar in the
threshold region as already discussed above. In comparison, using the
incorrect ``factorized rest frame'' amplitudes instead of the correctly
factorized ones leads to completely inconsistent results (green
crosses). Note that P-wave contributions to the total cross section
are suppressed by $v^2$ and the corresponding form factor is not taken
into account here. However, the conclusions just drawn apply as well
to the resummed and factorized matrix elements involving the
axial-vector current.
\subsubsection{P and CP behavior}
\label{sss:p_and_cp_behavior}
%
The \ac{HA} of the factorized process
in the DPA shown in \cref{eq:ME2_factorized_approx} 
allows to discuss the cross section for resonant $\tT{}$ production 
with different top/anti-top helicities.
This furthermore serves to validate our factorized approximation with
respect to P and CP transformations. 
For simplification, we disable at first the contribution of the s-channel
$Z$ exchange and only consider photon-initiated top-pair production.
The interference between $Z$ and photon exchange is well-studied and
causes the forward-backward asymmetry in top-pair production at lepton 
colliders~\cite{Jersak:1979uv,Jersak:1981sp,Jadach:1986xp}.
Disabling the $Z$ exchange allows us to concentrate on (the validation
of) the symmetry properties of the cross section under P
transformations.

\begin{figure}[htbp]
\centering
\includegraphics[width=\halfwidth]{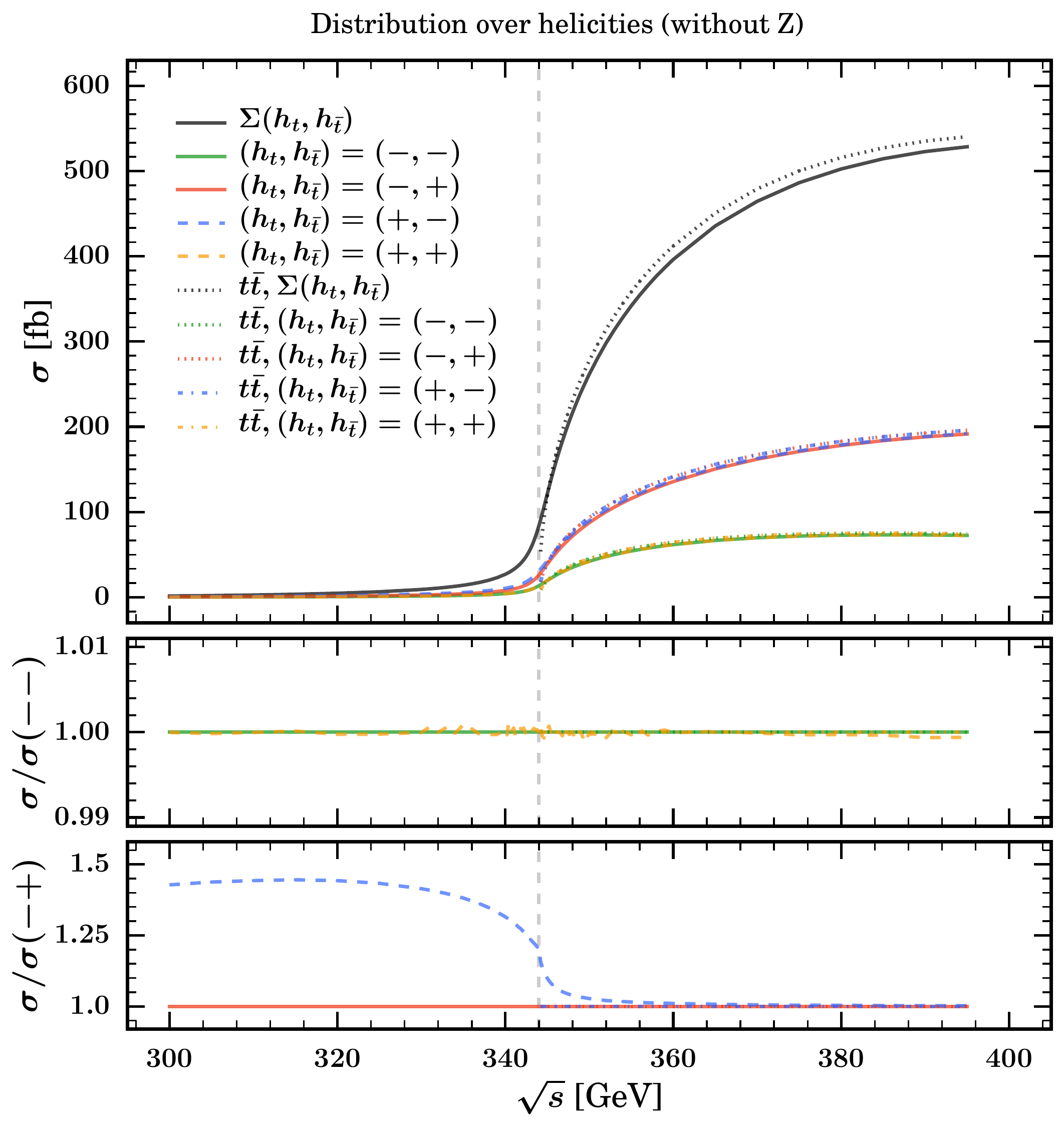}
\includegraphics[width=\halfwidth]{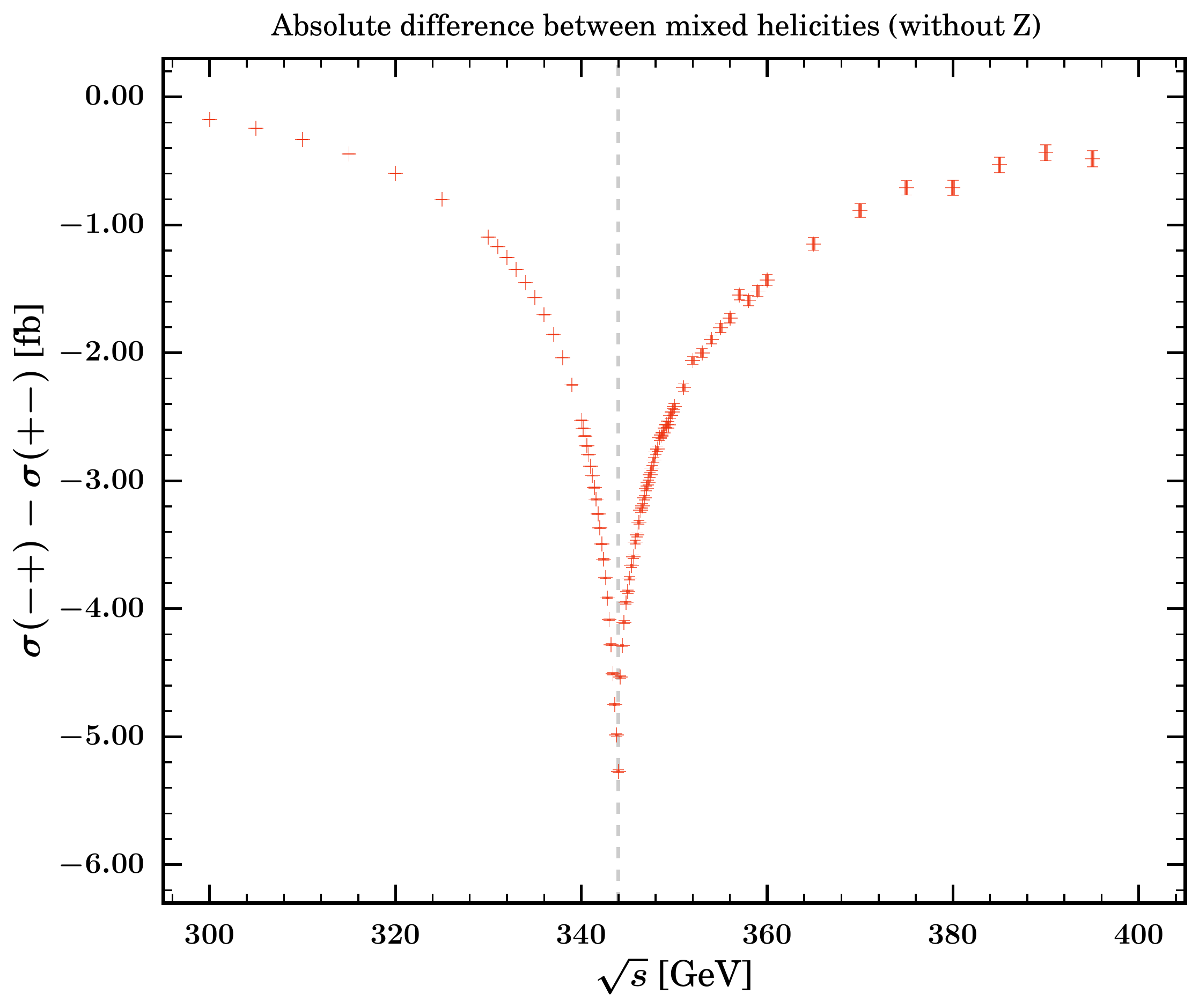}
\caption{In the left panel, we show the LO contributions to the total cross
  section from different top helicity configurations (colored) as well
  as their sum (black) for on-shell $\tT{}$ production (dotted and
  dash-dotted) and $\WbWb{}$ (solid and dashed) final states. 
  The $\WbWb{}$ cross sections are computed in the DPA and HA
  according to \cref{ss:factorization} and
  \cref{sss:helicity_correlations}, respectively. In the lower left
  panel we also show the ratios $\sigma(+,+)/\sigma(-,-)$  and
  $\sigma(+,-)/\sigma(-,+)$ for $\tT{}$ as well as $\WbWb{}$
  production, where the line color is according to the numerator of
  the plotted ratio. As usual, the dashed vertical line is located at
  $\sqrts = 2\mOneS{}$. In the right panel, the absolute difference
  $\sigma(-,+) - \sigma(+,-)$ is shown for $\WbWb{}$ production. Note
  that the absolute difference is slightly asymmetric around
  threshold. Above threshold it grows due to the constant relative
  error visible in the ratio plot and the growing cross section.}
\label{fig:infiniteZ_LO_helicities}
\end{figure}
%
In the left panel of \cref{fig:infiniteZ_LO_helicities}, we show the
distribution of the inclusive LO cross section as a function of \sqrts{} for the different top helicity
configurations as well as the sum over all helicities obtained from $\WbWb{}$ production in the DPA and 
from on-shell $\tT{}$ production.
Above \SI{360}{\GeV} the $\WbWb{}$ and $\tT{}$ cross sections are fairly
similar.
Especially the respective ratios of mixed top helicities
$\sigma(+,-)/\sigma(-,+)$ and equal top
helicities $\sigma(+,+)/\sigma(-,-)$ are both very close to unity.
The $\tT{}$ ratios are equal to one within the numerical (MC)
integration uncertainties for all energies. As the $\tT{}$ cross
section is given by the absolute square of the production matrix
element, we thus find that 
$|\MEprod{}|^2(-,-)=|\MEprod{}|^2(+,+)$ as well as
$|\MEprod{}|^2(+,-)=|\MEprod{}|^2(-,+)$.
This is, of course, to be expected as the electromagnetic production of
fermion pairs conserves parity (P).

When accounting for the top decay, however, only the combination
of charge conjugation and parity (CP) is a symmetry, due to the
left-handed coupling. This implies for the \ac{DPA} in the \ac{HA} that
\begin{align}
  \abs{\MEfact^\T{HA}}^2\Al(+,+\Ar) &=
    \abs{\propagators}^2 \abs{\MEprod^{+,+}}^2
    \abs{\MEdecT^{+}}^2 \abs{\MEdecTbar^{+}}^2 \no
    &\xrightarrow{CP}
    \abs{\propagators}^2 \abs{\MEprod^{-,-}}^2
    \abs{\MEdecTbar^{-}}^2 \abs{\MEdecT^{-}}^2= \abs{\MEfact^\T{HA}}^2\Al(-,-\Ar) \po
\label{eq:ME2_factorized_CP_pp}
\end{align}
Note that $t$ and $\bar{t}$ have been swapped due to the C conjugation in
\cref{eq:ME2_factorized_CP_pp}, but due to the equal helicities this
has no effect.
For mixed top helicities, this is not the case.
In fact,
\begin{equation}
  \abs{\MEfact^\T{HA}}^2\Al(+,-\Ar)\xrightarrow{CP}
  \abs{\MEfact^\T{HA}}^2\Al(+,-\Ar)\neq
  \abs{\MEfact^\T{HA}}^2\Al(-,+\Ar)\po
\label{eq:ME2_factorized_CP_pm}
\end{equation}
So just from the CP properties, we cannot infer the correct behavior of the
ratio of mixed helicities for $\WbWb{}$ production in the \ac{DPA}.
It is thus interesting to see that for high energies the ratio still
approaches unity and P becomes approximately a good symmetry.
In the right panel of \cref{fig:infiniteZ_LO_helicities}, we also show
the absolute difference between the contributions with mixed
helicities to $\WbWb{}$ production in the \ac{DPA}. It is remarkably
symmetric around threshold. In principle, one could work out the exact
analytic result for this (P-violating) difference, but this is beyond
the scope of our validation.

\begin{figure}[htbp]
\centering
\includegraphics[width=\halfwidth]{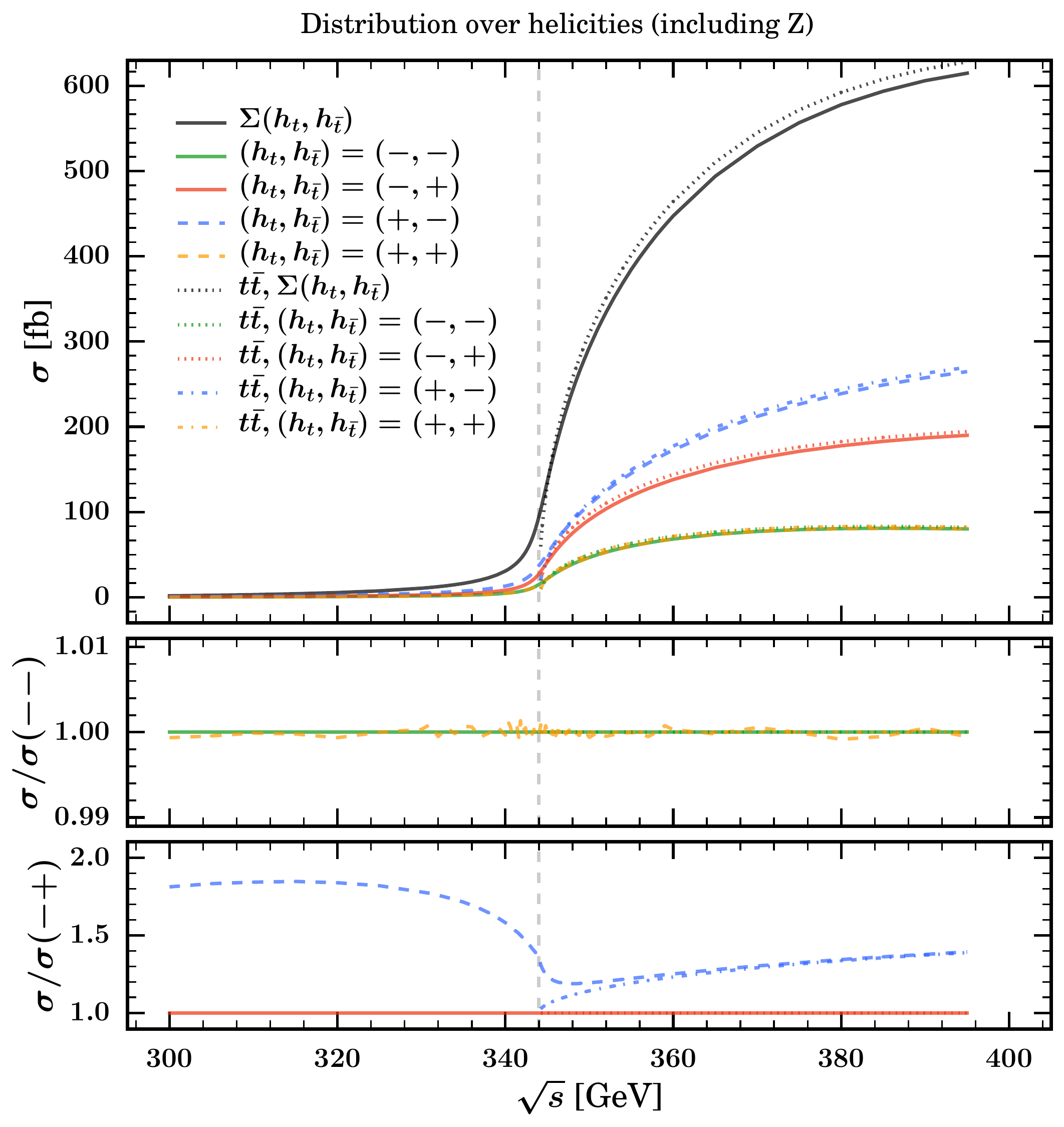}
\caption{All curves and color coding as in the left panel of
  \cref{fig:infiniteZ_LO_helicities}, but now with the effect of
  $Z$ exchange in the production process taken into account.}
\label{fig:LO_helicities}
\end{figure}

Finally, in \cref{fig:LO_helicities}, we also show the effect of
including the production channel via $Z$ exchange. 
As expected from \cref{eq:ME2_factorized_CP_pp}, equal top/anti-top
helicities in $\tT{}$ and $\WbWb{}$ production still give equal
results, as also shown in the upper ratio plot. However, now there is
a larger P-violating difference between the two different mixed
helicities for all energies that grows with energy. This difference is  
now also present for $\tT{}$ production.
We observe  that the $Z$ exchange enhances mostly the $(+,-)$
configuration, while the effect on the other helicity configurations
is comparatively small. This is also visible in the lower ratio plot
where  the $(+,-)$ configuration is compared to the $(-,+)$ one. 
As for the case of pure photon exchange, the contributions from  mixed
helicities are larger than from equal helicities. 
This can be understood from the fact that for massless quark production, 
mixed helicities are the only contributing configurations due to the
spin-1 intermediate gauge-bosons.
Hence, equal helicity contributions only arise due to the spin flip associated
with the top quark mass and are therefore less likely to occur at high energies.
Concerning the ratios, we see that for equal helicities they remain at
unity.  For mixed helicities the ratios for $\WbWb{}$ and $\tT{}$
production approach each other for c.m.\ energies above \SI{360}{\GeV}
as we already observed for pure photon exchange.

\subsection{NLO QCD corrections with WHIZARD}
\label{ss:whizard_nlo}
%
\whz~can compute fixed-order QCD-NLO corrections to various
processes in an automated manner.
In a recent publication~\cite{1609.03390}, some of us studied the QCD-NLO
corrections to $\eetT{}$ in the relativistic continuum with fully off-shell top 
and gauge boson decays.
This study also included a thorough validation against other MC generators that
are able to compute this process.
Thus, for fixed-order QCD corrections, we can build upon a well-tested
framework.

\whz at QCD-NLO uses the FKS subtraction
scheme~\cite{hep-ph/9512328,0908.4272}, both in the standard
approach and the resonance-aware extension~\cite{1509.09071}.
The FKS scheme uses a partition of the real phase-space into regions,
where only one divergent configuration exists.
In each of these regions, the divergence is regulated using
plus-distributions.
In the FKS scheme, we can make use of the optimized multi-channel
phase-space generator for the underlying Born kinematics, as the full
real phase-space factorizes into Born times radiation phase-space.
Starting from the Born phase-space configuration, real kinematics are
generated according to the specific kinematics of each singular region.

\whz has been interfaced to the One-Loop providers (OLPs)
\gosam~\cite{Cullen:2014yla} and \openloops~\cite{Cascioli:2011va} with the
BLHA interface~\cite{1308.3462}, and to
\recola~\cite{1211.6316,1605.01090} with a dedicated interface.
In addition to virtual matrix elements, we can use these interfaces to
obtain color- and spin-correlated, as well as tree-level matrix elements
(as alternative to \OMega).
Events can be generated at fixed-order or using the \powheg{} matching
scheme~\cite{Nejad:2015opa}. More details can be found
in~\Rcite{Weiss:2017qbj}.
%
%
\subsubsection{Setup of fixed-order corrections in WHIZARD}
\label{sss:fixed_order_corrections}
As motivated in \cref{ss:factorization}, we will include the form factor
in the $\tT{}$ production matrix element using the \ac{DPA}.
For the decay matrix elements we compute the full set of (relativistic) QCD-NLO corrections to the
on-shell top quark decay using the momentum projections dicussed in \cref{sss:on_shell_projection}. 
We can therefore achieve QCD-NLO precision for observables that probe the decay kinematics.
Our approach entails that we use the NLO width in the matrix elements and form factors  
of the matched computation when we account for the QCD-NLO corrections in the decay. 
In addition, we also employ the full $\WbWb{}$ fixed-order results at QCD-LO and
QCD-NLO.
We will now introduce a convenient graphical notation for the various components of our matched calculation 
that will be useful for the discussion of the matching procedure in \cref{s:matching}.

For the full $\WbWb{}$ process at QCD-LO, which includes all
non-signal and background processes and their interference, we
use the notation 
\begin{align}
  \XSLO    &= \abs{\diagram{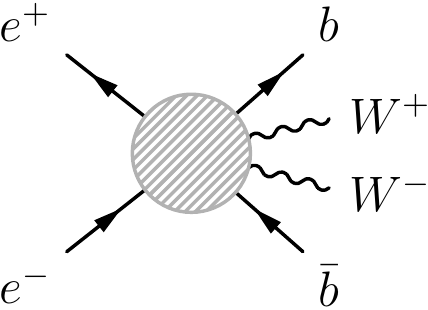}}^2 \,,
\label{eq:LOthreshold}
\end{align}
which represents the modulus square of the sum of all $\WbWb{}$ tree-level
diagrams and where the phase space integrations are implied. 

Concerning QCD-NLO contributions, $\diagramInText{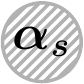}$ stands for the sum of all
virtual gluon exchange one-loop diagrams for the $\WbWb{}$  final state, and
$\Bl(a\left\lmoustache b\right.\Br) \equiv 2 \Re\ofIT{a \cdot
b^*}$. In this notation the QCD-NLO fixed-order cross section is represented as
\begin{align}
  \XSNLO &= \XSLO + \realOfDiagrams{LOfull}{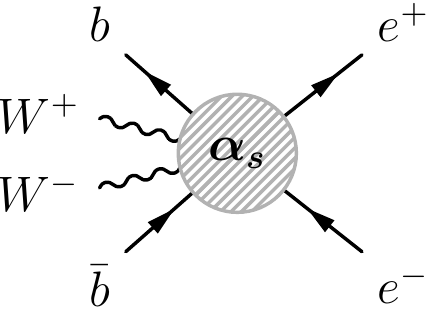}
                 +\abs{\diagram{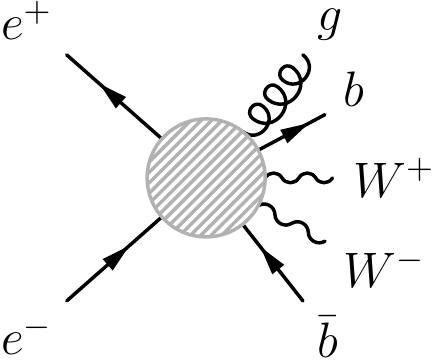}}^2\,,
\label{eq:NLO}
\end{align}
where in $\XSLO$ a consistent use of the QCD-NLO corrected parameters
such as the top quark mass and width is implied. The last term stands
for the real radiation corrections, and in this context   
all infrared cancellations between real and virtual corrections are
guaranteed by the KLN theorem. 

In the factorized DPA approach, we use for the QCD-LO cross section the notation
\begin{align}
  \XSLO^\T{fact} &= \abs{\diagram{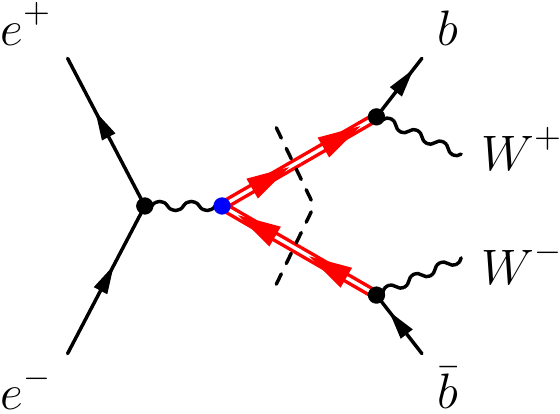}}^2\,,
\label{eq:factorized_LO}
\end{align}
which was already introduced in Fig.\ \ref{fig:factorized}.
The double lines denote the top and anti-top propagators and the
dashed lines indicate the factorized computations of the matrix
elements  with the on-shell projections. The corresponding cross
section including QCD-NLO corrections to the decay of top and anti-top
quarks is then represented by the terms 
\begin{align}
  \XSNLO^\T{fact} &= \XSLO^\T{fact}
                   + \abs{\diagram{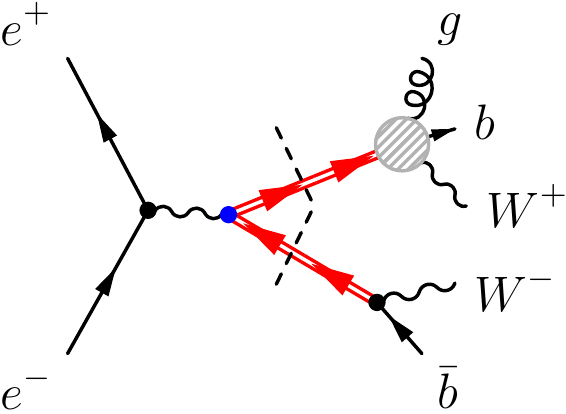}}^2
                   + \abs{\diagram{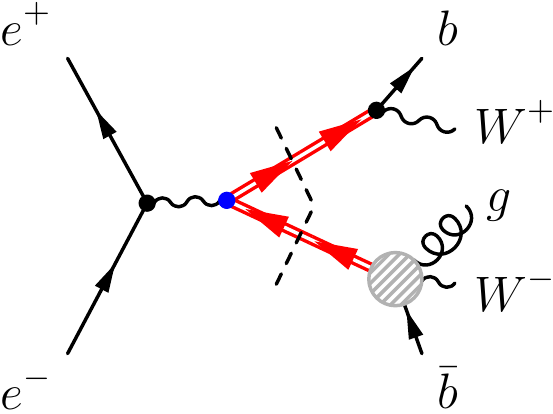}}^2 \no
                  &+ \realOf{\diagram{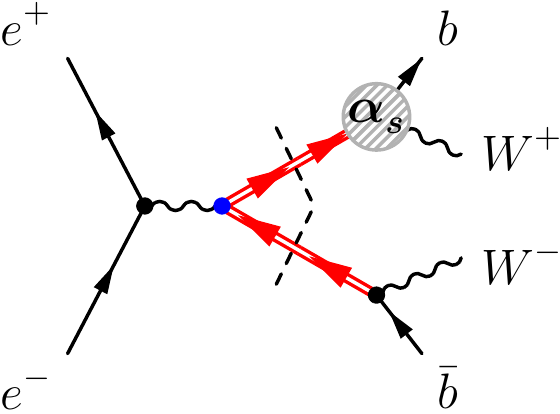}
                   + \lessspace\diagram{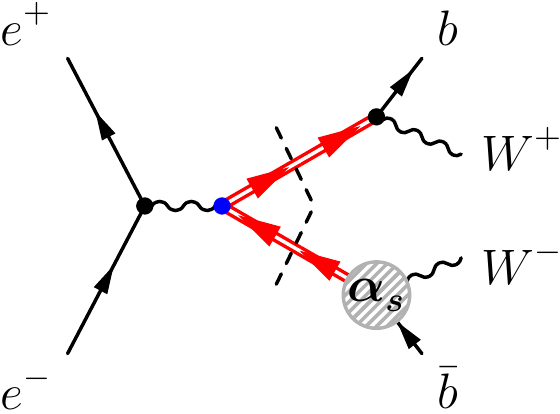}}
                     {\diagram{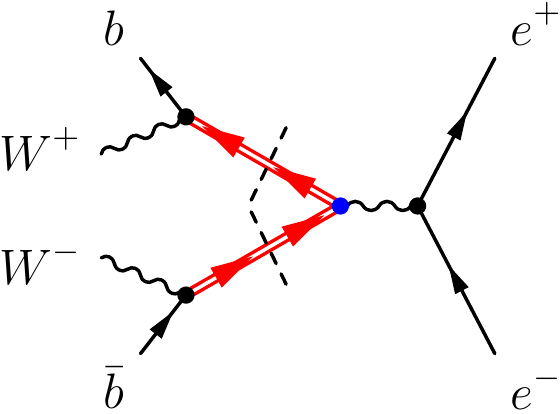}}\co
\label{eq:factorized_NLO}
\end{align}
applying our notations for the QCD-NLO fixed-order cross section
described above also for the top and anti-top decays. 
The diagrams in \cref{eq:factorized_NLO} can individually carry IR divergences which, however, cancel in the sum.
Note that in the factorized approach we omit the real final-state gluon interference
diagrams of the kind 
\begin{align}
  \realOf{\diagram{factorizedReal}}{\diagram{factorizedRealOtherLeg}}
\,,
\label{eq:realWithReal}
\end{align}
which correlate top and anti-top decays, as well as virtual gluon
corrections of the kind 
\begin{align}
\realOf{\diagram{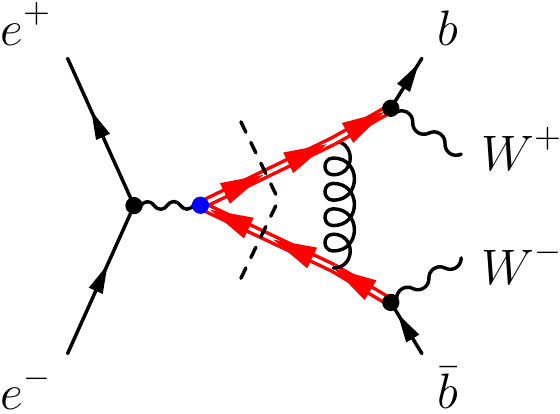}+\diagram{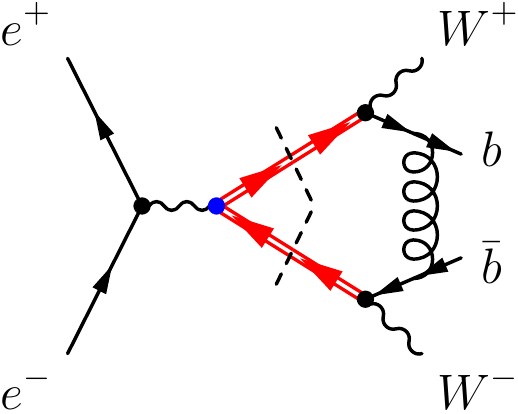}+\diagram{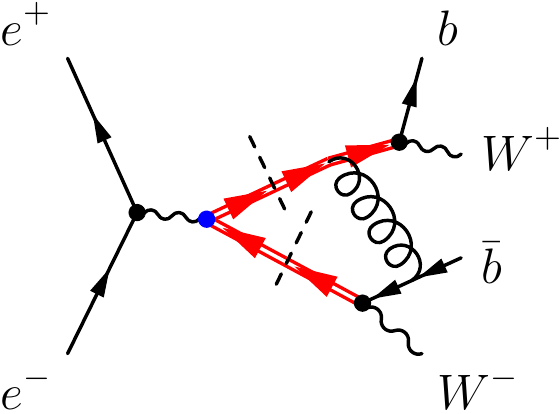}+\ldots}{
\diagram{factorizedLessDetailFlipped}} \po
\label{eq:realWithVirtual}
\end{align}
including diagrams involving e.g.\ gluon exchange between the final
state $b$ quark and the anti-top. As we already explained in
\cref{s:intro}, in the threshold region, which is where we employ  
the factorized approach, these kinds of corrections cancel at NLL
order in the total cross section~\cite{Fadin:1993dz,Melnikov:1993np}, 
or for acceptance cuts that  
do not resolve the $\tT{}$ double resonant portion of the $\WbWb{}$ 
phase space~\cite{1002.3223,1004.2188}. Such corrections, however, in 
general contribute to differential distributions at NLL order in 
the threshold region~\cite{Peter:1997rk} and have, eventually, to be 
included in the context of a more sophisticated approach. Thus the
approach we describe in this work provides differential predictions at
LL order in the threshold region. We emphasize, however, that via the
matching procedure we still include the full set of fixed-order relativistic QCD-NLO
corrections associated with the corresponding diagrams with offshell
top quarks in an exact way. At high energies we thus have fully differential
predictions at QCD-NLO precision.

\subsubsection{Modifications to standard FKS for factorized QCD-NLO}
\label{sss:fks_mod}
In the following, we review the three main modifications to the FKS
subtraction needed to cope with our factorized QCD-NLO computations.
We focus on top production but the statements also hold for analogous
processes where the \ac{DPA} to Born, Real, and Virtual contributions 
to unstable particle decays shall be computed at QCD-NLO.
We note that if one determines QCD-NLO corrections to production and
decay simultaneously, one encounters ambiguities in the real radiation
computation. This is due to the difference in the invariant mass of
the resonance in case the radiation, carrying momentum, occurs in the
production or the decay stages. Thus, if one is only interested in
fixed-order corrections, hybrid schemes have been devised, where the
\ac{DPA} is only applied to the virtual part~\cite{hep-ph/0005309}.
In the threshold region, however, the dominant corrections to the
production process are already encoded in the form factors. They
contain the resummed Coulomb singular terms, as already discussed in
\cref{eq:factorized}, and we do not consider real corrections
to the production as (ultra)soft radiation off the nonrelativistic top quarks is of higher order.
\paragraph{On-shell generation of the real phase-space}
Like the tree-level matrix element, the (decay) matrix elements with real gluon emission have to be
evaluated using on-shell projected momenta.
To generate this phase-space, we use the same mappings as in
resonance-aware FKS.
In this approach, the real emission is generated in such a way that the
invariant mass of the respective resonance is kept at its Born value,
which removes mismatches between the real matrix-element and its soft
approximation.
Thus, starting from an already on-shell projected Born momentum
configuration, obtained as described in \cref{sss:on_shell_projection}, we apply
this mapping to obtain an also on-shell projected real phase-space
point.
Note that, to ensure correct subtraction of soft divergences, also the
real-emission FKS variables $\xi$ and $y$ need to be computed in the
on-shell projected Born system.
We stress that the on-shell momenta only enter the matrix elements and
their subtraction terms but not the phase-space Jacobian.
For the latter as well as for event generation, the physical (and in general 
off-shell concerning the top quarks)
phase-space is used, which is generated alongside the on-shell case.
\paragraph{Decay subtraction}
The IR divergences in the factorized real corrections all originate from the $t
\to b W g$ matrix element.
It consists of two Feynman diagrams.
One in which the gluon is emitted from the top quark and another one in
which it is emitted from the bottom quark.
Divergences can only occur in emissions from particles with on-shell
momenta and zero width.
Therefore, in the full $\WbWb{}$ matrix element, emissions from internal
top quarks do not yield divergences, as they are regularized by the width and the virtuality.
However, in the factorized approach, the gluon emission from the top
quark is a singular contribution as there is no top width insertion in the virtual top line
that emerges after gluon radiation. It 
therefore needs to be treated by the FKS subtraction.
We call this additional singular region a \textit{pseudo-ISR} region
because its underlying kinematics in the decaying top quark decay is
similar to the case of QCD initial state radiation.
This way, each singular pair index $(b,g)$  and $(\bar{b},g)$ is
associated with a pseudo-ISR tuple $(b,g)^*$ and $(\bar{b},g)^*$, in
which the gluon radiation is emitted not from the bottom, but from the top
quark.
This implies that in the corresponding singular region, the FKS
phase-space contribution
\begin{equation*}
  d_{ij} = 2 \left(p_i \cdot p_j\right) \frac{E_i E_j}{\left(E_i + E_j\right)^2}
\end{equation*}
is evaluated with $p_i \to p_t = p_b + p_W$.
\paragraph{Omission of interference terms}
In the real matrix element, we omit interference terms between gluon
emissions from different top quark legs, cf. \cref{eq:realWithReal}.
In consequence, we remove these interference contributions from the
color-correlated Born matrix element.
The same reasoning applies to the corresponding virtual corrections
and their subtractions. This means that the loop matrix elements we consider do not
include diagrams with virtual gluon exchange between quarks associated with
different top legs, cf. \cref{eq:realWithVirtual}. The dominant contributions from these gluons are already included in the (Coulomb) resummed form factors. 
Therefore, also in the soft part of the virtual subtraction
terms, we leave out all terms corresponding to gluon exchange between
different top quark legs. The absence of these interference terms
allows to split up the FKS regions into two disjoint subsets of
singular pairs, as depicted in \cref{tab:fks_factorized}.
\begin{table}[htbp]
  \caption{Singular regions in standard FKS for the full process $e^+
  e^- \to W^+ W^- b \bar{b}$ (left table) and in modified FKS for the
  factorized process (right table). The latter split up into interference-free
  subsets and involve pseudo-ISR regions. $\alpha_r$ is the index for
  the singular regions. External particles are labelled $1\, (e^+)$, $2\,
  (e^-)$, $3\, (W^+)$, $4\, (W^-)$, $5\, (b)$, $6\, (\bar{b})$, $7\,
  (g$, for the real radiation). The singular FKS pairs consist of
  emitter and radiated particle, and the asterisk denotes the special
  configurations, where the emitter is the intermediate top or
  anti-top quark. For more details see \Rcite{Weiss:2017qbj}.}
  \vspace{1em}
  \centering
  \begin{tabular}{cc}
    \begin{tabular}[t]{c c c}
     \toprule{}
     $\alpha_r$ & emitter & singular pairs \\
     \midrule{}
     1 & 5 & $\left\lbrace(5,7), (6,7)\right\rbrace$ \\
     2 & 6 & $\left\lbrace(5,7), (6,7)\right\rbrace$ \\
     \bottomrule{}
    \end{tabular}
    \qquad
    \begin{tabular}[t]{c c c c}
     \toprule{}
     $\alpha_r$ & emitter & pseudo-ISR & singular pairs \\
     \midrule{}
     1 & 5 & no  & $\left\lbrace(5,7), (5,7)^*\right\rbrace$ \\
     2 & 5 & yes & $\left\lbrace(5,7), (5,7)^*\right\rbrace$ \\
     3 & 6 & no  & $\left\lbrace(6,7), (6,7)^*\right\rbrace$ \\
     4 & 6 & yes & $\left\lbrace(6,7), (6,7)^*\right\rbrace$ \\
     \bottomrule{}
    \end{tabular}
  \end{tabular}
  \label{tab:fks_factorized}
\end{table}


%
\section{Validation for the inclusive cross section}
\label{s:validation}
%
In order to validate the \whz{} implementation of the combination of
our factorized matrix element approach and the nonrelativistic form
factors, we compare in this section the numerical MC results
for the inclusive cross section with available analytic
calculations in the threshold region obtained as described in 
\cref{ss:sigmaincl}. 
To avoid contributions from unphysical phase space regions contained
in the nonrelativistic analytic results (which come from the expansion
in inverse powers of the top quark mass), we apply a cut \DeltaM{}  on
the invariant mass of the reconstructed top momenta of the form
\begin{align}
  \abs{\sqrt{(\pWp + \pb)^2} - \mOneS} \le \DeltaM{} \MT{and}
  \abs{\sqrt{(\pWm + \pB)^2} - \mOneS} \le \DeltaM{} \po
\label{eq:DeltaMcut}
\end{align}
While  in \whz  the cut is implemented exactly as shown in \cref{eq:DeltaMcut}, 
in the analytic calculation
we implement a cut on the \emph{nonrelativistic}
invariant mass variables as explained in \cref{ss:sigmaincl}, 
see \cref{eq:sigmaCut}.

The difference in the implementation of the cut is one source of disagreement
between the MC and the analytic results.
In the threshold region (and for reasonably small cuts), this difference 
is, however, of higher order.
For the purpose of validation, in the following we only discuss comparisons 
involving the dominant vector-current induced cross section and the 
correponding S-wave form factor. The axial-vector current together with its 
(P-wave) form factor only contributes beyond NLL to the inclusive cross 
sections we consider here.

We also want to reiterate that the results discussed in this section
only involve the $\tT{}$-double-resonant contributions contained in
the factorized matrix element approach on the side of the MC
calculation, which (near threshold) corresponds to the double-resonant 
nonrelativistic calculation in \cref{eq:sigmaCut}. 
We thus omit the contributions related to non-$\tT{}$-resonant $\WbWb{}$ production 
discussed in \Rcites{1002.3223,1004.2188,Ruiz-Femenia:2014ava},
but note that they are included in our final matched predictions through
the full QCD-LO and QCD-NLO $\WbWb{}$ cross section calculations, as
described in \cref{s:matching}. Furthermore, in this validation section 
we use the \ac{EHA} in \whz{} and consistently treat the top decay 
at LO in all analytic and MC results.
\subsection{Reconstructed top invariant mass scans}
\label{ss:validation_delta_m_t_scans}
\begin{figure}[htbp]
\centering
\includegraphics[width=\halfwidth]{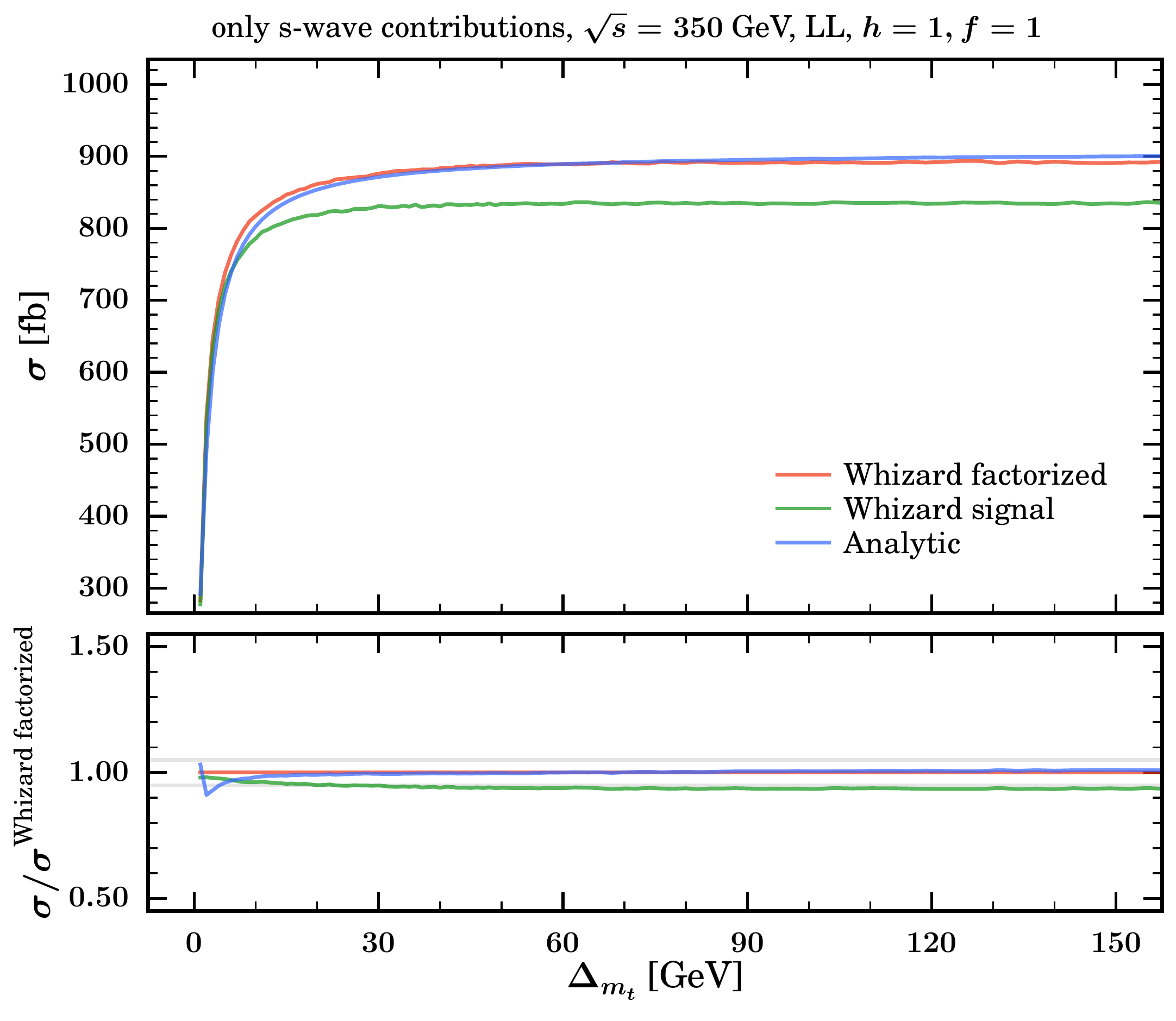}
\includegraphics[width=\halfwidth]{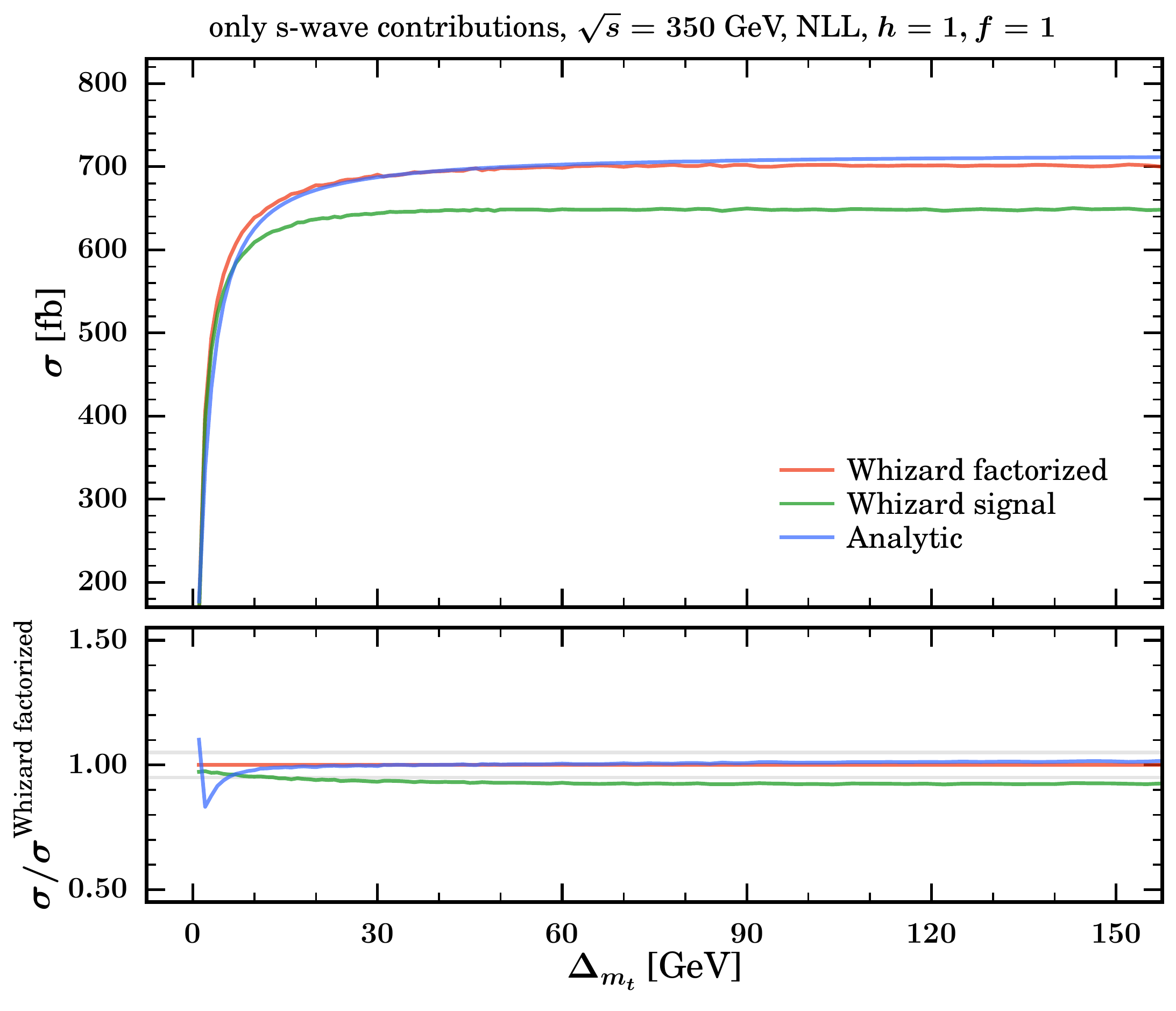}
\caption{Comparison of analytic results (blue) with the implementation in
  \whz{} with the factorized (red) and the signal-diagram approach (green) for
  $\sqrt{s}=\SI{350}{\GeV}$ using a LL or NLL form factor.
  For better orientation, we indicate here and in \cref{fig:validation_dm_scan_330} 
  the $\pm\SI{5}{\percent}$
  range in the lower ratio plots with horizontal gray lines.}
\label{fig:validation_dm_scan_350}
\end{figure}
\begin{figure}[htbp]
\centering
\includegraphics[width=\halfwidth]{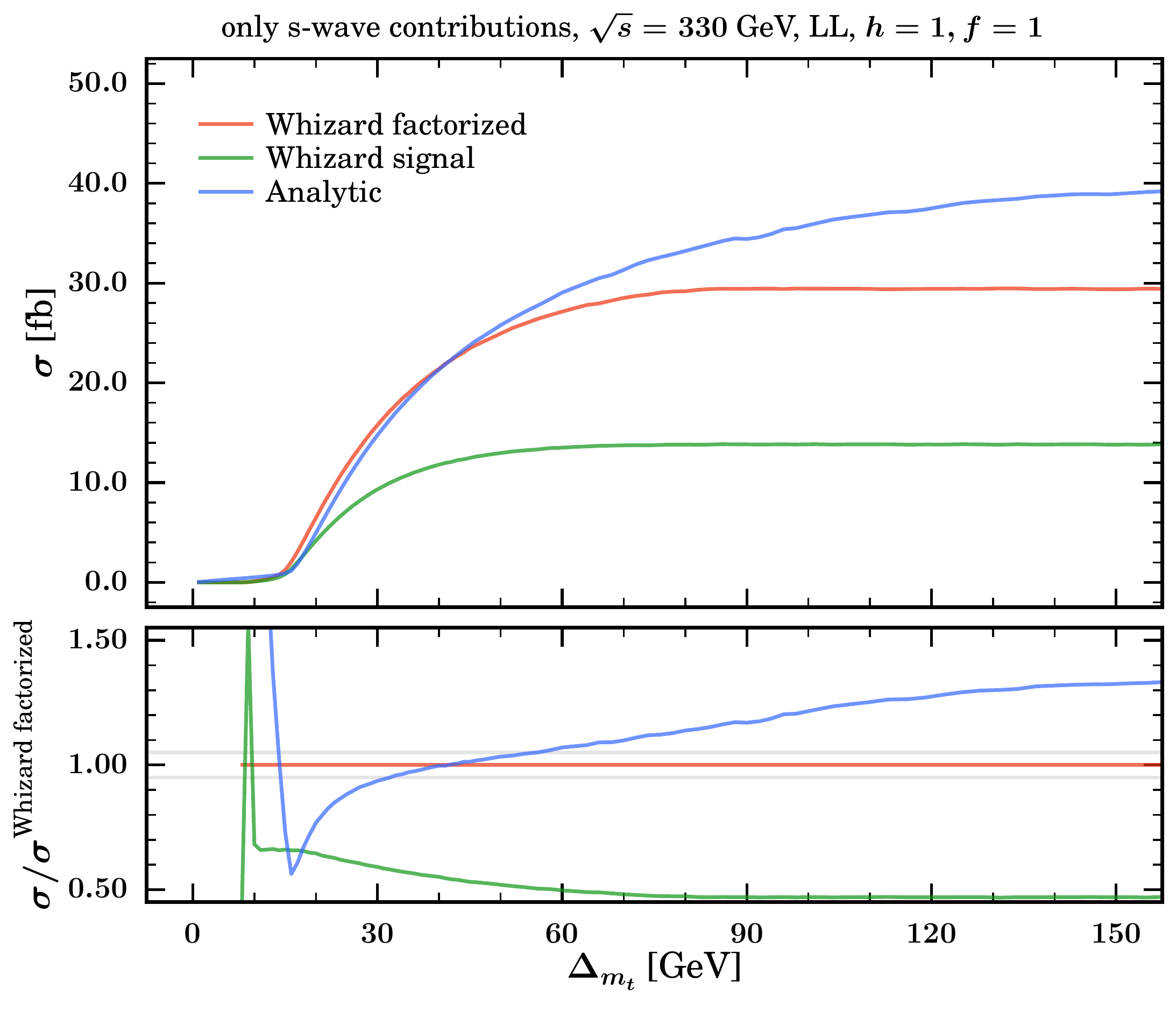}
\includegraphics[width=\halfwidth]{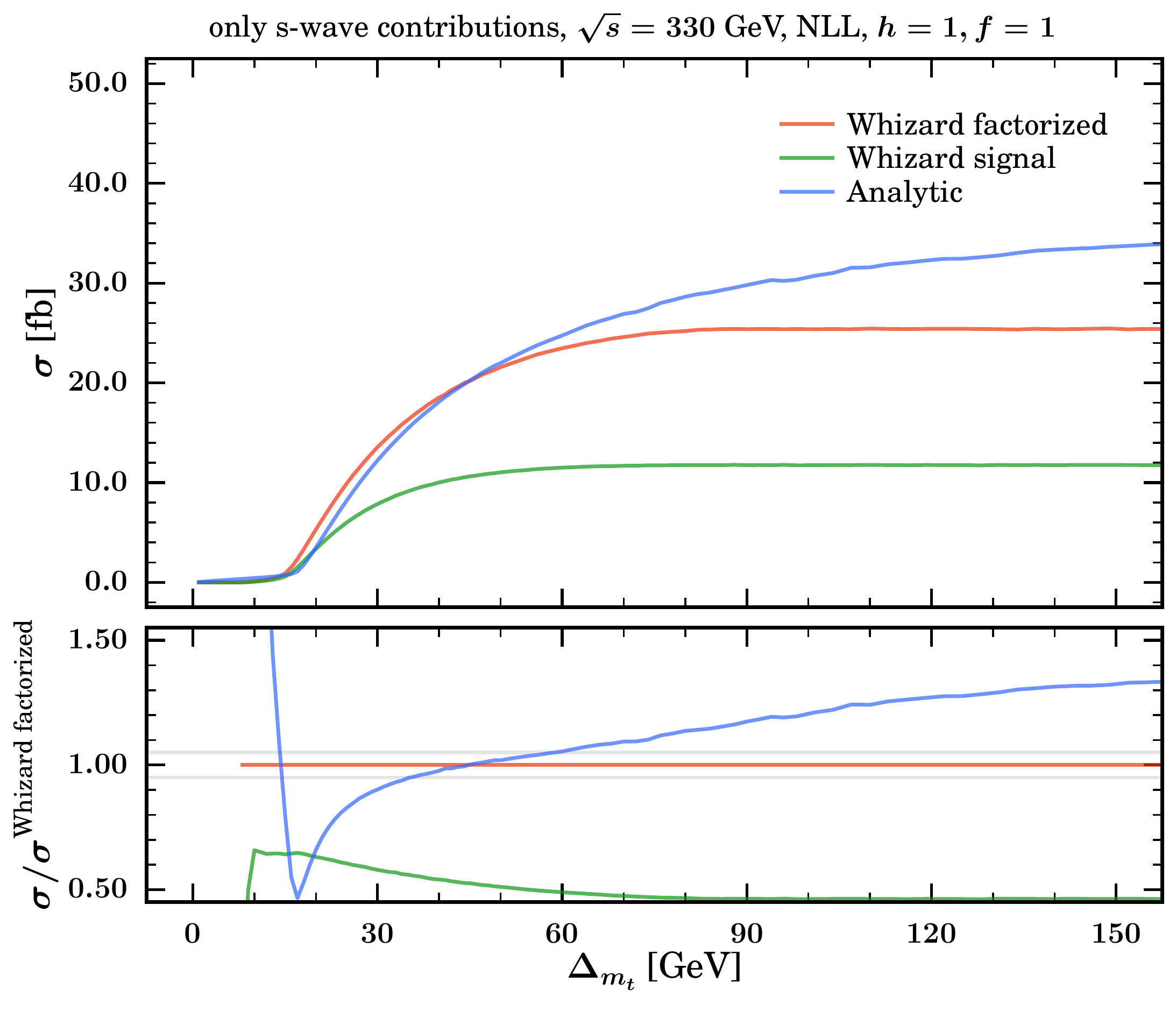}
\caption{Comparison of analytic results (blue) with the implementation in
  \whz{} with the factorized (red) and the signal-diagram approach (green) for
  $\sqrt{s}=\SI{330}{\GeV}$ using a LL or NLL form factor.}
\label{fig:validation_dm_scan_330}
\end{figure}
In \cref{fig:validation_dm_scan_350}, we show the cross section as a function 
of \DeltaM{} using the form factor at LL and NLL order for $\sqrts=\SI{350}{\GeV}$, 
which is about $6$~GeV above the toponium peak position.
As expected, the ratios of the factorized \whizard{} (red) and the
analytic results (blue) are nearly independent of the used form
factor, because in both calculations the top decay factorizes by
construction. 
The differences for small \DeltaM{} originate from a 
different implementation of the cuts as discussed below.
At $\sqrts=\SI{350}{\GeV}$, both approaches yield nearly the same results for 
all values of \DeltaM{} above $10$~GeV. The maximal deviation is about one 
percent.
For comparison, we have also shown the corresponding results based on
the signal diagram (green), which we already discussed to be inconsistent. We
see that it yields results that are too small by up to 5\% for all values of \DeltaM{}. 
For \DeltaM{} below
$10$~GeV the analytic results fall off below the \whizard{} results,
and the relative difference reaches $\sim 10$\% for cuts around $1$ to
$2$~GeV. This disagreement is due to the approximate implementation 
of the invariant mass cuts, \cref{eq:DeltaMcut}, in the analytic result. The 
latter uses an expansion in inverse powers of \DeltaM{}, see
\cref{eq:LambdaApprox}, and is therefore unreliable for tight
invariant mass cuts. The relatively good agreement for very high
cuts of $80$~GeV and beyond shows that for energies above the
threshold the numerical effects of the unphysical phase space regions
contained in the nonrelativistic calculation are relatively small.

In \cref{fig:validation_dm_scan_330} the analogous results are shown
for $\sqrts=\SI{330}{\GeV}$, which is about $6$~GeV below the toponium
peak position.  
In this kinematic regime the inclusive cross section is
already very small and the concept of  $\tT{}$-double-resonant $\WbWb{}$
production loses its meaning because it is not possible to have top
and anti-top quarks on-shell at the same time. Here, off-shell effects
and non-resonant processes are important. Indeed, also the MC
integration of \whz{} does not find double-resonant configurations
within a \SI{5}{\GeV} window at this energy.  The analytic
calculation, which is based on the concept of determining
$\tT{}$-double-resonant phase space configurations by 
expansions in $v$ and inverse powers of the invariant mass cut, cf. e.g.
\cref{eq:DeltaMcut}, is therefore not expected to provide a very precise 
description.
This is confirmed in our comparison shown in \cref{fig:validation_dm_scan_330}, 
where we see that the  analytic calculation only works well in a narrow region 
of $\DeltaM{}$ around $40$~GeV, where the precise location of this region should be 
considered accidental. For lower cuts the analytic computation becomes 
unstable and shows an unphysical (though in absolute numbers tiny) rise for 
values below $\sim\SI{15}{\GeV}$.
At high values of \DeltaM{} the analytic prediction is substantially larger than 
the factorized calculation indicating that the relative size of the 
contributions from unphysical portions of the phase space in the 
nonrelativistic analytic calculation is particularly large for energies below 
threshold.  
The \whz{} computation, on the other hand, correctly stabilizes for \DeltaM{} 
of \SI{80}{\GeV} and larger as the physically correct (relativistic) 
phase-space does not allow for larger invariant masses at this energy.
We do not want to leave unmentioned that the signal-diagram calculation is, as 
expected from \cref{sss:around_threshold}, completely unreliable at this energy.
We also remark again that for all cases the ratios are independent of 
which approximation is used for the form factor.
\subsection{Center-of-mass energy scans}
\label{ss:validation_sqrts_scans}
\begin{figure}[htbp!]
\centering
\includegraphics[width=0.41\textwidth]{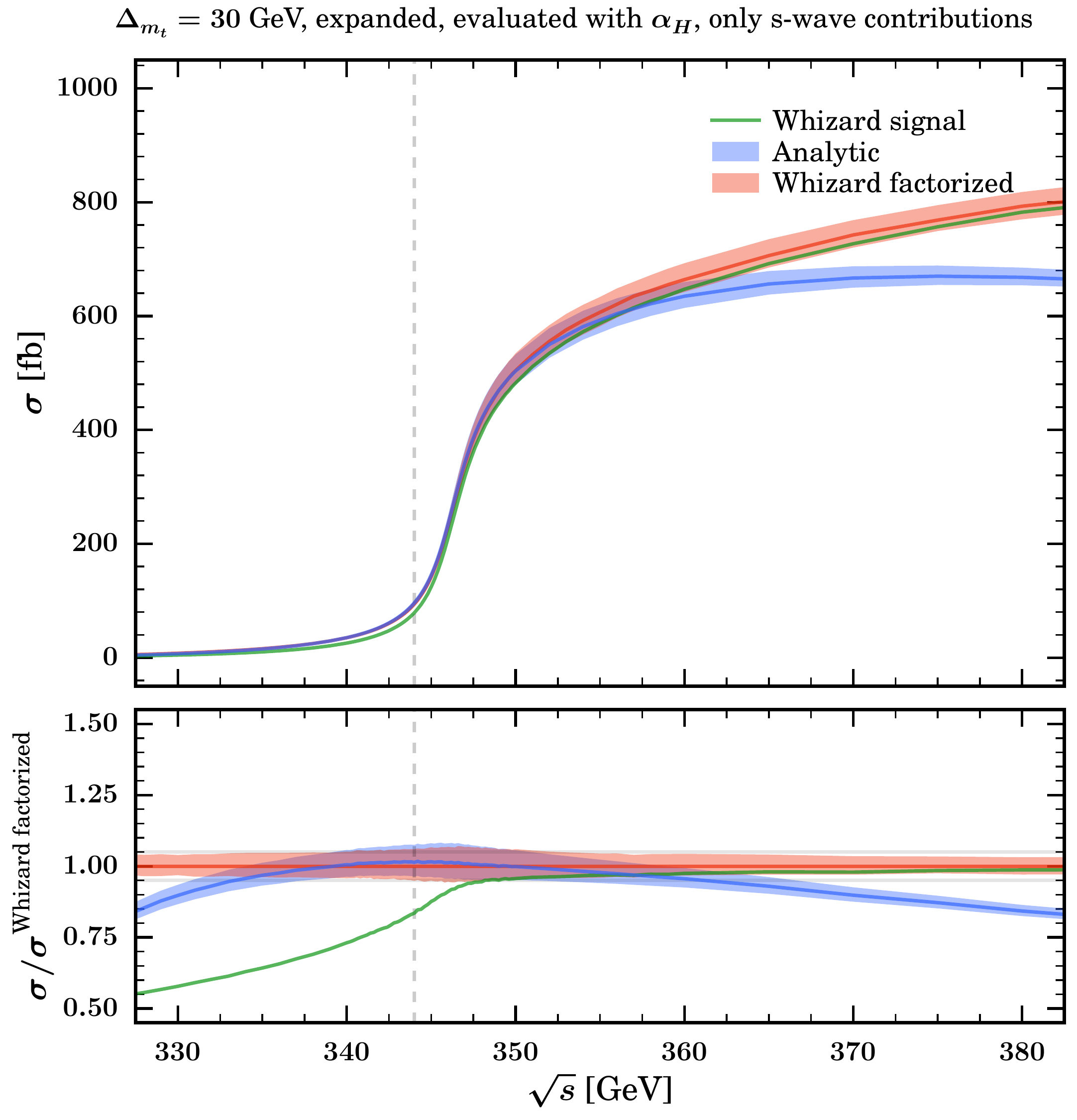}
\includegraphics[width=0.41\textwidth]{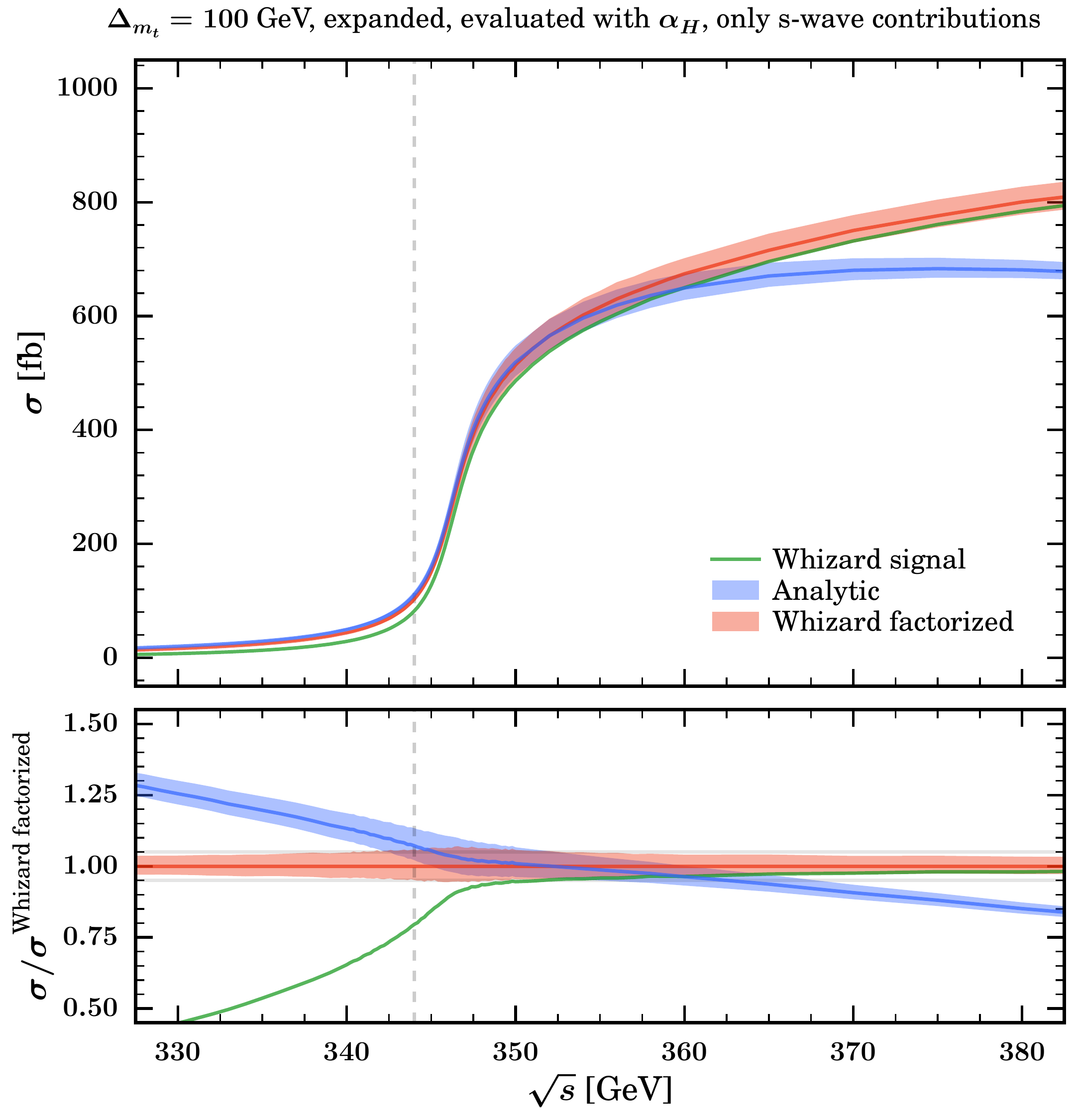}\\
\includegraphics[width=0.41\textwidth]{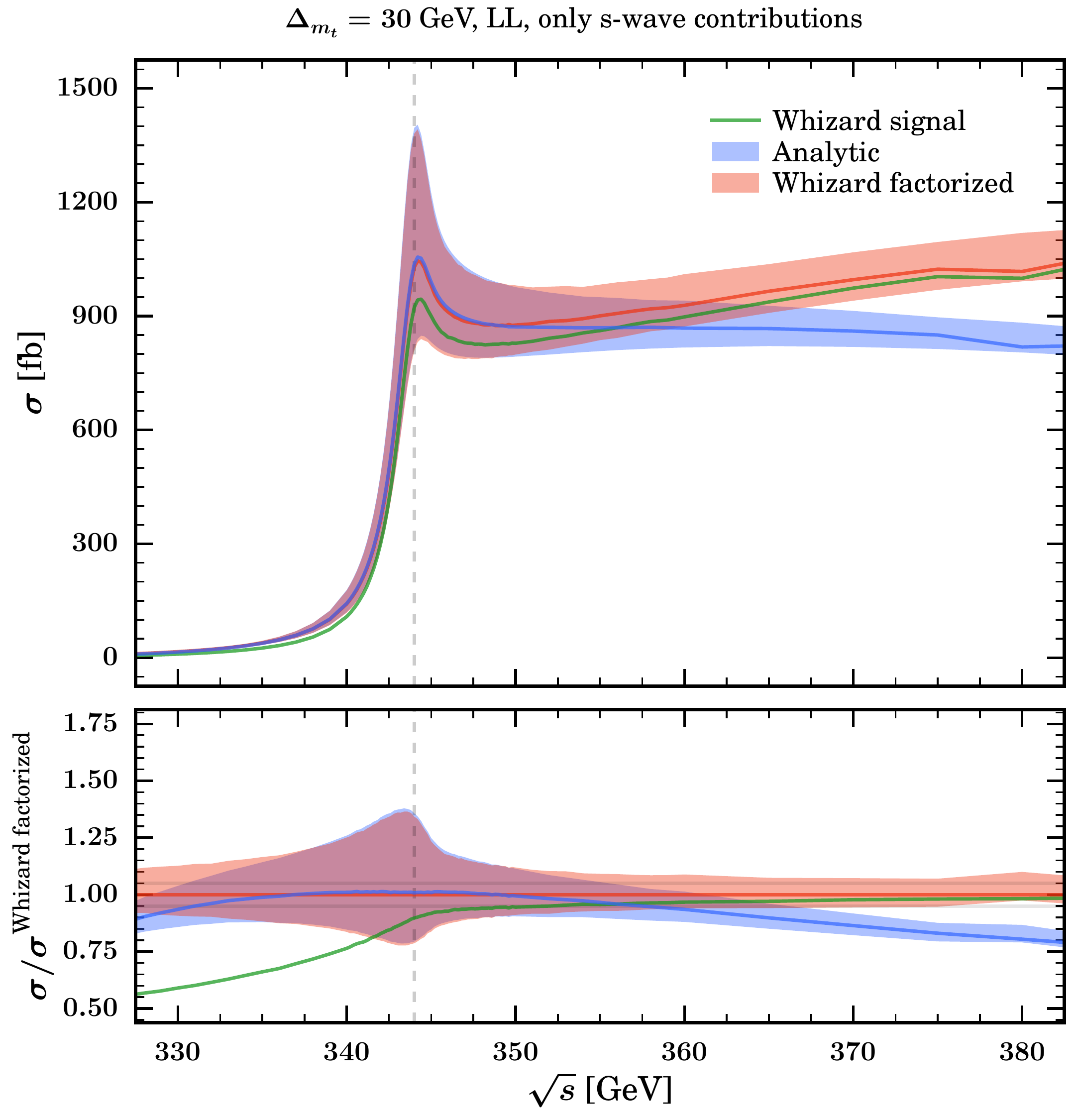}
\includegraphics[width=0.41\textwidth]{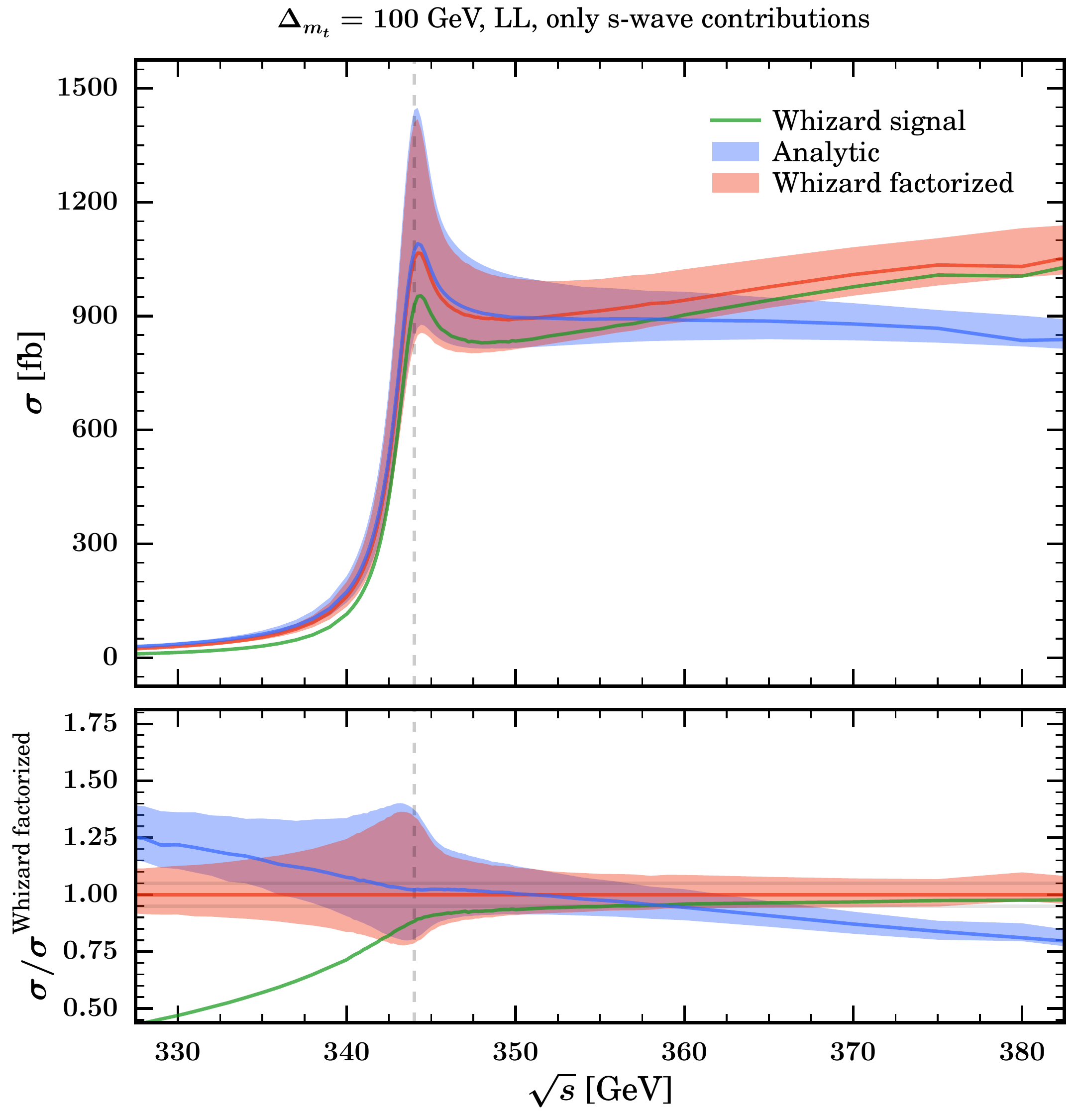}\\
\includegraphics[width=0.41\textwidth]{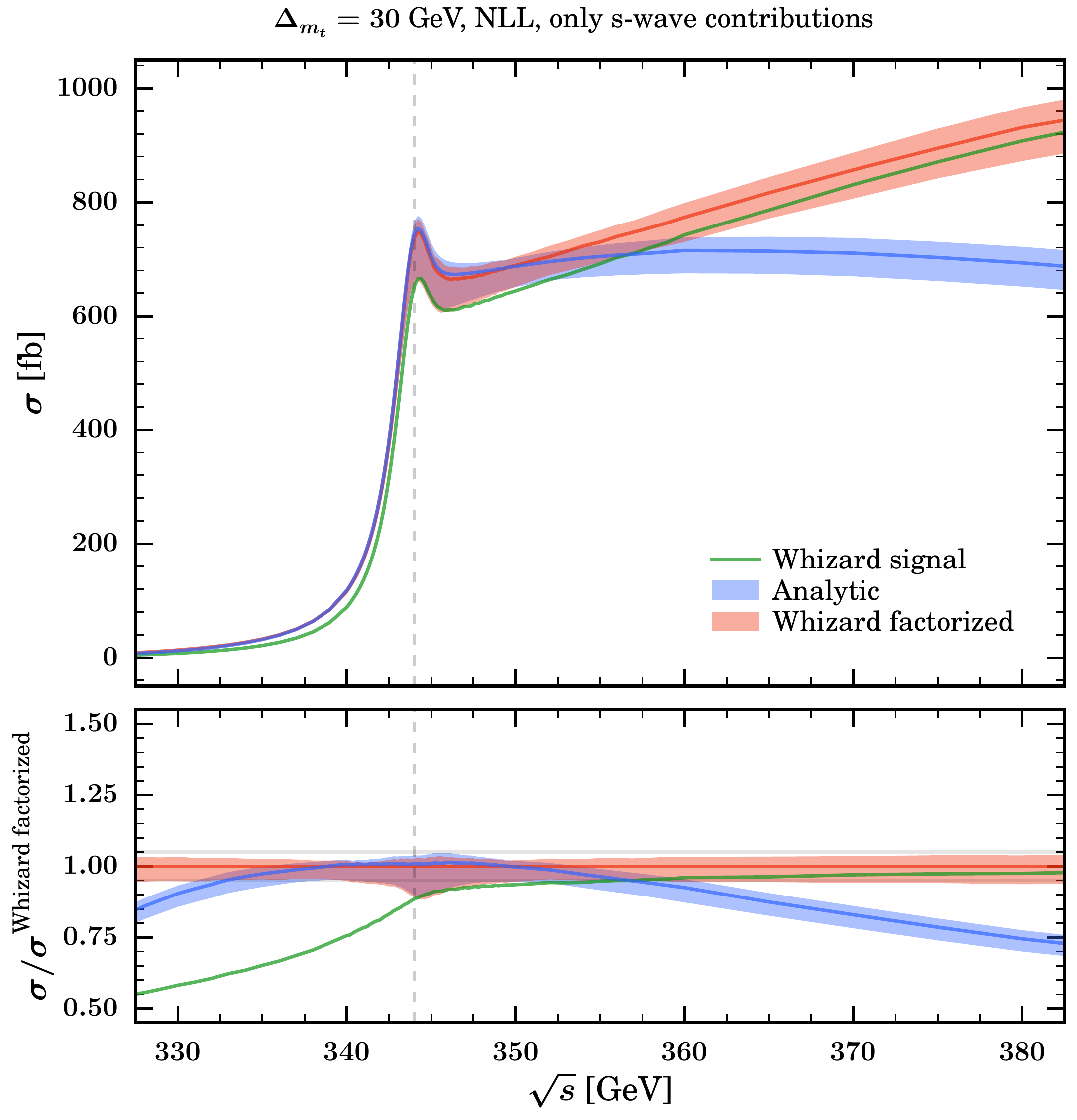}
\includegraphics[width=0.41\textwidth]{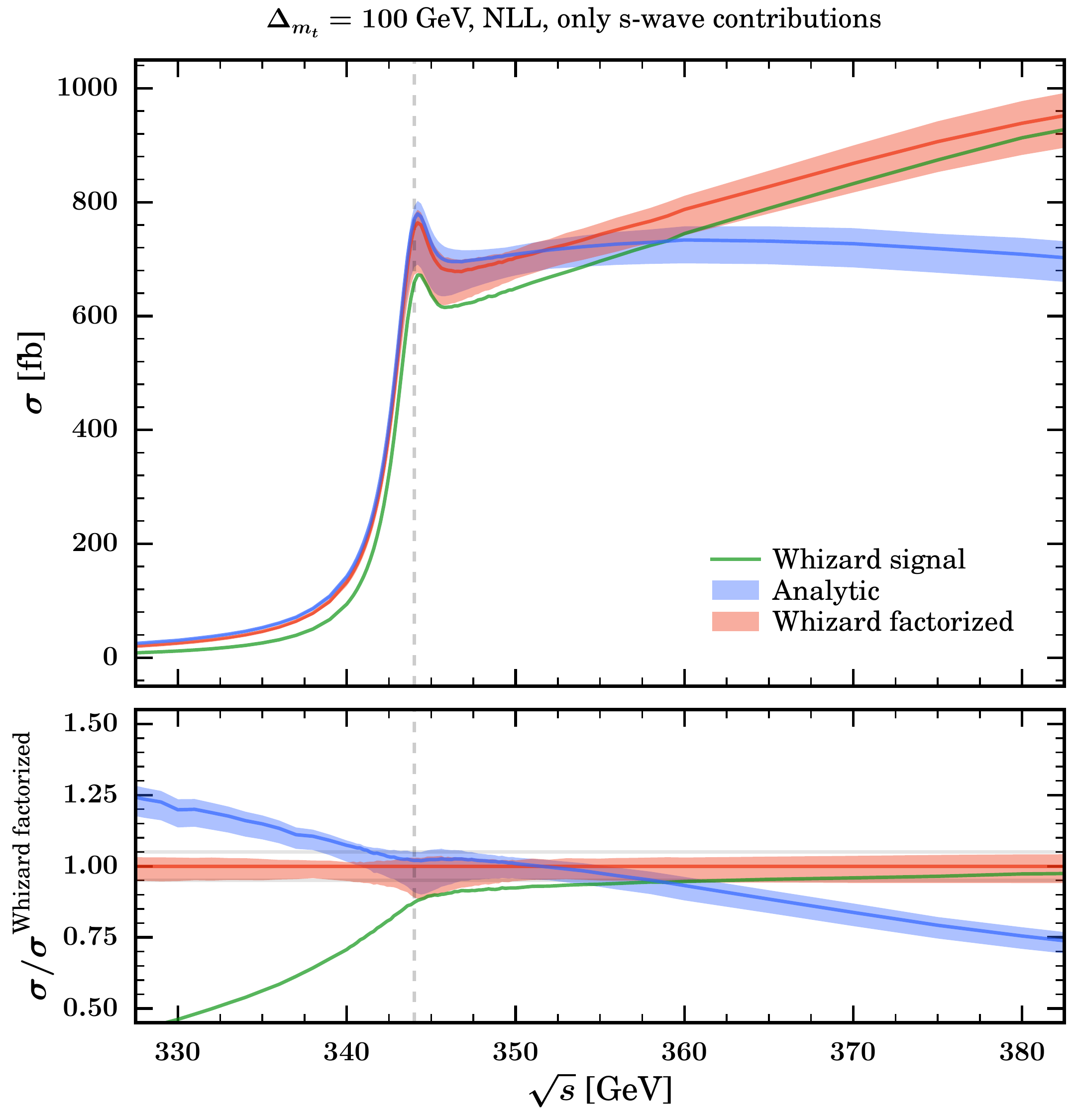}
\caption{Comparison of analytic results with the implementation in \whz{} with
  the factorized and the signal-diagram approach for $\DeltaM =
  \SI{30}{\GeV}$ (left panels) and $\DeltaM = \SI{100}{\GeV}$ (right panels) using
  an expanded, LL or NLL form factor in the upper, middle, and lower
  row, respectively. The bands correspond to the envelope of the scale
  variations mentioned in the text.
  }
\label{fig:validation_dm_fixed}
\end{figure}
In \cref{fig:validation_dm_fixed}, we show the inclusive cross section over 
\sqrts{} for fixed values of the invariant mass cut \DeltaM{}.
Continuing on the considerations of the last section, we employ a
\emph{moderate} ($\DeltaM = \SI{30}{\GeV}$) and a \emph{loose} cut
($\DeltaM = \SI{100}{\GeV}$).
Cross sections for a \emph{tight} cut ($\DeltaM = \SI{15}{\GeV}$) are shown in
\cref{app:additional_validation_results}.
As explained above, our nonrelativistic analytic computation is only applicable for moderate cuts.
To also check the implemented scale variations with the
constraints discussed in \cref{ss:resummation_and_scales}, we have
produced four curves for each cross section corresponding to the corners of the 
$h$-$f$ region defined by \cref{eq:hfvar}:
\begin{align}
\label{eqn:hfchoices}
  \Bl( h,f \Br) &= \Bl( 2, 1         \Br) \co &
  \Bl( h,f \Br) &= \Bl( 2, \frac 1 2 \Br) \co \no
  \Bl( h,f \Br) &= \Bl(\frac 1 2, 2  \Br) \co &
  \Bl( h,f \Br) &= \Bl( \frac 1 2, 1 \Br) \po
\end{align}
The scale variation bands shown in \cref{fig:validation_dm_fixed} correspond to 
the envelope of the four associated curves.
We have checked that (for the inclusive cross sections based on the analytic calculation) 
this procedure usually gives a very good approximation to the scale variation 
bands one would obtain by scanning over the complete selected $h$-$f$ region 
(as displayed in Fig.~2 of \Rcite{1309.6323}).
In addition to the results based on the LL and NLL form factors, we also show 
the corresponding inclusive cross sections using the $\ord{\ALstrong}$ expanded 
NLL (S-wave) form factor given in \cref{eq:FFexpNLL} and evaluated 
with $\alpha_s=\AShard{}$.

Yet again, the ratios shown in the respective lower panels depend only mildly 
on the approximation used for the form factor for energies above the
peak position. However, they have a rather strong dependence on
\DeltaM{} below the peak region due to relativistic off-shell
contributions, e.g.\ in the top and anti-top  
Breit-Wigner propagators, which are contained in the factorized \whz{} 
calculation but missing in the analytic one, where the nonrelativistic
approximation is employed. These off-shell effects have a much 
larger relative impact in the region below the peak position, where
the cross section becomes small. 

For $\DeltaM=\SI{30}{\GeV}$, we observe perfect agreement between the
analytic computation and \whz{} with the factorized approach within a
window around threshold of at least \SI{10}{\GeV}.
For $\DeltaM=\SI{100}{\GeV}$, this range is reduced significantly due
to the unphysical relativistic off-shell contributions mentioned in
the previous paragraph that arise for$\DeltaM>\SI{30}{\GeV}$ in the
analytic calculation.
Notably, the  behavior above threshold is not strongly affected
but the ratio of analytic over factorized \whz{} results falls off for c.m.\
energies above about $360$~GeV with approximately the same slope for
both invariant mass cuts. This is due to (not systematically
controlled) higher-order relativistic effects (e.g. associated with the 
relativistic production current or the top quark propagators) in the difference 
of the factorized computation and the nonrelativistic analytic result. However, we 
emphasize that this is not problematic as in our fully matched calculations, as explained in 
\cref{s:matching}, the factorized results do not contribute at these 
energies due to the switch-off procedure. In addition to the
shown validation plots, we have also cross checked at individual 
phase-space points that the implementations of the expanded,
LL and NLL form factors are consistent within the numerical precision.

Overall, we have tested that our nonrelativistic form factors are correctly and
consistently embedded in \whz{}.
The differences to the purely nonrelativistic analytic calculation according to 
\cref{ss:sigmaincl} are understood and we can rely on the implementation in 
\whz{} with the factorized approach. 
This yields reliable results for all \DeltaM{} values and in fact allows 
for fully differential predictions including threshold resummation. 
%
%
\section{Matching}
\label{s:matching}
%
In this section, we discuss our approach to combine (match) the
nonrelativistic cross section based on factorized matrix elements
with (N)LL threshold-resummed form factors and (N)LO top/anti-top quark
decays into $\WbWb{}$ (called  \XSresummed{}) with the full fixed-order
QCD-(N)LO cross section for $\WbWb{}$ production including all
irreducible background processes and interferences (called
\XSqcd{}). Within the approximations explained already in previous
sections, we maintain all relevant interference terms between full and 
nonrelativistic factorized matrix elements in \XSresummed{}
and we keep terms beyond the corresponding order counting wherever
suitable from a practical point of view. The essential point is that
our matching procedure avoids any double counting of terms 
simultaneously contained in the two components at their respective order.

Before discussing the details of the matching procedure, let us first
remind the reader that the resummed form factors are computed based on
the assumption that $v\sim\ALstrong\ll 1$. In a matched computation,
which shall provide a smooth description from the threshold region up
to high energies, this counting becomes more and more inappropriate
with increasing c.m.\ energy \sqrts{} until the point where it is no
longer meaningful and provides wrong results. This means that the
threshold resummations by themselves do not contain any natural
mechanism to smoothly transition to the relativistic counting.   
To construct a matching approach that provides smooth predictions it
is therefore mandatory to introduce a \emph{switch-off function}
\switch{}. We note that the matching procedure devised in \Rcite{1007.0075}
did not involve a switch-off function because the threshold-resummed
form factors were combined with the full QCD-LO matrix element for
$\WbWb{}$ production only.
They thus argued that their approach is strictly correct at leading order, 
and that, formally, the QCD corrections resummed in the nonrelativistic
form factors constitute terms beyond this level of approximation.  

In our matching procedure the switch-off function is unity
in the threshold region, vanishes in the relativistic region where the
nonrelativistic calculations cannot be trusted and is monotonically
falling everywhere. The implementation that we are using is specified
in \cref{ss:switchoff}. As the detailed shape of the switch-off
function is not unique, the matching procedure entails an additional
source of theoretical uncertainties, which we examine. We implement our 
switch-off function by multiplying it to the strong couplings that enter the 
form factors. In this way higher orders in $\ALstrong$ resummed in the form factors are naturally stronger 
affected by the switch-off than lowers orders.

In order to avoid double counting of terms contained in \XSqcd{} and \XSresummed{}
we also have to 
define an \emph{expanded nonrelativistic} cross section, called \XSexpanded{}, that is constructed 
like \XSresummed{}, but contains the form factors expanded in \ALstrong{} to the 
appropriate order:
The explicit expressions for the expanded NLL form factors are given in 
\eqs{FFexpNLL}{FFAexpNLL}. At LL the required expanded form factors are 
trivially $F^\mathrm{exp}_{V,\mathrm{LL}}=F^\mathrm{exp}_{A,\mathrm{LL}}=1$.

With these basic ingredients
our matching procedure can be formulated schematically
by the master formula 
\begin{align}
  \XSmatched \;=\; \XSqcd\of{\ASfirm}
  \;+\; \XSresummed\of{\switch\,\AShard,\;\switch\,\ASsoft,\;\switch\,\ASusoft} 
  \;-\; \XSexpanded\of{\switch\,\ASfirm}\co
  \label{eq:matched_simple}
\end{align}
with \ALstrong{} evaluated at the hard ($\mu_\T{H}$), firm
($\mu_\T{F}$) (the name "firm" signifying a scale in the geometric
mean between the soft and hard scale), soft ($\mu_\T{S}$) and
ultra-soft scales ($\mu_\T{US}$) 
\begin{align}
  \alpha_\T{H} &= \alpha_{s}\of{ \mu_\T{H} = h\mOneS}\co &
  \alpha_\T{F} &= \alpha_{s}\of{ \mu_\T{F} = h\mOneS\sqrt{\nustar}} \co\no
  \alpha_\T{S} &= \alpha_{s}\of{ \mu_\T{S} = h\mOneS f\nustar   }\co &
  \alpha_\T{U} &= \alpha_{s}\of{\mu_\T{US} = h\mOneS (f\nustar)^2} \po
  \label{eq:scales}
\end{align}
By subtracting the expanded cross section defined with the firm
coupling \ASfirm{}, which is also used for \XSqcd{}, we remove
contributions that are simultaneously contained in the relativistic
and nonrelativistic cross sections. The firm scale $\mu_\T{F}$ is
defined as the geometric mean between the soft and the hard scales. In
the threshold region it provides a hybrid scale optimized for the
hard and low-energy nonrelativistic higher order corrections (in the
$v$ expansion) contained in \XSqcd{} and not accounted for in
\XSresummed{}. Away from the threshold region
$\mu_\T{F}$ is very close to the hard scale, which is an appropriate
choice for the fixed-order expansion. From the fixed-order point of
view, $\ASfirm{}$ is therefore a reasonable choice 
with a safe IR behavior.\footnote{We note that in the recent N$^3$LO
  threshold-resummed total cross section calculations~\cite{1506.06864}   
  (which did not account for a systematic resummation of velocity
  logarithms), \ALstrong{} was employed only at one single
  renormalization scale chosen between the firm and the hard scales.}
For comparison, we discuss the numerical
impact of choosing \ASfirm{} over \AShard{} in
\cref{ss:fixed_order_results}. Returning to the matching, to avoid
double counting we define the expanded cross section at the firm
scale, and we maintain the canonical scales in \XSresummed{} at
threshold according to \cref{s:resummation}. The switch-off function \switch{} 
guarantees that we obtain
only \XSqcd{} in the high-energy continuum. The exact contributions in
\XSqcd{}, \XSresummed{} and \XSexpanded{} depend on the order of interest and 
are discussed in detail in the next subsection. 

We emphasize again that the master formula shown in
\cref{eq:matched_simple} is schematic in order to illustrate the
underlying principles of our matching procedure. In practice, we implement the
matching at the amplitude level as described in more detail below, which leads to a
matching formula that is substantially more involved than \cref{eq:matched_simple}.

\subsection{Contributions in the matched cross section}
\label{ss:contributions}
%
Our matching procedure is set up such that the cancellation of IR
divergences between real and virtual QCD-NLO corrections is maintained
and that the dominant interference terms between nonrelativistic
resummed and non-resonant or background contributions are
accounted for. We discuss all contributions in a diagrammatic way
using the notation introduced in \cref{sss:fixed_order_corrections},
omitting phase-space factors and the exact choice of parameters (which
follow the guidelines described before). We emphasize that all shown
components represent gauge-invariant contributions at the amplitude
level, cf.\ \cref{ss:gauge_invariance}.

We start by discussing the matching of fixed-order QCD-LO and LL
threshold-resummed cross sections. As already mentioned in
\cref{ss:factorization}, the LL threshold-resummed cross section is
obtained in the DPA according to  \cref{eq:ME_factorized}, where the
production matrix elements are multiplied with the corresponding
nonrelativistic LL form factors $F$  given in
\eqs{FF_vec_LL}{FF_ax_LL}. To this end, we define
$\tilde{F}\equiv{F}-1$, i.e.\ all terms contained 
in $\tilde{F}$ are $\order{\ALstrong}$ and thus of higher order from the
relativistic fixed-order point of view. In our notation we can rewrite
the LL threshold cross section as 
\begin{align}
  \XSLL &= \abs{\FFOnePlus\diagram{factorizedLessDetail}}^2 &\no
        &= \abs{\FFOne\diagram{factorizedLessDetail}}^2 +
        \realOf{\FFLLmOne\diagram{factorizedLessDetail}}
               {\diagram{factorizedLessDetailFlipped}
                \lessspace{}\FFOne{}\extraspace{}} \no
        &\quad+ \abs{\FFLLmOne\diagram{factorizedLessDetail}}^2\co
  \label{eq:LL}
\end{align}
where the ``1'' term corresponds to tree-level $\tT{}$ production
without treshold resummation. It is now straightforward to include the
full QCD-LO cross section by replacing the ``1'' terms by the full QCD-LO
amplitude. Since the QCD-LO contributions do not
contain \ALstrong{} corrections, no double-counting occurs at this
point. The expanded contribution that we have to handle according to
the scheme of \cref{eq:matched_simple}, is represented by the sum of the first (square) and the second (interference) term (with factor 1) in \cref{eq:LL}. 
The fully matched QCD-LO+LL cross section then reads
\begin{align}
  \XSLOLL = \XSLO
    &+ \realOfDiagrams[\FFLLmOne]{factorizedLessDetail}{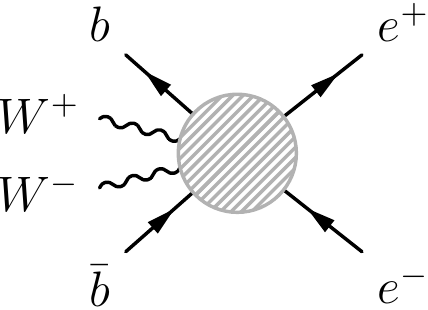} 
    + \abs{\FFLLmOne\diagram{factorizedLessDetail}}^2 \co
\label{eq:LOLL}
\end{align}
where the scale settings and the switch-off function \switch{} are
implemented as described in \cref{eq:matched_simple} and we employ the
top width $\Gamma_t$ at QCD-LO. We note that the matched
result takes into account the interference corrections involving the
LL threshold resummed and the full QCD-LO amplitudes. The latter
includes all $\tT{}$-double-resonant, single-top and background
diagrams. These interference corrections constitute an important part of the 
NLL electroweak corrections at threshold and are important to achieve
top mass measurements with uncertainties of better than $50$~MeV. 
The use of the \ac{DPA} ensures that the amplitude phases are treated properly, cf.\ \cref{sss:interference_terms}.

As an intermediate next step let us now consider including in addition
the full QCD-NLO cross section, \XSNLO, as well as the NLL form
factors, while still keeping the top and anti-top decays at QCD-LO in the
factorized calculation. As \XSNLO{} contains terms of
$\order{\ALstrong}$, for \XSexpanded{} we now have to use the
expansion of the form factors to $\order{\ALstrong}$, given in
\cref{eq:FFexpNLL} and (\ref{eq:FFAexpNLL}). According to this, the matched cross
section has the form
\begin{align}
  \XSNLONLL^\T{LOdecay} = \XSNLO &+
        \realOfDiagrams[\textcolor{blue}
          {\left(\FFNLL - \FFexpNLL\right)}\hspace{-1.3em}]
          {factorizedLessDetail}{LOfullFlipped}\no
          &+ \abs{\FFNLLmOne\diagram{factorizedLessDetail}}^2 \co
 \label{eq:NLONLL_LOdecay}
\end{align}
where still $\tilde{F}_\T{NLL}\equiv{F}_\T{NLL}-1$, and we remind the
reader that now $F_\T{NLL}-\FFexpNLL\neq \tilde{F}_\T{NLL}$ due to the
$\order{\ALstrong}$ terms contained in $\FFexpNLL$. 

Although the expression in \cref{eq:NLONLL_LOdecay} appears
straightforward in principle, it has a problem concerning making a
consistent choice for the top quark width. From the point of view of
the factorized computation, we are using a QCD-LO description of the
top decays, so a consistent choice for the top width parameter
appearing in the matrix element expressions for the
$\absIT{\tilde{F}_\T{NLL}}^2$ term as well as in the interference term
is the QCD-LO width. On the other hand, the Coulomb singular
contribution to \XSNLO{}, that is supposed to be subtracted by the
$\FFexpNLL$ term,  requires a QCD-NLO width. Thus, also the
subtraction term should be evaluated with the QCD-NLO width. The
apparent conflict is an artifact of trying to match two computations
that treat the top decay at different orders. This problem can only be
resolved by incorporating the QCD-NLO decay also in the factorized
parts of the cross section.

To proceed, we first observe that it is quite difficult to add any
real and virtual corrections to the interference term in
\cref{eq:NLONLL_LOdecay}. This is because the IR divergences that
arise when soft gluon corrections are added to the factorized
amplitude and the full $\WbWb{}$ matrix element differ from each
other. At the NLL precision we are aiming for, however, we do not have
to consider these corrections because the interference term itself
represents a NLL correction. Therefore, within the approximations
adopted in this work, it only remains to add the 
QCD-NLO top decay corrections to the completely factorized
$\absIT{\tilde{F}_\T{NLL}}^2$ term in \cref{eq:NLONLL_LOdecay}. We 
thus arrive at the following form of our final matching formula, which
combines QCD-NLO fixed-order and NLL threshold cross section
predictions, cf.\ \cref{eq:factorized_NLO}:
\begin{align}
  \XSNLONLL &= \XSNLO + \realOfDiagrams[\textcolor{blue}
      {\left(\FFNLL - \FFexpNLL\right)}\hspace{-1.3em}]
      {factorizedLessDetail}{LOfullFlipped}\no
      &+\; \abs{\FFNLLmOne\diagram{factorizedLessDetail}}^2 
      +\; \abs{\FFNLLmOne{}\diagram{factorizedReal}}^2
      +\; \abs{\FFNLLmOne{}\diagram{factorizedRealOtherLeg}}^2
      \no
      &+ \realOf{\FFNLLmOne{}\extraspace{}\left(\diagram{factorizedVirtual}
                        + \lessspace\diagram{factorizedVirtualOtherLeg}\right)}
        {\diagram{factorizedLessDetailFlipped}\lessspace{}\FFNLLmOne{}\extraspace{}}
        \po
\label{eq:NLONLL}
\end{align}
\subsection{Switch-off function}
\label{ss:switchoff}
\begin{figure}[htb]
\centering
\includegraphics[width=\standardwidth]{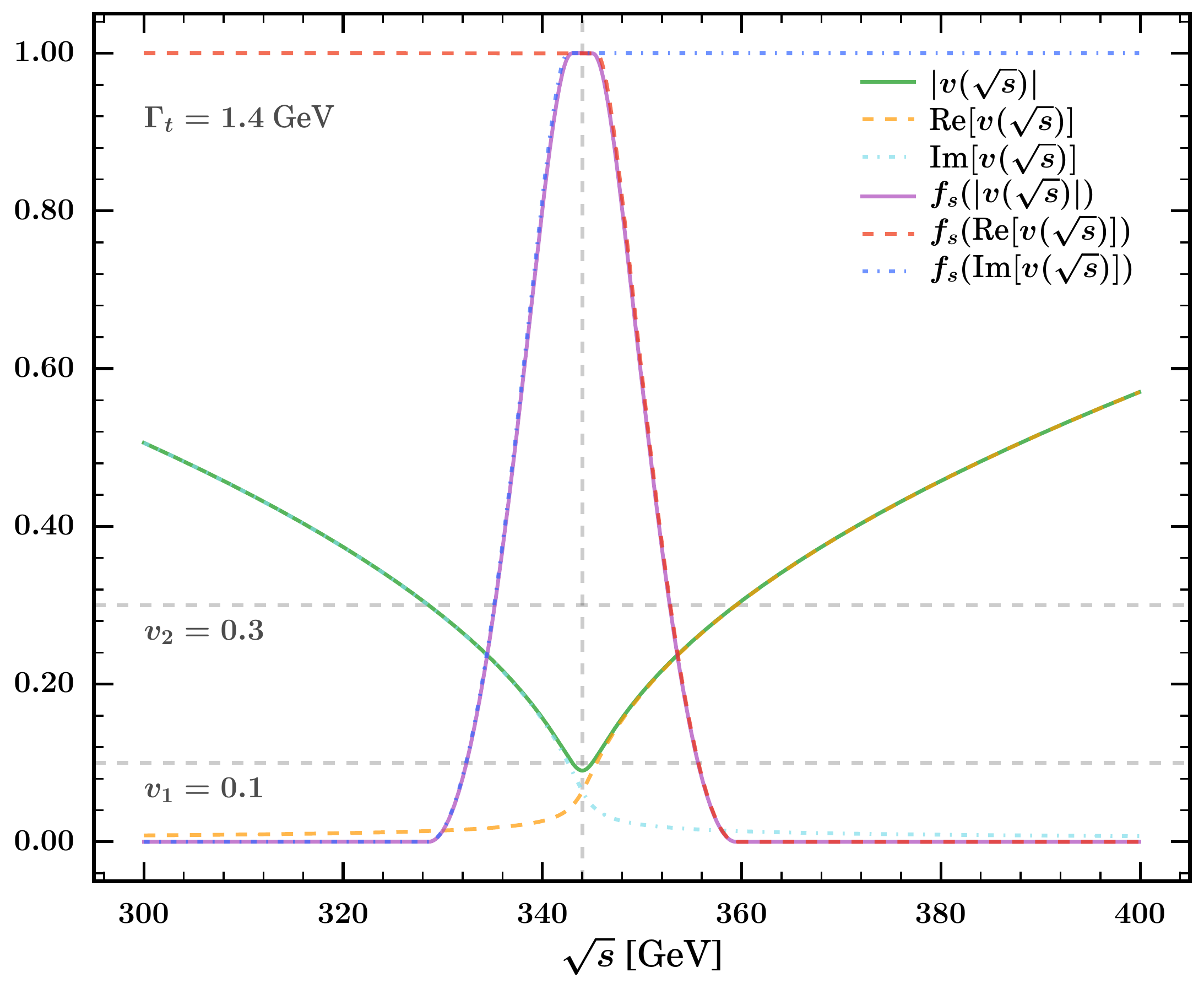}
\caption{Switch-off function $\switch$ according to \cref{eq:switch_smoothstep} together with the different choices for its argument:
the absolute value, the real, and the imaginary part of the (quasi) velocity $v_\T{1S}$. In addition to the vertical line at $2\mOneS{}$, we display the two matching parameters $v_1$ and $v_2$ used for the plot as horizontal lines.}
\label{fig:switch-off}
\end{figure}

As noted before, we need a definite way to switch off the
resummations encoded in the form factors at center-of-mass energies
away from the threshold region where the threshold resummation is not
meaningful, and where we only want to use the relativistic fixed-order
predictions. For this we define the switch-off function $\switch{}$,
which is a monotonic function of a $\sqrt{s}$-dependent velocity
parameter $v_s$. It satisfies the basic requirements
\begin{align}
  \switch\left(v_s\Al(\sqrt{s}\approx 2\mOneS{}\Ar)\right) = 1
  \MT{and} \switch(v_s \sim 1) = 0\co 
  \label{eq:switch-req}
\end{align}
which means that $\switch{}(v_s)$ is unity in the threshold region and
vanishes in the region where the relativistic fixed-order predictions
are sufficient. Above threshold $\switch{}(v_s)$ should be
monotonically decreasing and below threshold monotonically
increasing. To this end we define the complex velocity,
cf. \cref{eq:nustar}, 
\begin{align}
  v_\T{1S} 
  = \sqrt{\frac{\sqrts{} - 2 \mOneS + \I
    \Gamma_{t}^*}\mOneS}\po
  \label{eq:velocity1S}
\end{align}
The dependence on the 1S mass is motivated 
to ensure that \switch{} is unity in the peak region.
Thus, we center the switch-off around the 1S mass and not around the
pole mass. $\switch{}$ is a function of a real parameter, for which
we may use the imaginary part, real part or modulus of $v_\T{1S}$ as
the argument. All three options are shown in \cref{fig:switch-off}.
We see that the real and imaginary parts of $v_\T{1S}$ are
roughly mutual reflections with respect to $2\mOneS{}$ with a slight
asymmetry due to the $+\I\Gamma_{t}^*$ term in
\cref{eq:velocity1S}. To switch off the resummation for large values
of \sqrts{}, it would be sufficient to only use the real part of $v_\T{1S}$.
However, since our matched calculations cover also the kinematic
region far below threshold where the fixed-order predictions are
valid, we have to ensure that the threshold resummations are also
switched off reliably in that region. It is therefore natural to
define $v_s\equiv\absIT{v_\T{1S}}$ as the velocity parameters for the
switch-off function $\switch$. We note that due to the width term that
enters the definition of the velocity parameter, $\absIT{v_\T{1S}}$
adopts its minimal value $\sqrt{\Gamma_{t}/\mOneS}\sim 0.1$ at
$\sqrt{s}=2\mOneS{}$, so $v_s$ never vanishes. 

The concrete form of \switch{} is in principle not very important as
it has to be varied anyway to estimate the matching uncertainty,
and as long as the essential parameters to control the switch-off
behavior are implemented. We define these parameters as $v_1$ and
$v_2$, which satisfy $0< v_1<v_2<1$, and impose the  
conditions $\switch{}(v_s<v_1)=1$ and $\switch{}(v_s>v_2)=0$.
There are two more practical issues relevant for devising \switch{}.
Firstly, \switch{} should be not only continuous but also continuously
differentiable like any physical cross section.
This property disqualifies simply using a linear interpolation for
$v_1\le v_s\le v_2$. 

On the other hand, when using polynomials of high order for \switch{} that
at first switch off slowly but transition too quickly, one risks to
introduce unphysical wiggles in the cross section when $v_1$
and $v_2$ are varied. After experimenting with different types of
switch-off functions, cf.\ \cref{s:alternative_switch_off_functions},
we decided to use the following polynomial interpolation, which
satisfies continuity and continuous differentiability at both
$v_s=v_1$ and $v_s=v_2$:
\begin{align}
  \switch(v_s) = \begin{cases}
    1                          & v_s < v_1 \\
    1 - 3 \left(\frac{v_s-v_1}{v_2-v_1}\right)^2 - 2
      \left(\frac{v_s-v_1}{v_2-v_1}\right)^3  & v_1 \leq v_s \leq v_2 \\
    0                          & v_s > v_2
    \end{cases}\po
  \label{eq:switch_smoothstep}
\end{align}
It represents a cubic Hermite interpolation that in
the context of interpolations is known to belong to the family of \emph{smoothstep} functions.
For illustration, we plot \switch{} in \cref{fig:switch-off}  as a
function of $v_s=|v_\T{1S}|$ (which is our default) as well as a function of the real and imaginary parts of $v_\T{1S}$.
The matching parameters are set to $v_1=0.1$ and $v_2=0.3$, and displayed as horizontal
lines. Their intersections with $v_s$ define start and end points of the
transition region.

From \cref{fig:switch-off} we can see that $v_1=0.1$ should be considered
as the lower limit for reasonable $v_1$ values. Going lower would cut into the 
threshold region around $2\mOneS{}$ and thus artificially reduce the
threshold peak. On the other hand, it is not obvious how to devise a
meaningful upper limit of $v_2$. Here, we consider values up to
$v_2=0.4$, which leads to a fairly quick switch-off and to a
reasonable behavior of the matched calculation. In any case, more
concrete conclusions on this matter require the analysis of the
matching procedure including corrections beyond NLL and
QCD-NLO~\cite{Widlmatching} which is beyond the scope of this paper.

Let us finally remark that it would in principle also be possible to
implement the switch-off function by multiplying it with the NRQCD cross 
sections in \cref{eq:matched_simple} instead of the strong couplings. Both approaches differ by how the resummations contained in the form factors are treated. At the orders considered in our studies we found that multiplying the
couplings leads to a somewhat smoother transition.

\subsection{Theoretical uncertainties}
\label{ss:theoretical_uncertainties}
For our examinations of the theory uncertainties we will perform variations of the different matching and renormalization scales $\mu_\T{H}$, $\mu_\T{F}$, $\mu_\T{S}$ and $\mu_\T{US}$, cf.\ \cref{eq:scales}. We will vary coherently over
the scale multipliers $h$ and $f$, where $h$ parametrizes the relative
variation of the hard scale and $f$ the relative variation of the
subtraction velocity scale $\nu$ as discussed in
\cref{ss:resummation_and_scales}. Since an entire scan over the
$h$-$f$ area defined in \cref{ss:resummation_and_scales} is in practice not feasible for fully matched predictions, we sample the four corners of the $h$-$f$ area, defining the set
\begin{align}
  \T{HF} = \Cl\{
  \Bl( h,f \Br) &= \Bl( 2, 1         \Br) \co &
  \Bl( h,f \Br) &= \Bl( 2, \frac 1 2 \Br) \co \no
  \Bl( h,f \Br) &= \Bl(\frac 1 2, 2  \Br) \co &
  \Bl( h,f \Br) &= \Bl( \frac 1 2, 1 \Br) \Cr\} \po
  \label{eq:scale_variations}
\end{align}
The envelope of the four associated curves is expected to give a good approximation of the result of a full scan, see also \cref{ss:validation_sqrts_scans}.
Besides the scale variations, we also vary the matching (velocity) parameters $(v_1,v_2)$ in the interval $[0.1, 0.4]$, while keeping $v_2\approx v_1+0.2$.
In practice we will study four different $(v_1,v_2)$ matching
parameter choices with $v_1\in\set{0.1,0.15}$ and
$v_2\in\set{0.3,0.4}$. Concerning the reliability of the obtained
variation bands, we remind the reader that it has been shown in
\Rcite{1309.6323} that for the total cross section the naive NLL scale variation band does not envelope the NNLL prediction.\footnote{This is also true for the NLO
  and NNLO fixed-order threshold predictions, see
  e.g.\ \Rcite{Hoang:2000yr,1506.06864}.} 
While for our matched predictions we also include additional
electroweak and relativistic corrections, we have to assume that this
also applies to our computation. Because the NLL scale variations for
the threshold-resummed predictions are highly asymmetric with respect
to the $(h,f)=(1,1)$ central values in the region around the peak, 
cf.\ \cref{fig:validation_dm_fixed}, as a more conservative estimate
of the theoretical uncertainties in this work 
we furthermore symmetrize the scale variations by computing the
upper and lower error band envelopes. Hence, for each $\sqrt{s}$ value
we are using the prescription 
\begin{align}
  \sigma_\T{upper} &= \max\of{\max_{i\in\T{HF}}{\sigma_i} \co\;
              \sigma_0 + (\sigma_0-\min_{i\in\T{HF}}{\sigma_i})}\,,\no
  \sigma_\T{lower} &= \min\of{\min_{i\in\T{HF}}{\sigma_i} \co\;
              \sigma_0 - (\max_{i\in\T{HF}}{\sigma_i} - \sigma_0)}\,,
\label{eq:symm}
\end{align}
where $\sigma_0\equiv\sigma(h=1,f=1)$ is the default cross
section. The obtained bands are then by construction symmetric around
the default cross section and are always enveloping and broadening
the original error band. Note that we carry out this procedure for
each $(v_1,v_2)$ matching parameter choice. To obtain our final
uncertainty band for the fully matched QCD-NLO+NLL predictions, we
take the envelope of all of these bands.

%
\section{Inclusive results}
\label{s:inclusive_results}
\subsection{Fixed-order results}
\label{ss:fixed_order_results}
\begin{figure}[bhtp]
\centering
\includegraphics[width=\halfwidth]{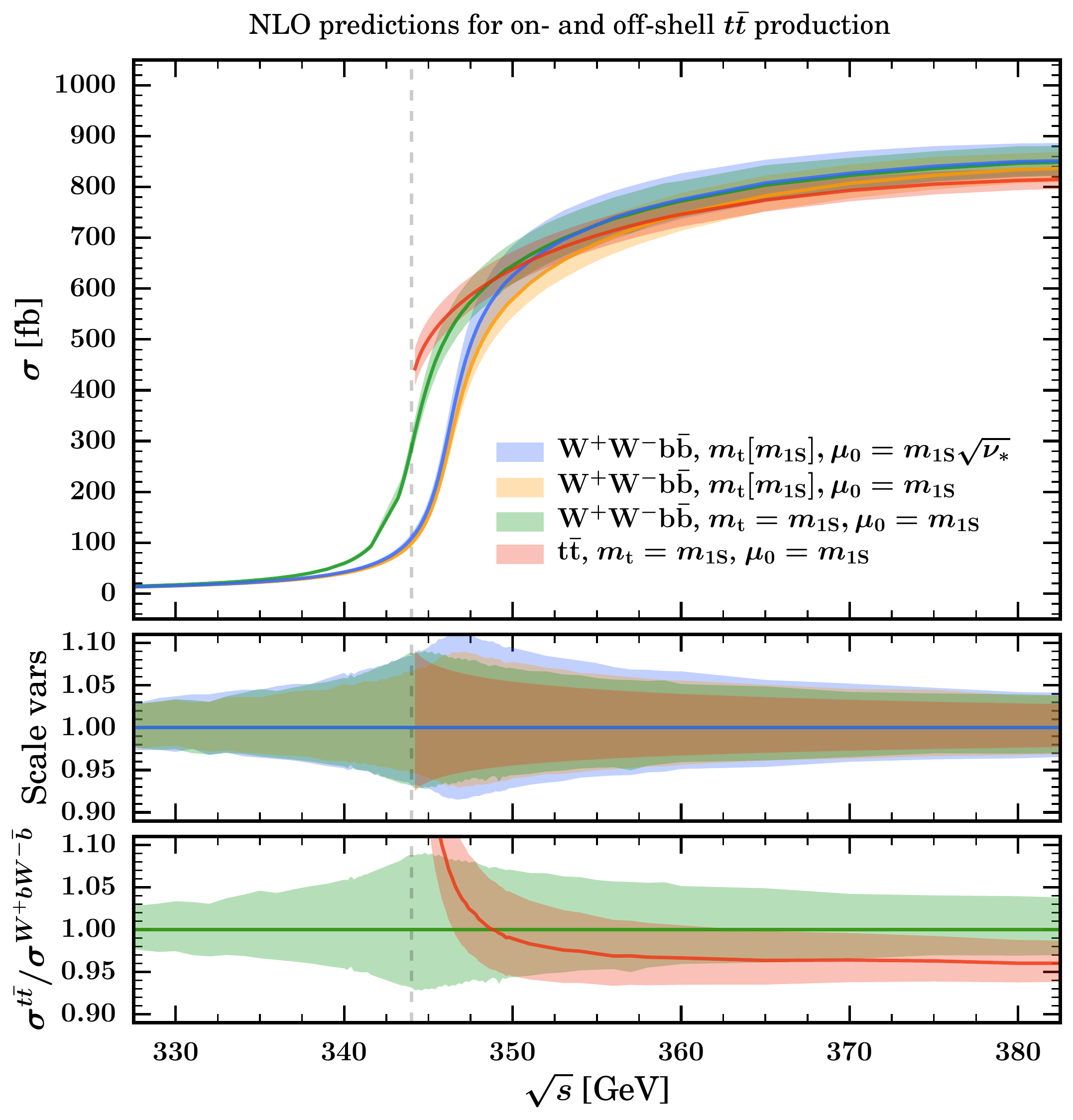}
\includegraphics[width=\halfwidth]{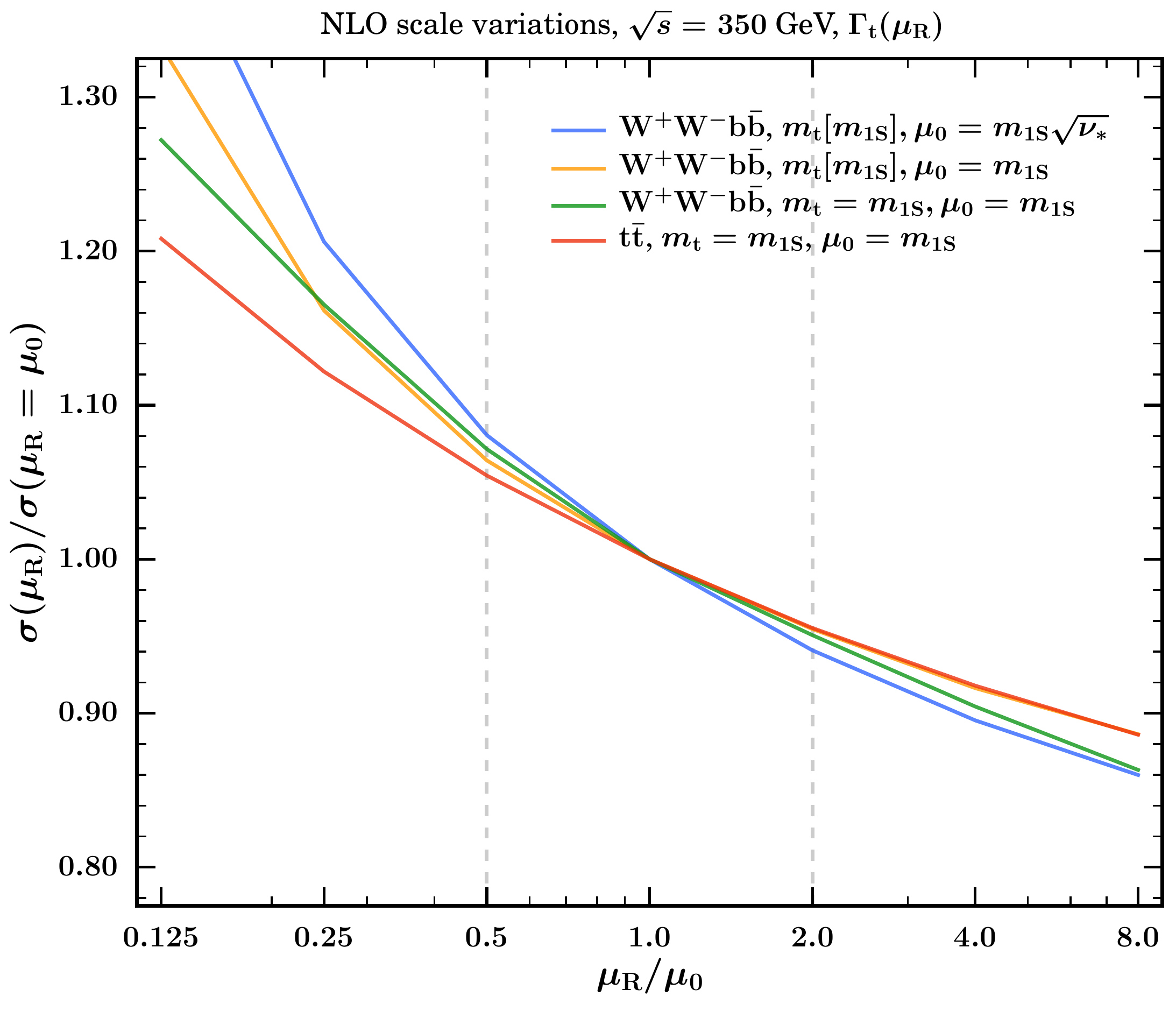}
  \caption{Total QCD-NLO cross sections for $\WbWb{}$ and on-shell $\tT{}$
  production as a function of $\sqrt{s}$ in a region around threshold
  (left panel) and of the renormalization scale $\mu_\T{R}$ for
  $\sqrt{s}=350$~GeV (right panel). 
  The error bands in the left panel arise from $\mu_\T{R}$
  variations with $0.5 \le \mu_\T{R}/\mu_0 \le 2$,  where $\mu_0$ denotes the reference (default) scale. In the lower panels
  of the $\sqrt{s}$ scan we show the relative size of the scale
  variations of each choice of $\mu_0$ as well as the ratio of $\tT{}$
  over $\WbWb{}$ cross sections for $\mu_0=\mOneS$. The different
  versions for the $\WbWb$ cross sections show the impact of using the
  pole mass scheme (setting $\mpole=\mOneS$) rather than the 1S mass
  scheme and using either $\mOneS$ or $\mOneS\sqrt{\nustar}$ as the
  default renormalization scale. The details of the scale variations
  are explained in the text.}
\label{fig:nlofull}
\end{figure}

%
In the left panel of \cref{fig:nlofull}, we show QCD-NLO predictions
for the total cross sections for $\WbWb{}$ (blue, orange, green) and
on-shell $\tT{}$ (red) production as a function of $\sqrts$ and the
renormalization scale $\mu_{\T{R}}$. 
For the off-shell production, we show the effect of using the pole 
mass scheme (green, setting $\mpole=\mOneS$) versus the 1S mass scheme
(blue, orange, setting $\mpole=\mpole[\mOneS]$ using \cref{eq:mPoleRelation}) and of using different choices for the default scale
$\mu_0$. The error bands shown arise from $\mu_\T{R}$ variations with
$0.5 \le \mu_\T{R}/\mu_0 \le 2$ using $\mu_0=\mOneS$ (orange, green, red)
and $\mu_0=\mu_\T{F}=\mOneS\sqrt{\nu_*}$ (blue). As expected, using
the 1S mass scheme, which involves computing the pole mass as a
function of $\mOneS$, leads 
to a visible energy shift of the cross section by around $2$~GeV, indicating the size of the ground state toponium binding energy.
The size of the scale variations is quite
similar for both mass schemes. This is noteworthy, as we have varied
$h$ as well as $f$ according to \cref{eq:scale_variations} -- both of which affect $\DeltaMM$ as shown in \cref{eq:mPoleRelation} --  to
obtain the results in the 1S mass scheme, where we have identified $h$
with $\mu_\T{R}/\mu_0$. On the other hand, for the pole mass
calculation, there is no $f$ dependence and only the $h$ variation can
be performed. So we see that the $f$ variation represents only a minor
effect and does not lead to a larger scale variation for the full
QCD-NLO cross section.  

As already discussed in \cref{s:matching} 
the default renormalization scale choice for the QCD-NLO fixed-order cross section within our matched prediction is
the firm scale $\mu_0=\mu_\T{F}\equiv\mOneS\sqrt{\nustar}$ (blue), which is more sensitive to the threshold dynamics.
The default scale yields slightly larger cross sections 
and scale variations in the threshold region than the hard
renormalization scale $\mu_0=\mOneS$.

To assess the impact of the off-shell effects at QCD-NLO, we also show
the cross section for on-shell $\tT{}$ production. We see that on-shell
$\tT{}$ production yields results that are \SIrange{3}{4}{\percent} below
the $\WbWb{}$ cross section for energies above $\sim\SI{355}{\GeV}$.
Concerning the high-energy behavior, where the deviations are dramatically 
larger because additional (not \tT{}-related) resonant channels for $\WbWb{}$ 
production open up, we refer to the examinations in \Rcite{1609.03390}. Below
\SI{350}{\GeV}, i.e. in the threshold region, the $\tT{}$ cross section
rises rapidly over the $\WbWb{}$ cross section and even doubles the latter 
at the  threshold point ($\sqrts=2 \mOneS$, vertical dashed line). The behavior for the displayed energy range is slightly different from the one at QCD-LO discussed in the
context of the purely factorized signal diagram in
\cref{sss:around_threshold} where the cross
section for $\tT{}$ was larger than the one for $\WbWb{}$ production
in a much wider range.

In the right panel of \cref{fig:nlofull}, we show the relative scale
variations for fixed $\sqrts=\SI{350}{\GeV}$ in the range $0.125 \le
\mu_\T{R}/\mu_0 \le 8$. This is much wider than the default variation
which is indicated by the vertical dashed lines. In this case, we
set $f=1$ and just varied $\mu_\T{R}$ via $h$ as
defined above. The qualitative scale variation behavior is the same
for all shown curves. Overall, the variations are slightly asymmetric
with stronger dependence for lower $\mu_\T{R}$ while within our
default variation range the scale variation is linear to a good
approximation. We emphasize that it is important that the QCD-NLO
width is computed with the same renormalization scale that it is used
in the QCD-NLO cross section computation to achieve a consistent
normalization of total cross section results, see
also~\Rcite{1609.03390}.

\subsection{Matched results}
\label{ss:matched_results}
\begin{figure}[htbp]
\centering
\includegraphics[width=\textwidth]{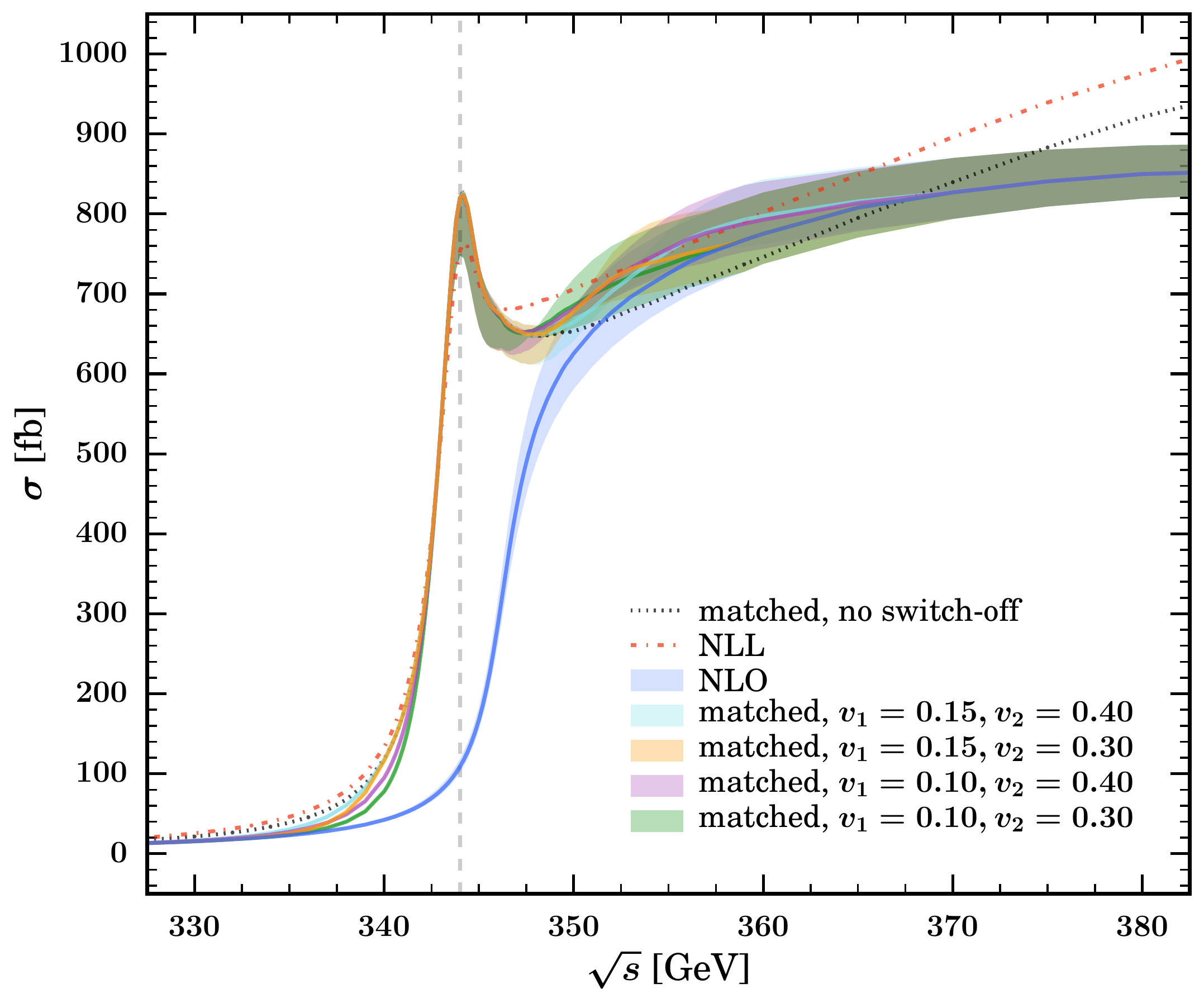}
\caption{Total cross section employing the \ac{NLL} form factors in combination   with QCD-NLO decay and full QCD-NLO contributions to $\WbWb{}$ production as described in \cref{eq:NLONLL}.  
  We show four different matching parameters with the
  cyan, orange, purple and green bands, the matched cross section
  without switch-off function as black dotted line, the pure NRQCD \ac{NLL} as red
  dash-dotted line as well as the pure fixed-order QCD-NLO result for
  $\WbWb{}$ production as blue band.  The bands correspond to $h$ and $f$ variations (unsymmetrized) as described in \cref{ss:theoretical_uncertainties}.
  }
\label{fig:matching-NLL}
\end{figure}

In \cref{fig:matching-NLL}, we show the fully matched result for the
total $\WbWb{}$ cross section over the c.m.\ energy $\sqrt{s}$ (light
blue, orange, purple, green) employing \ac{NLL} threshold resummation
via the S- and P-wave form factors, QCD-NLO top and anti-top quark
decays and fixed-order $\WbWb{}$ production at QCD-NLO, as described in
\cref{eq:NLONLL}. The error bands arise from the correlated $h$ and
$f$ variations as described in \cref{ss:theoretical_uncertainties},
see also \cref{ss:resummation_and_scales} and \cref{eq:scales}. The
matched cross section is displayed for four different sets of the
matching parameters $(v_1,v_2)$ with $v_1\in\set{0.1,0.15}$ and
$v_2\in\set{0.3,0.4}$. For comparison we have also displayed the pure
\ac{NLL} threshold-resummed cross section without invariant mass cuts
(dash-dotted red, using default scale setting), the matched cross
section without switch-off function, i.e.\ setting $\switch=1$ 
(dotted black, referred to as ``no switch-off'' and using default scale
setting), and the pure QCD-NLO fixed-order cross sections (blue
band). 

We observe, that the fully matched cross section has the desired
properties: In the threshold region (the $\sim\SI{5}{\GeV}$ window
around $\sqrts=2\mOneS{}$) and in the continuum
($\sqrt{s}\gtrsim\SI{370}{\GeV}$) as well as the regions 
below threshold ($\sqrt{s}\lesssim \SI{330}{\GeV}$) we recover the
threshold-resummed and the relativistic fixed-order result,
respectively. In the intermediate/matching regions, where the details
of the matching procedure are relevant, we see a quite stable
transition behavior. We note that our best prediction in the threshold
region is not the pure QCD \ac{NLL} threshold-resummed cross section
(dash-dotted red), but the matched  
QCD-NLO+NLL version of \cref{eq:NLONLL} as it in addition accounts for
off-shell-top, single-top as well as background contributions
including their interference with 
the threshold-resummed matrix elements associated with the form factor.
Phenomenologically, we observe that in the important peak region the matched
cross section is 5-10\% larger than the pure QCD \ac{NLL} prediction, which
shows the importance of the non-$\tT{}$-resonant effects.
Interestingly,
these corrections agree in sign and also roughly in size with the known 
NNLL/NNLO corrections at threshold~\cite{Hoang:2000yr,hep-ph/0107144,hep-ph/0209340,Pineda:2006ri,1309.6323,1506.06864}.

It is also conspicuous that the difference between the different
choices of matching parameters is non-negligible. 
In the cross-over region $\sqrt{s}\approx\SI{348}{\GeV}$ around the local
minium above the peak (examined in more detail below) 
and also the shoulder region below the peak it exceeds by far the scale variations.
We see that at the level of our approximation, renormalization scale
variation alone is not sufficient to estimate the remaining
theoretical uncertainties in these matching regions and that one has
to also account for different viable choices of the matching
parameters when estimating the remaining theoretical uncertainties.

\begin{figure}[htbp]
\centering
\includegraphics[width=\halfwidth]
  {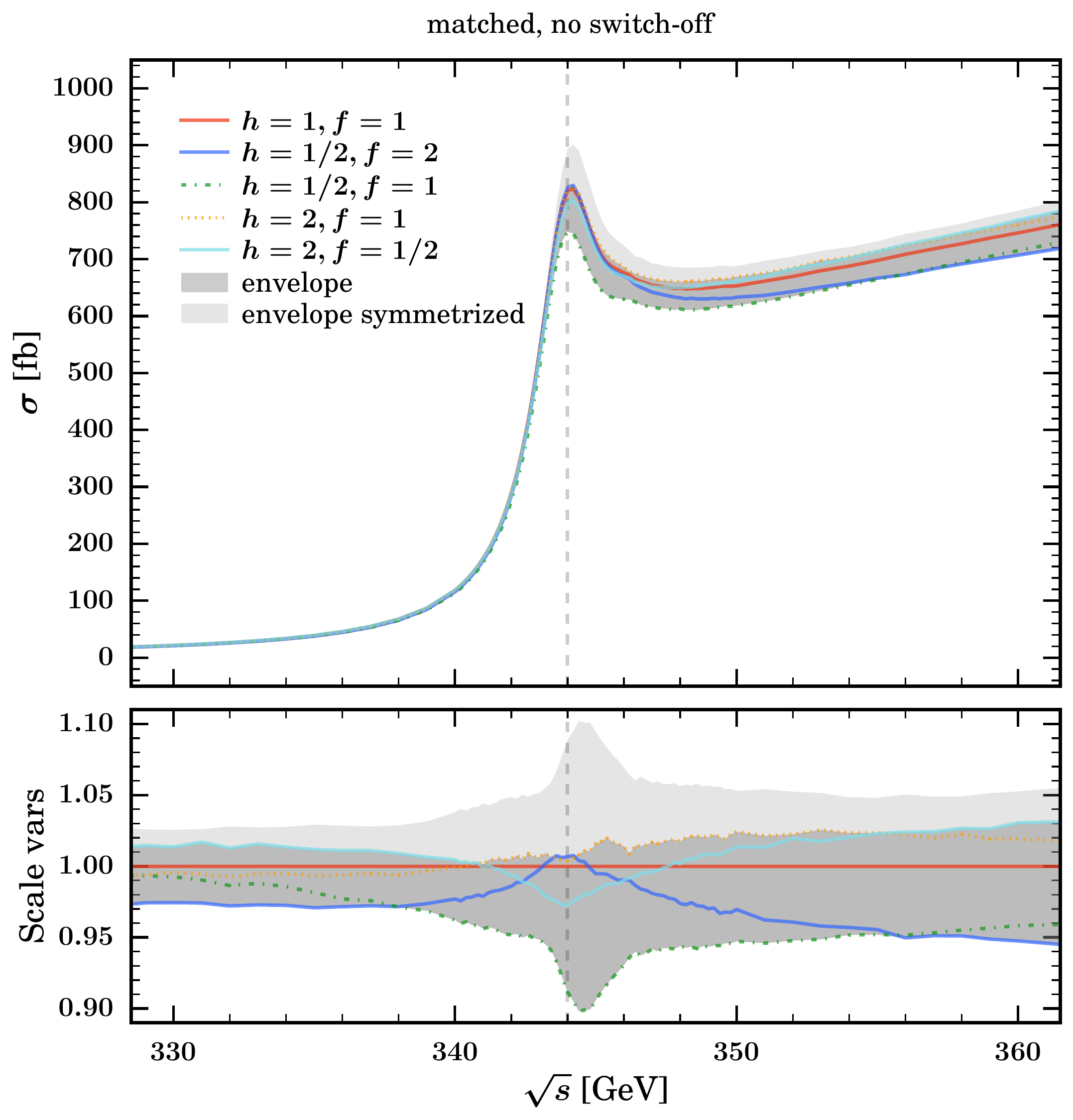}
\includegraphics[width=\halfwidth]
  {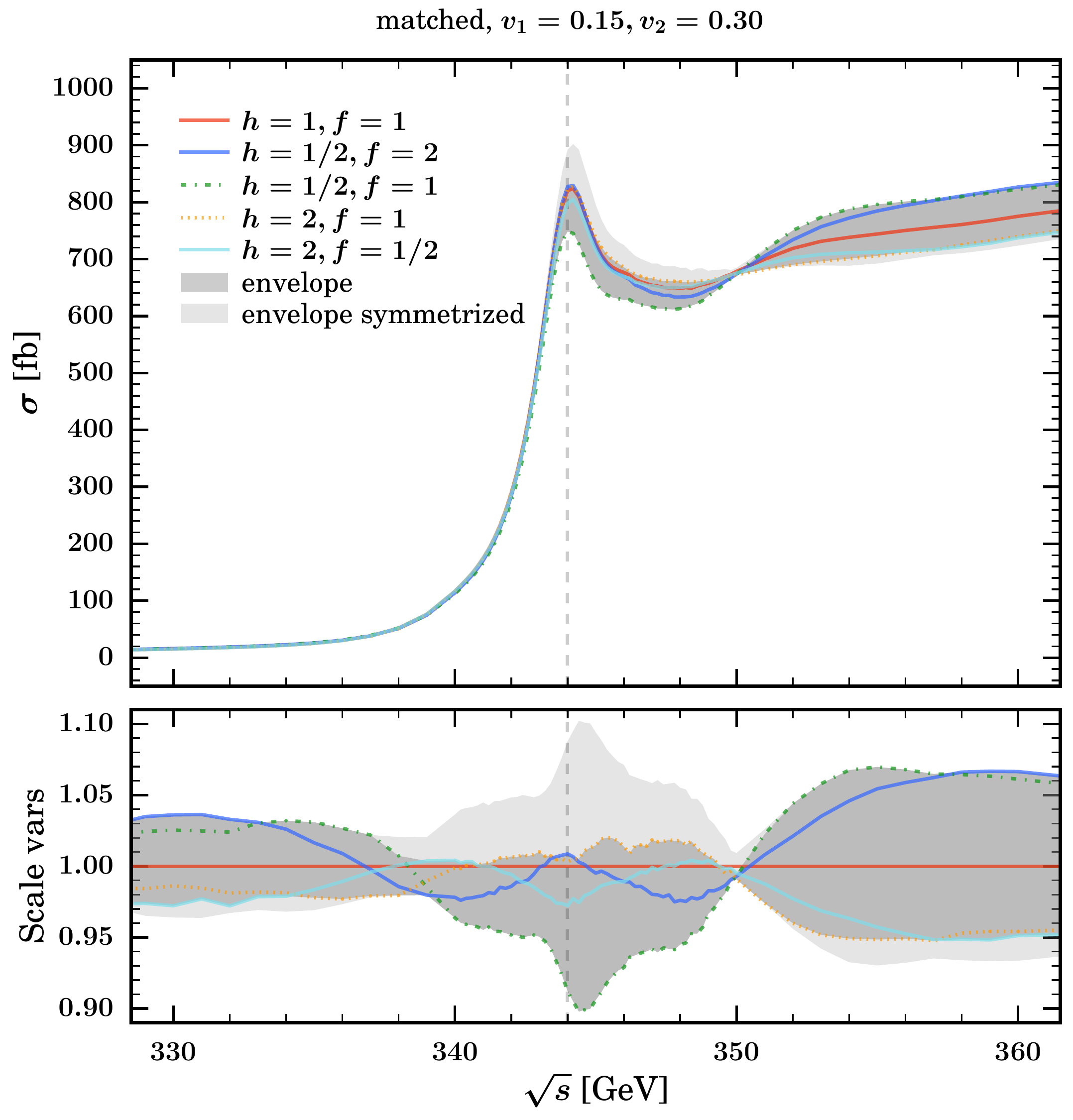}
  \caption{Inclusive cross section according to different scale ($f$, $h$) 
choices without switch-off function (left panel) and with switch-off
between  $v_1=0.15$, $v_2=0.30$ (right panel). In addition to the
lines that can be identified using the legend, we show the envelope
(dark gray) as well as the symmetrized envelope (light gray) according
to \cref{ss:theoretical_uncertainties}.} 
\label{fig:matched_nlofull_nlodecay_symm_scalevars}
\end{figure}

An interesting aspect of the threshold-resummed NLL calculations is that
the renormalization scale variations are quite asymmetric with respect to the default
scales, see \cref{fig:validation_dm_fixed} and also the discussion in \Rcite{1309.6323}.
This behavior is therefore also present in our matched results and examined in 
more detail in \cref{fig:matched_nlofull_nlodecay_symm_scalevars}. Let us start 
the discussion with
the left panel, where we show the individual behavior of the different
$(h,f)$ scale choices with the switch-off
function $f_s=1$, i.e.\ using no switch-off. We see that the maximal variation envelope
is obtained by a nontrivial interplay of the different $(h,f)$ scale choices. 
For example, away from the threshold (either below or
above) $(h,f)=(1/2,2)$ provides the largest deviation from the
default, while in the threshold region it is $(h,f)=(1/2,1)$. We also
see that in the threshold region the default result is close to the
maximal result obtained for $(h,f)=(2,1)$ visualizing once more the
asymmetry mentioned above. We have also shown in light gray the
symmetrized envelope that has been computed with the procedure outlined
in \cref{ss:theoretical_uncertainties} and which represents our (more
conservative) theory error estimate.

\begin{figure}[thb]
\centering
\includegraphics[width=\textwidth]
  {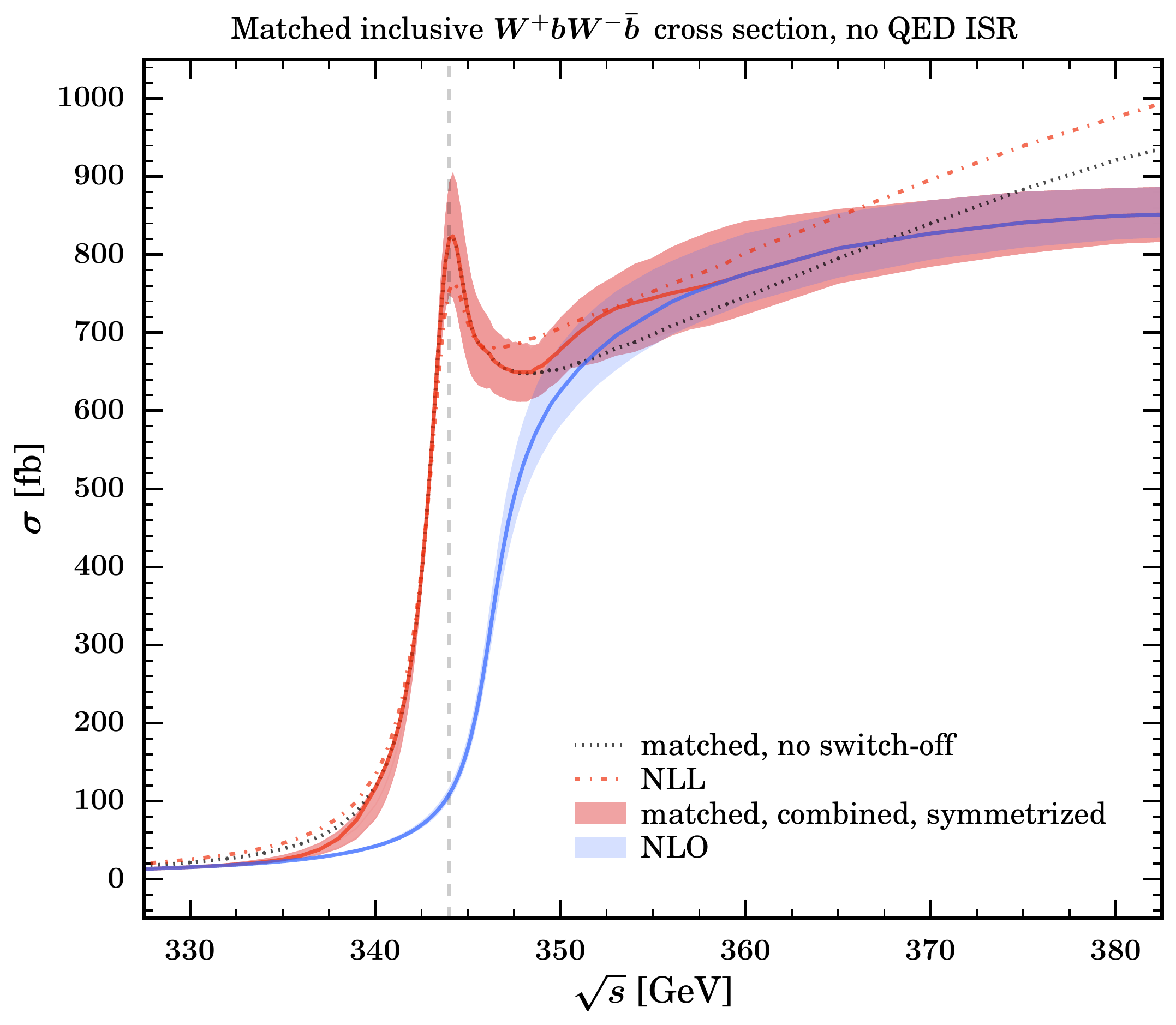}
\caption{Inclusive cross section according to our matching presription
  by combining the symmetrized scale variation envelopes for different
  matching parameters. The red line corresponds to the default parameter 
  setting: $h=1$, $f=1$, $v_1=0.15$, $v_2=0.3$. Blue band and other lines as in
  \cref{fig:matching-NLL}.}
\label{fig:matched_nlofull_nlodecay_symm_comb}
\end{figure}

In the right panel of
\cref{fig:matched_nlofull_nlodecay_symm_scalevars} we show the results
with the same $(h,f)$ scale choices, but when the switch-off
function $f_s$ is different from the identity, as described in
\cref{ss:switchoff}. We see an interesting cross-over behavior which
is a result of the opposite scale behavior of the \ac{NLL}
threshold-resummed calculation (which has a quite complicated shape),
and the fixed-order QCD-NLO calculation in the continuum (which simply
follows the scale behavior of $\alpha_\T{F}$). 
The QCD-NLO scale variations are as one would naively expect and as we
have already shown in \cref{fig:nlofull}:
For smaller (larger) renormalization scales (controlled by $h$), $\alpha_s$
increases (decreases) and thus the cross section increases (decreases).
Here the $f$ variation represents only a very minor additional
modification as it only results in changing the pole mass $\mpole$
value for our approach to implement the 1S mass scheme, see
\cref{ss:mass}. The \ac{NLL} scale variations in the threshold region,
however, are more involved and have a quite strong $f$ dependence,
which leads to an opposite behavior concerning the $(h,f)$
variations. This artificially creates a very small scale variation in
a cross-over region, where the cross sections for all $(h,f)$ settings happen to agree at an energy that depends on the actual values of the matching
parameters $v_1$ and $v_2$, as is clearly visible in the right panel of \cref{fig:matched_nlofull_nlodecay_symm_scalevars} at $\sqrt{s}=350$~GeV. Note that the
symmetrization of the envelope barely affects the continuum region and
also does not remove the cross-over behavior. It is therefore crucial
to also account for variations of the matching parameters  $v_1$ and
$v_2$ to obtain a reasonable estimate of the theoretical uncertainties.

This is finally shown in
\cref{fig:matched_nlofull_nlodecay_symm_comb}, which displays the
fully matched total $\WbWb{}$ production cross section at QCD-NLO+NLL including 
the full combination of renormalization scale and matching parameter
variations (red). The result represents our best prediction for the
$\WbWb{}$ production total cross section at QCD-NLO+NLL order and is
valid in all kinematic regions. By performing the symmetrization
procedure, we believe to have a reliable estimate of the theory
uncertainty in the sense that the next order result with respect to
QCD corrections is expected to have at least a substantial overlap
with the current uncertainty band. Of course, this does not include
known classes of large electroweak corrections, such as initial state
radiation, which have to be considered on top of this. Parts of these
effects will be discussed in the next section.

%
\subsection{QED initial state radiation}
\label{s:isr_and_polarization}
%
Within the \whz{} framework it is straightforward to combine our fully matched predictions
for $\WbWb{}$ production with \ac{ISR}, beamstrahlung and beam energy
spread or to account for the polarization of the colliding
electron-positron pair. Since \ac{ISR}, beamstrahlung and beam energy
spread involve a convolution of hard cross sections at collision
energies $\sqrt{\hat s}\leq \sqrts$, where using our fully matched
predictions provide a substantially improved description, we will
exemplarily only discuss the effects of \ac{ISR} in the following. We
leave the examination of  beamstrahlung, beam energy spread and (the at leading electroweak
order much simpler) polarization effects to future work and simulation
studies of the experimental collaborations.
\begin{figure}[htb]
\centering
\includegraphics[width=.8\textwidth]{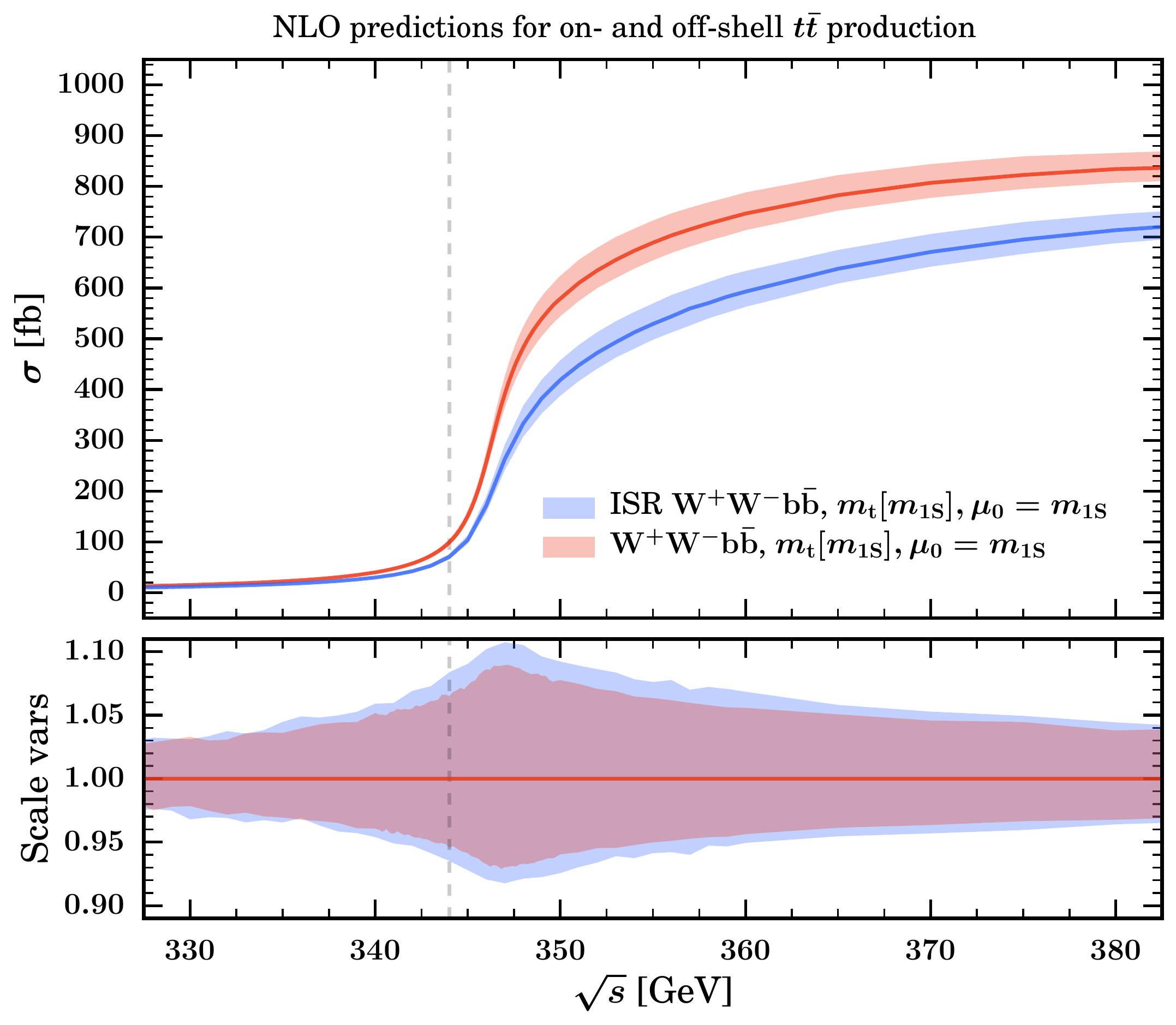}
\caption{Inclusive cross section for the process, $e^+e^-\to \WbWb$ at 
QCD-NLO. The 1S mass, \mOneS{}, has been used
  as central renormalization scale and the bands correspond to the usual ($h$) 
variations by a factor between $1/2$ and $2$. The blue and red band show the 
process with and without QED ISR, respectively.}
\label{fig:isr_fixorder}
\end{figure}

While \ac{ISR} is the QED all-order photon radiation in the
initial state that substantially alters the energy of the hard
interaction, beamstrahlung ist the classical coherent radiation from
the highly collimated charge density clouds of the lepton collider
beam bunches. These are steered in \whz{} with the usual
\prog{Sindarin} syntax known from LO studies. In this section we
demonstrate that our matching approach can be combined with a varying
hard collision energy $\sqrt{\hat s}\leq \sqrts$ and we show how this
affects the threshold and peak behavior. For the full QCD-NLO part of
the matched cross section with polarization, we use the BLHA extension
of the \whz-\openloops{} interface, that allows to pass squared
amplitudes from \openloops{} to \whz{} exclusive in the helicity of
the initial states and that has been validated in \Rcite{1609.03390}. 

Though the more drastic effect on the threshold shape arises from the
combination of ISR, beamstrahlung and beam energy spread, which
together completely wash out the 1S peak (cf. e.g.~\Rcite{1411.7318}),
a coherent analysis of these effects or of using polarization to
achieve the reduction of background contributions is left to the
experimental collaborations. We note that for example beamstrahlung
effects that closely model the environment of the corresponding linear
collider setup can be easily simulated with the \whz{} subpackage
\prog{Circe2}. Linear collider beam spectra are available within the
\whz{} framework for the legacy TESLA project, for ILC, for CLIC, as
well as for CEPC. 

For the inclusion of ISR effects, we first compare in
\cref{fig:isr_fixorder} the full relativistic QCD-NLO fixed-order
result for the off-shell (with respect to the top quarks) process $e^+e^- 
\to \WbWb$ with (blue) and without ISR (red).
The ISR effects
have been included as usual in \whz{} in the structure function
approach in a completely collinear setup, resumming soft photons to
all orders and hard-collinear photons up to third order in $\ALweak$. 
Although the QCD corrections for the final
state completely factorize with the initial state structure function
convolution, the reduction of the effective energy $\sqrt{\hat{s}}$
affects the different components differently and leads to a shift of
phase space points from different energies, 
which can cause additional nontrivial effects. 
We see in \cref{fig:isr_fixorder} that this is indeed so because ISR 
leads to the well-known overall relative 
reduction of the cross section, but at the same time also to a 
visible enhancement of the relative scale variation band. 
\begin{figure}[htbp]
\centering
\includegraphics[width=.45\textwidth]
                {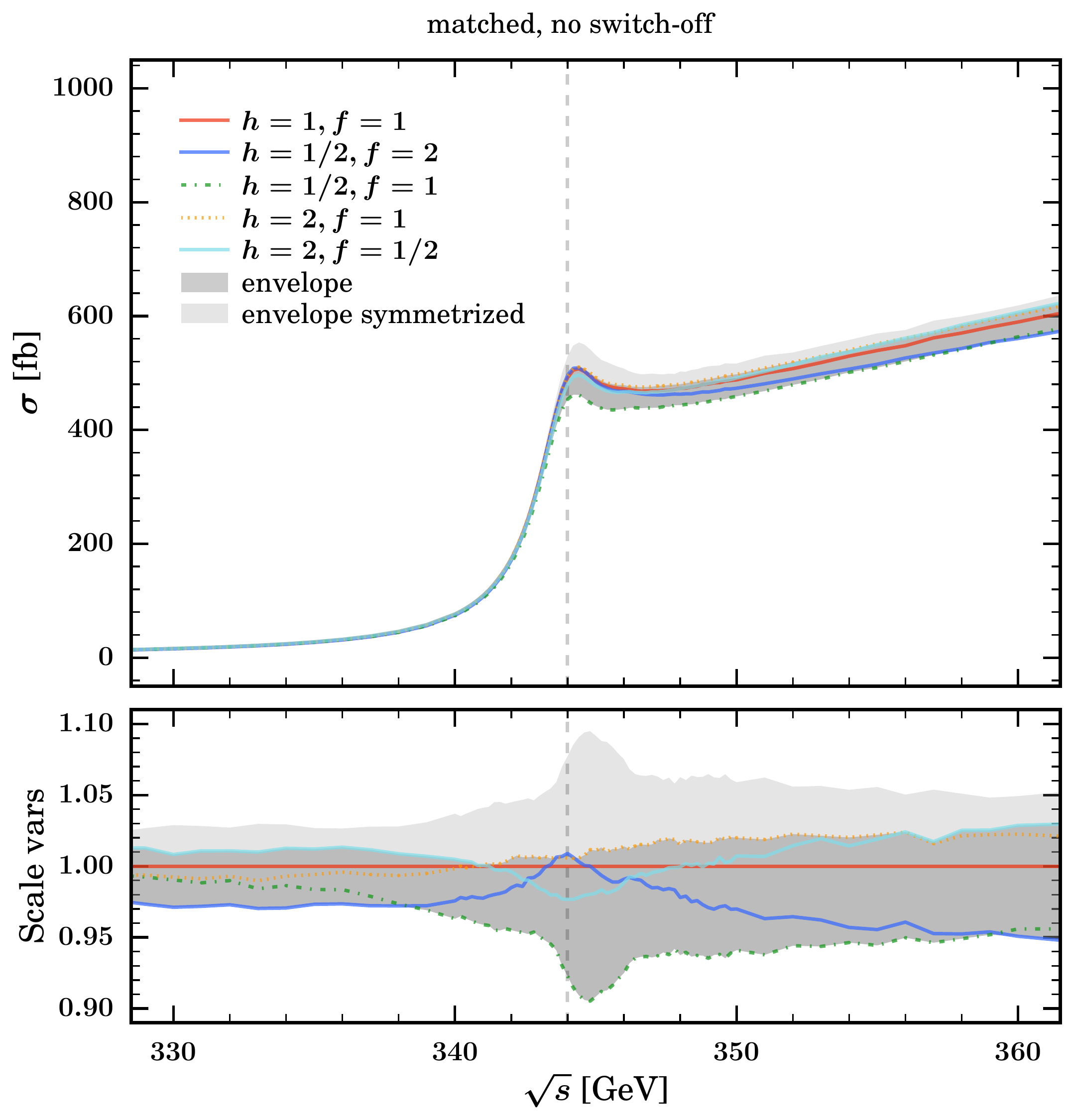}
                \raisebox{.2cm}{
\includegraphics[width=.52\textwidth]
  {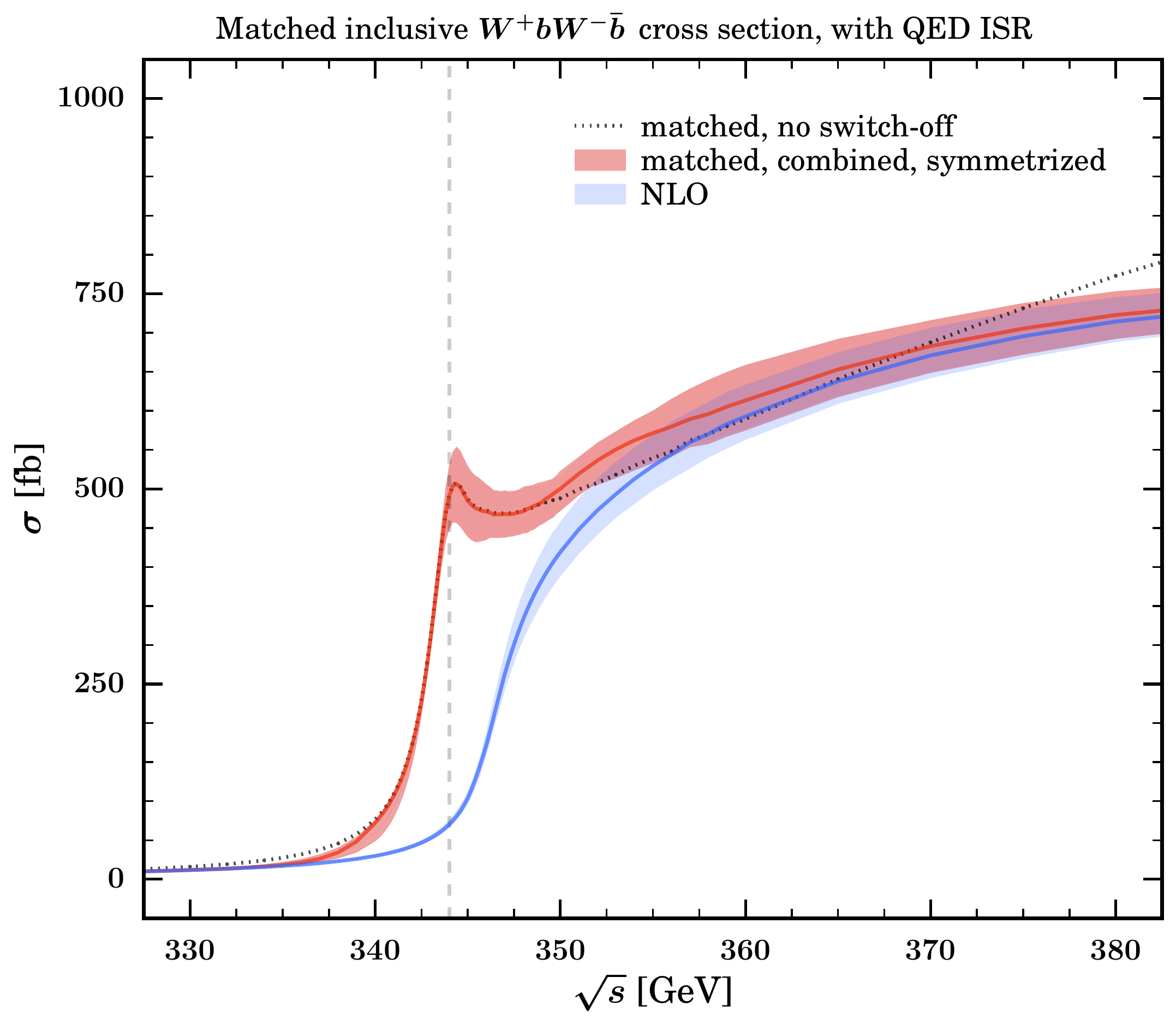}}
  \caption{Left: Inclusive cross section according to different scale choices
  for the case of no switch-off function for the process with full QED
  ISR. Same as
  \cref{fig:matched_nlofull_nlodecay_symm_scalevars}. Right: inclusive
  cross section with ISR structure function according to the matching
  description by combining the symmetrized scale variation envelopes
  of different variation parameters. Same as
  \cref{fig:matched_nlofull_nlodecay_symm_comb}.}
\label{fig:matched_nlofull_nlodecay_symm_scalevars_isr}
\end{figure}

In the left panel of
\cref{fig:matched_nlofull_nlodecay_symm_scalevars_isr}, we show the
matched QCD-NLO+NLL cross section as a function of $\sqrt{s}$ with ISR 
and without switch-off (i.e.\ $f_s=1$ and referred to as ``no switch-off''). 
Comparing to the analogous result without ISR
in \cref{fig:matched_nlofull_nlodecay_symm_scalevars}
we see, as expected, that the overall cross section is reduced and that
the peak is less pronounced, however still clearly visible. 
(Note that the peak is washed out only after beamstrahlung is included as well.)
Again the ISR, which shifts events from higher energies to
the threshold via the radiative return, has a nontrivial effect 
on the relative scale variation, which can be seen to be 
slightly smaller than for the case without ISR. This is just opposite
to the way how ISR affects the uncertainties for the QCD-NLO cross section 
prediction as shown in \cref{fig:isr_fixorder}.

In the right panel of
\cref{fig:matched_nlofull_nlodecay_symm_scalevars_isr} we show 
as a final result the fully matched QCD-NLO+NLL (red) and QCD-NLO
fixed-order (blue) total cross sections  
including QED ISR  and symmetrized scale variations, 
and where also the switch-off function and
the corresponding variations of the matching parameters are applied.
As a reference we also display the
matched QCD-NLO+NLL cross section without switch-off function for the default 
scale setting (dotted black).
Similar to the observations we made above, we again see that 
the convolution with the ISR structure function and the resulting
radiative return effects  
lead to nontrivial relative modifications of the corresponding cross section
predictions obtained without ISR effects and shown in 
\cref{fig:matched_nlofull_nlodecay_symm_comb}.
We in particular observe that with ISR effects the fully matched QCD-NLO+NLL cross section 
fully merges into the QCD-NLO fixed-order prediction only for 
c.m.\ energies beyond $360$~GeV, while this happens already above $353$~GeV without ISR.
This arises because the toponium peak enhancement of the fully matched calculation
also has an effect for c.m.\ energies above the peak position. 
The results also show the importance of using the fully matched predictions 
in the intermediate region between the toponium peak and the continuum region 
above $360$~GeV.

We note that a coherent study of \ac{ISR} and beamstrahlung effects 
based on fully matched predictions will contribute to a more refined  
classification of which energy regions have to be
considered pure continuum or pure threshold region. Such a detailed
study is especially important for the planned \SI{380}{\GeV} stage of \ac{CLIC}.
This is, however, beyond the scope of this paper.

%
\section{Differential results}
\label{s:differential_results}
After having validated and discussed the fully matched inclusive cross
section results for $e^+e^- \to \WbWb$ as a function of \sqrts{},
we can now start exploiting the full power of our MC implementation,
namely by analyzing differential distributions. We note that the
discussion of the differential distributions carried out here based on
our fully matched approach is not intended to be exhaustive and mainly
serves as a proof of principle. It is clear that a number of
distributions provide alternative means of measuring the top mass, but
we postpone a more systematic exploration of such possibilities to
future work. We would also like to remind the reader that at the
current level of implementation, final state interactions due to (ultra)soft
gluon exchange involving top and anti-top decay products are not
included beyond the level of the QCD-NLO corrections, as discussed in
the text following \cref{eq:factorized_NLO}. This means that in the
threshold region the distributions do strictly have LL precision only
and the uncertainties shown should be interpreted with a grain of
salt, particularly for kinematic thresholds visible in 
distributions where soft gluon exchange
involving top and anti-top and their decay products can play an important
role. But we also note, that in many cases the missing NLL corrections
may not represent significant contributions. 

For the analysis of the generated events, we use a custom
\textsc{Rivet}~\cite{1003.0694} analysis. Partons are clustered with
the generalized $k_\T{T}$ algorithm (\texttt{ee\_genkt} in
\fastjet)~\cite{0802.1189,1111.6097} with $R=0.4$ and $p=-1$.
A minimal jet energy of \SI{1}{\GeV} is required. We assume a perfect
$b$-tagging efficiency including the charge. Thus a $b$-jet
($\bar{b}$-jet), \jb(\jbbar), is a jet containing a
$b$($\bar{b}$) quark. We always require at least two jets during
the analysis. For distributions of observables that are identical for 
$t\lrarrow\bar{t}$, $b\lrarrow\bar{b}$ and
$W^+\lrarrow W^-$, we only show results for $t$, $b$ and $W^+$,
respectively. If not stated otherwise, the results are obtained at
$\sqrts{}=2\mOneS=\SI{344}{\GeV}$. Keep in mind that this is slightly
below the kinematical threshold $\sqrts=2\mpole$, thus the preferred
kinematical Born level situation, as far as $\tT{}$ production is
concerned, is one with one on-shell and one slighly off-shell top
propagator.
We stress that the uncertainty bands shown in this analysis correspond to 
the scale variations as described in \cref{eq:scale_variations} only, 
i.e.\ they have not been
symmetrized as described in \cref{ss:theoretical_uncertainties}. The 
indicated darker solid lines in the plots correspond to our default scale choice $h=f=1$,
so the effect of the symmetrization can be easily seen from the results we show.

\subsection{Top quark observables}
\label{ss:top_observables}
\begin{figure}[htbp]
\centering
\includegraphics[width=\halfwidth]{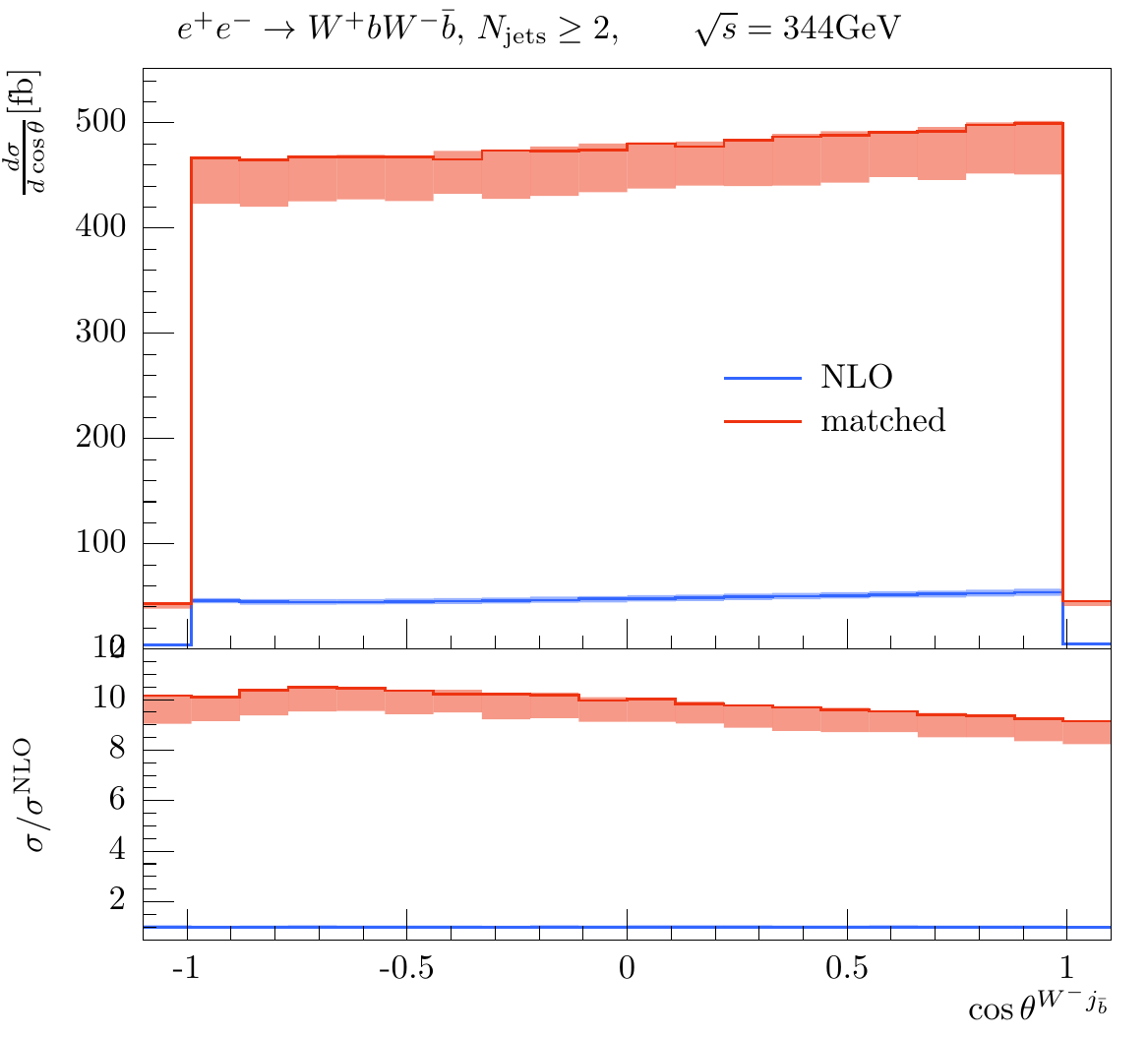}
\includegraphics[width=\halfwidth]{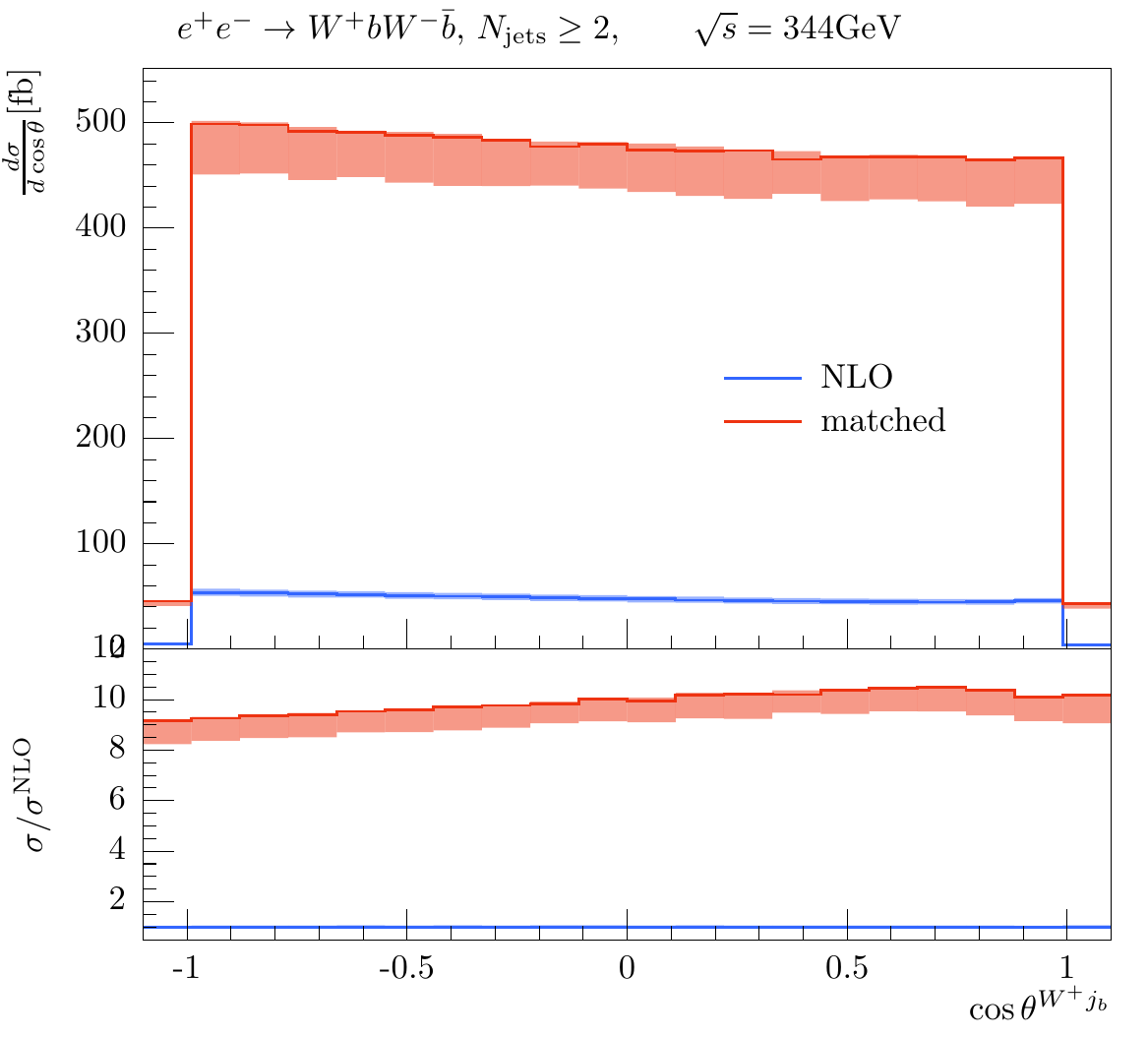}
\caption{Distribution over the top (left panel) and anti-top polar angle (right panel)
  for $\sqrts{}=2\mOneS=\SI{344}{\GeV}$.
  In blue we show the fixed-order cross section \XSNLO{}, while 
  in red we show  \XSmatched{} according to
  \cref{eq:NLONLL}.
  In the lower panel, we show the ratio of $\XSmatched/\XSNLO$.
  The bands correspond to the scale variations as described in
  \cref{eq:scale_variations}. They have not been
  symmetrized as proposed in \cref{ss:theoretical_uncertainties}.
  }
\label{fig:top_obs1}
\end{figure}
\begin{figure}[htbp]
\centering
\includegraphics[width=\halfwidth]{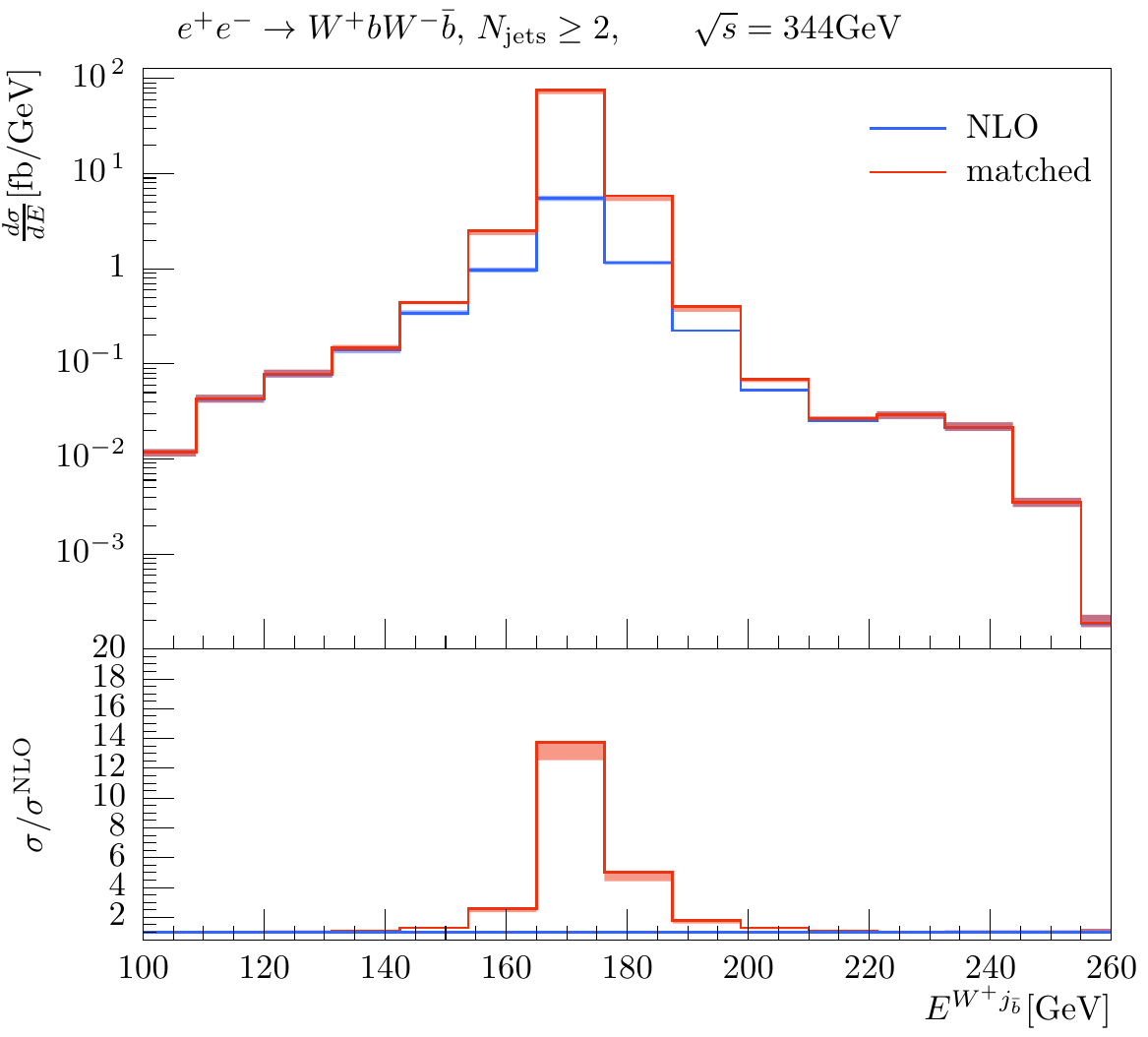}
\includegraphics[width=\halfwidth]{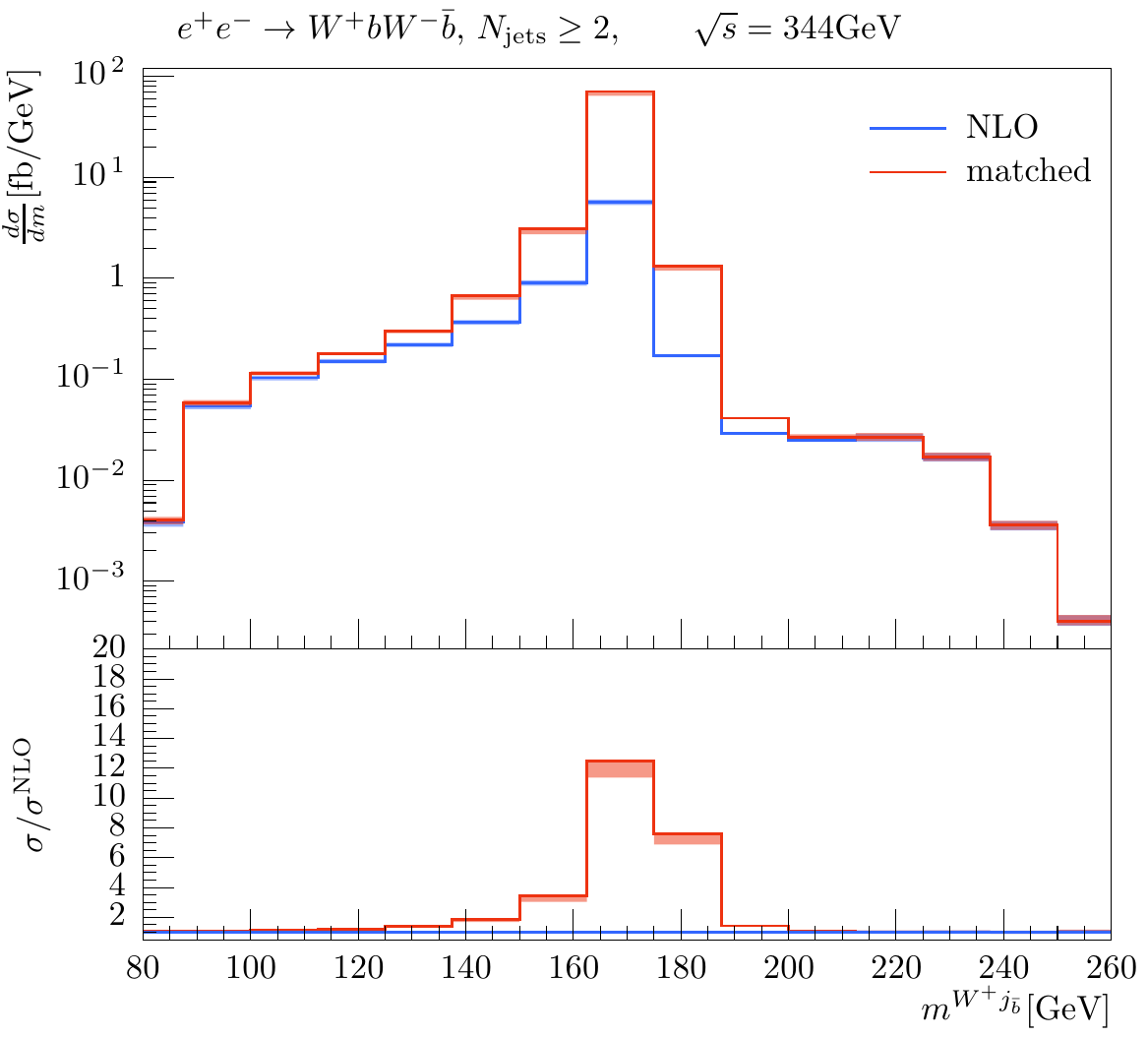} \\
\caption{Top energy (left panel) and invariant mass distribution (right panel),
  respectively. Plot descriptions are the same as in~\cref{fig:top_obs1}.}
\label{fig:top_obs2}
\end{figure}
\begin{figure}[htbp]
\centering
\includegraphics[width=\halfwidth]{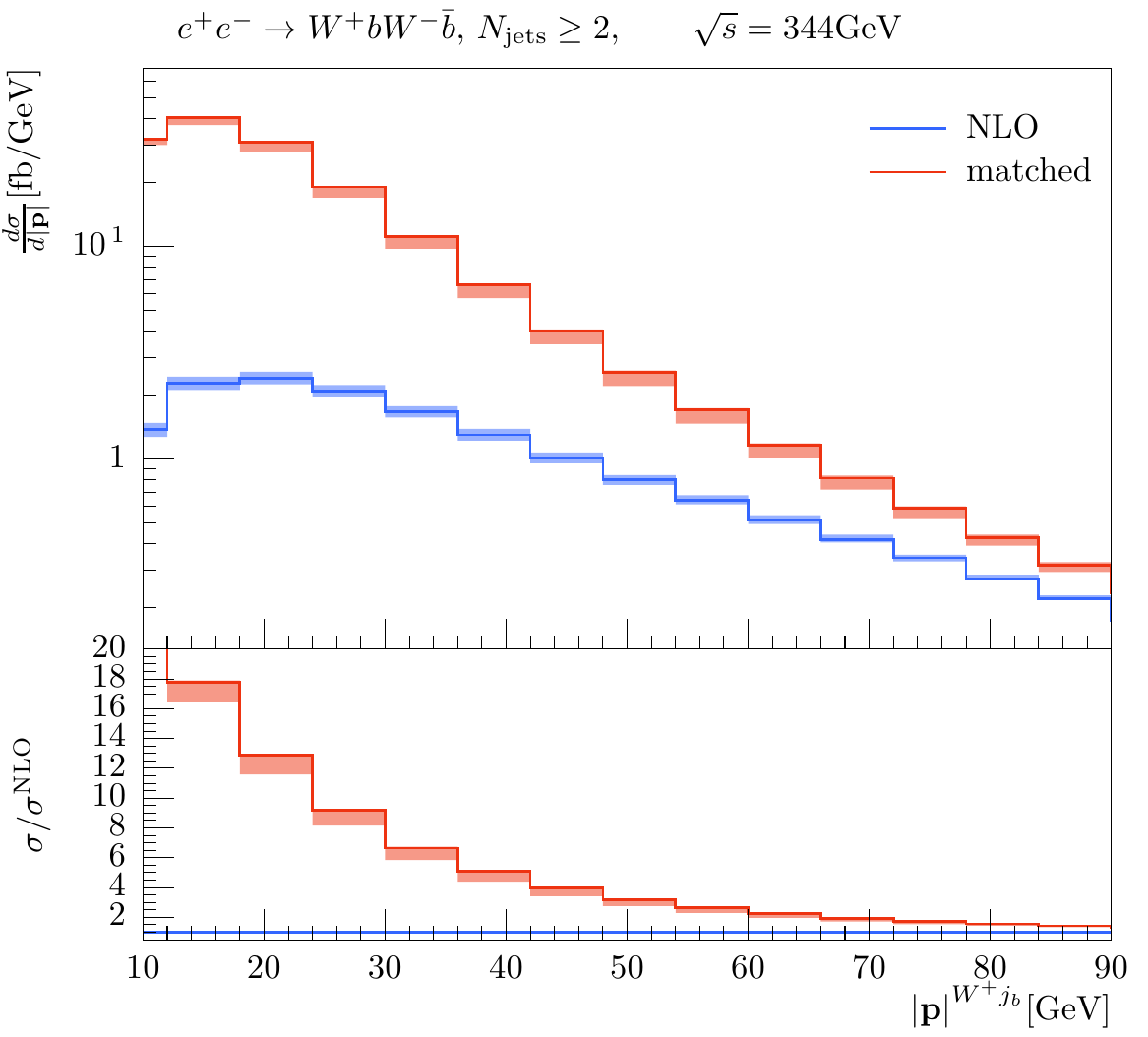}
\includegraphics[width=\halfwidth]{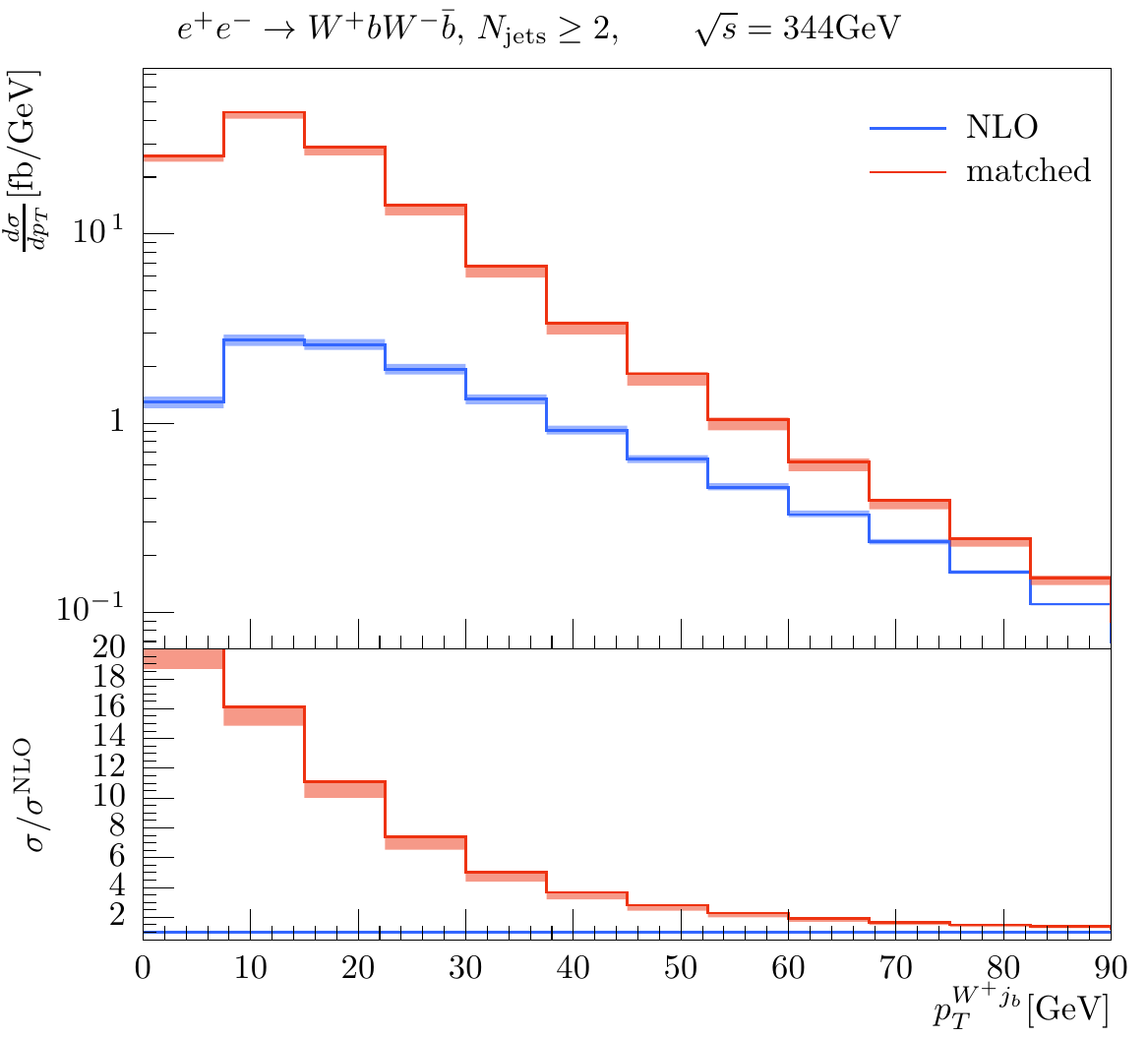}
\caption{Top three-momentum (left panel) and transverse momentum distribution (right panel),
  respectively. Plot descriptions are the same as in~\cref{fig:top_obs1}.}
\label{fig:top_obs3}
\end{figure}

\begin{figure}[htbp]
\centering
\includegraphics[width=\standardwidth]{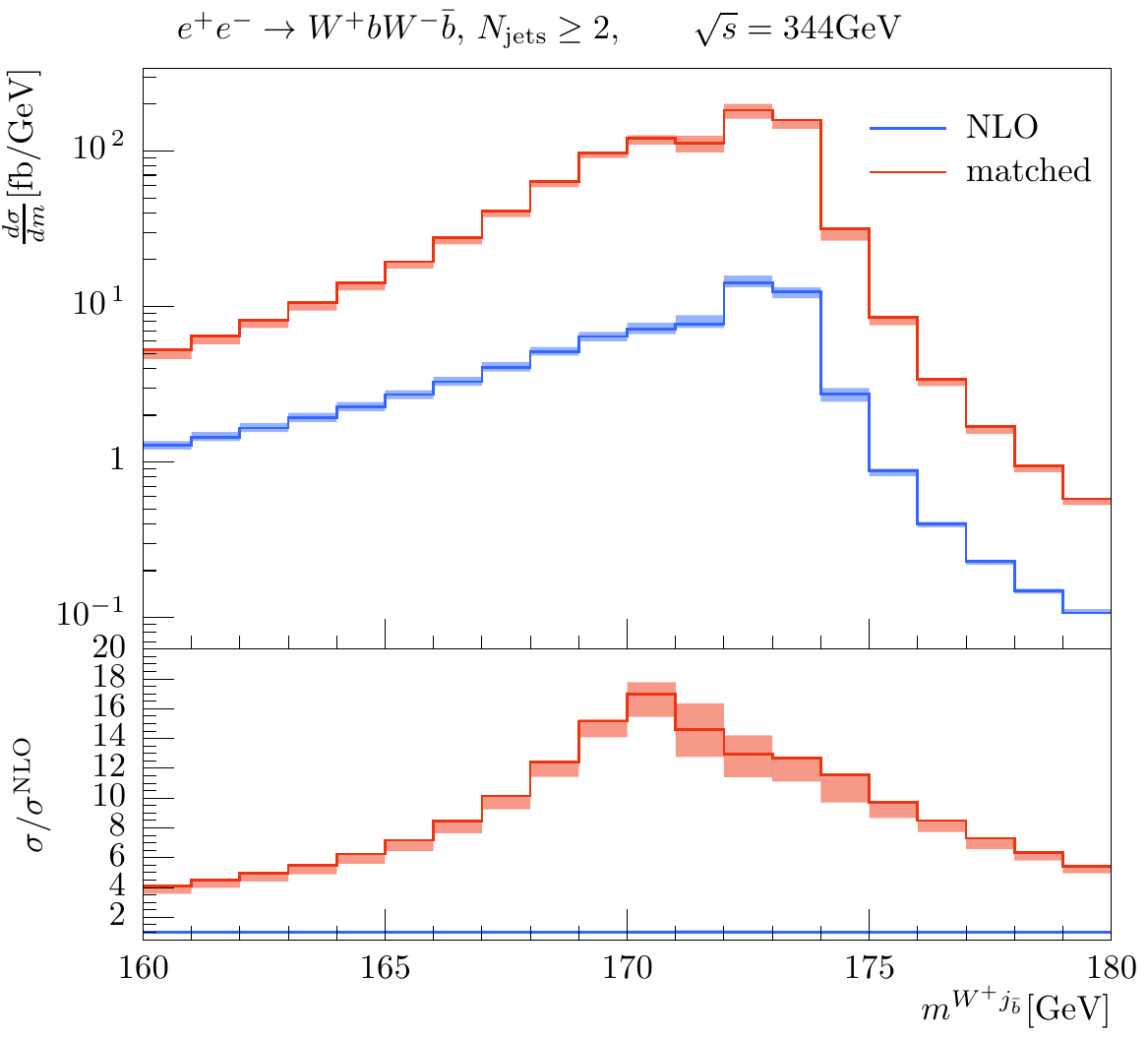}
  \caption{Invariant mass distribution of reconstructed top quarks close
  to the mass peak with fine binning.
  Lines, bands and panels as in \cref{fig:top_obs1}.}
\label{fig:top_inv_mass}
\end{figure}
We start the discussion of differential distributions with a few classic
top observables, which can already be defined for the on-shell $\eetT{}$
process.
In Figs. \ref{fig:top_obs1} to \ref{fig:top_obs3}, we show the top and
anti-top polar angle as well as the top energy, invariant mass,
three-momentum and transverse momentum distributions using the fully matched QCD-NLO+NLL (red) 
and the QCD-NLO fixed order calculations (blue). In the following discussion we
also refer to them simply as  \XSmatched{} and \XSNLO{}, respectively. 

The polar angle distribution shown in \cref{fig:top_obs1} is fairly flat already at NLO.
This is expected, as the forward-backward asymmetry for top-pair
production at lepton colliders is quite small at
threshold~\cite{Jersak:1979uv} due to the $v^2$-suppression of the P-wave contribution.
Note that $\XSmatched/\XSNLO$ shows a slight slope opposite to the polar
angle distribution and thus flattens the distribution even further.
The energy of the reconstructed top quark $E^{W^+\jb}$ displayed in the left 
panel of \cref{fig:top_obs2} 
peaks strongly in the region close to \mpole,
corresponding to resonant (close to mass shell) top quarks with no velocity.
The fully matched description enhances this peak by a factor of $\sim14$,
while adding very little to the off-shell (concerning the top quarks) 
configurations already present in the fixed-order calculation.
The invariant mass of the $W^+$-$b$-jet system $m^{W^+\jb}$ shows a 
very similar behavior.
As we are including all irreducible backgrounds to $\WbWb{}$ to QCD-NLO, 
there are still contributions for $\DeltaMM>\SI{30}{\GeV}$ at the
per cent level from \XSNLO{} in \XSmatched.
At this point, we remind the reader that $\XSmatched-\XSNLO$ contains, 
apart from the interference terms, only double-top
propagator contributions according to \cref{eq:NLONLL}.
Thus, it corresponds approximately to a Breit-Wigner distribution,
which falls off quicker than $\XSNLO$, especially for larger
$m^{W^+\jb}$ as seen in \cref{fig:top_obs2}.
Finally, in \cref{fig:top_obs3}, left panel, we show the distribution
of the reconstructed top three-momentum
$\absIT{\mathbf{p}}^{W^+\jb}$, which is known to be a key
observable in understanding the dynamics at threshold~\cite{Jezabek:1992np,hep-ph/9904468}.
As expected, low three-momenta are preferred both in \XSNLO{} and in
\XSmatched, but we observe a strong enhancement of low momenta due to
the threshold resummation, leading to an enhancement of a factor of over $\sim17$
below \SI{20}{\GeV} that flattens to a factor below 2 above \SI{70}{\GeV}.
The projection to the transverse plane results in a very similar
distribution in $p_\T{T}^{W^+\jb}$, which we show in the right panel
of \cref{fig:top_obs3}. As noted earlier, we omit the histograms for
$E$, $m$, $\absIT{\mathbf p}$ and ${p}_\T{T}$ of the ${W^-\jbbar}$ system, as
they are nearly identical to their ${W^+\jb}$ counterparts.

In \cref{fig:top_inv_mass}, we show a more finely binned distribution of
the reconstructed top invariant mass $m^{W^+\jb}$.
It peaks in the bin \SIrange{172}{173}{\GeV}, which is in the vicinity
to, but below the pole mass $\mpole=\SI{173.124}{\GeV}$, indicating a
shift of the visible physical mass peak due to QCD effects compatible with 
observations made in \Rcite{Fleming:2007xt,Butenschoen:2016lpz,Hoang:2017kmk}
for boosted top quarks. 
It is interesting to see, though, that $\XSmatched/\XSNLO$ is maximal
slightly below the peak in the \SIrange{170}{171}{\GeV} bin.
This is related to the threshold resummation which entails that 
the dominant kinematic configuration is associated with top quark decays
emerging from the would-be toponium resonance, which implies 
two slightly off-shell tops. At the peak of the
$m^{W^+\jb}$ distribution, in contrast, one top quark  
propagator is predominantly on-shell and the other one is
slightly below the top-mass shell.

\subsection{Top decay product observables}
\label{ss:top_decay_products}
\begin{figure}[htbp]
\centering
  \includegraphics[width=\halfwidth]
  {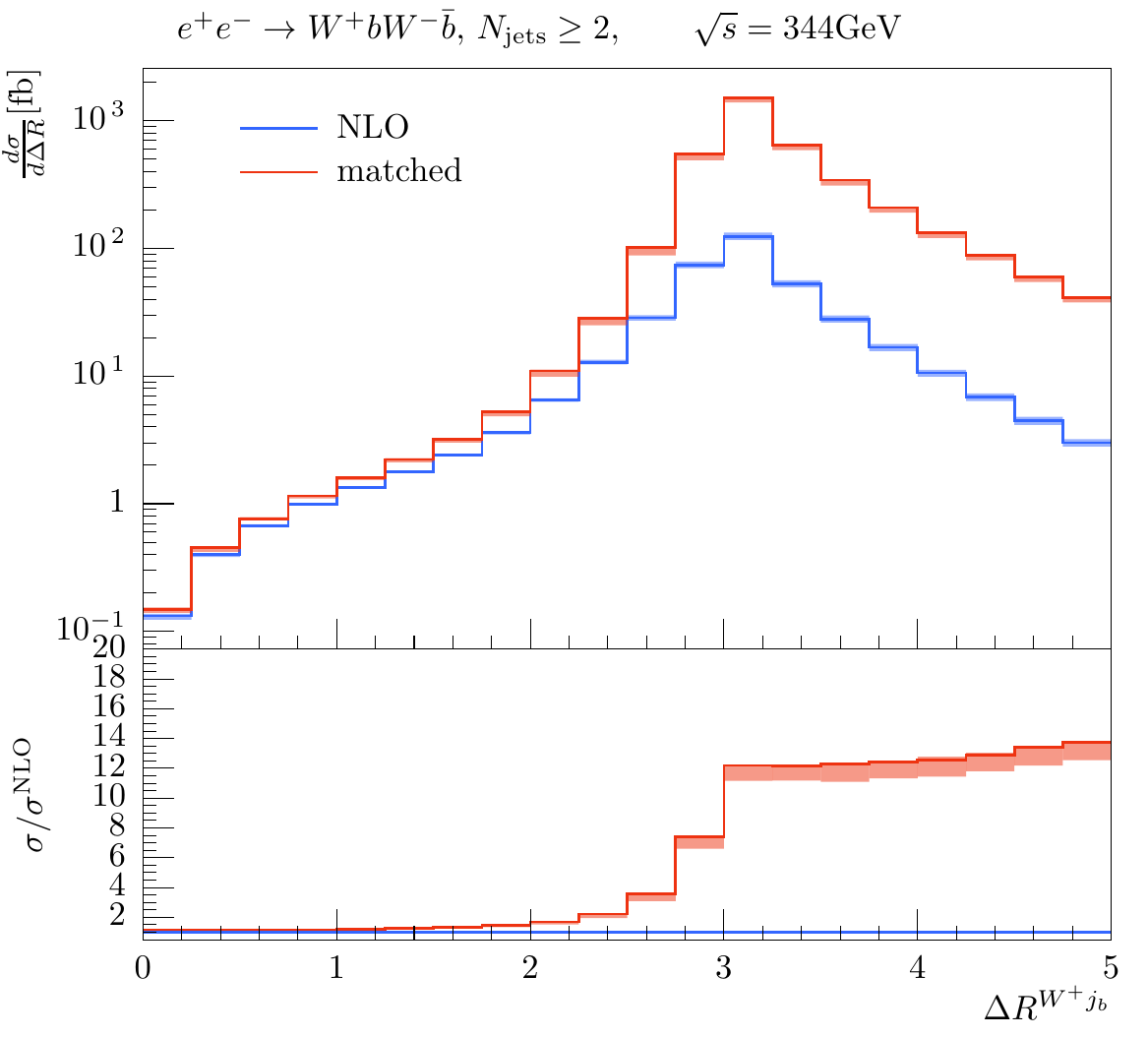}
  \includegraphics[width=\halfwidth]
  {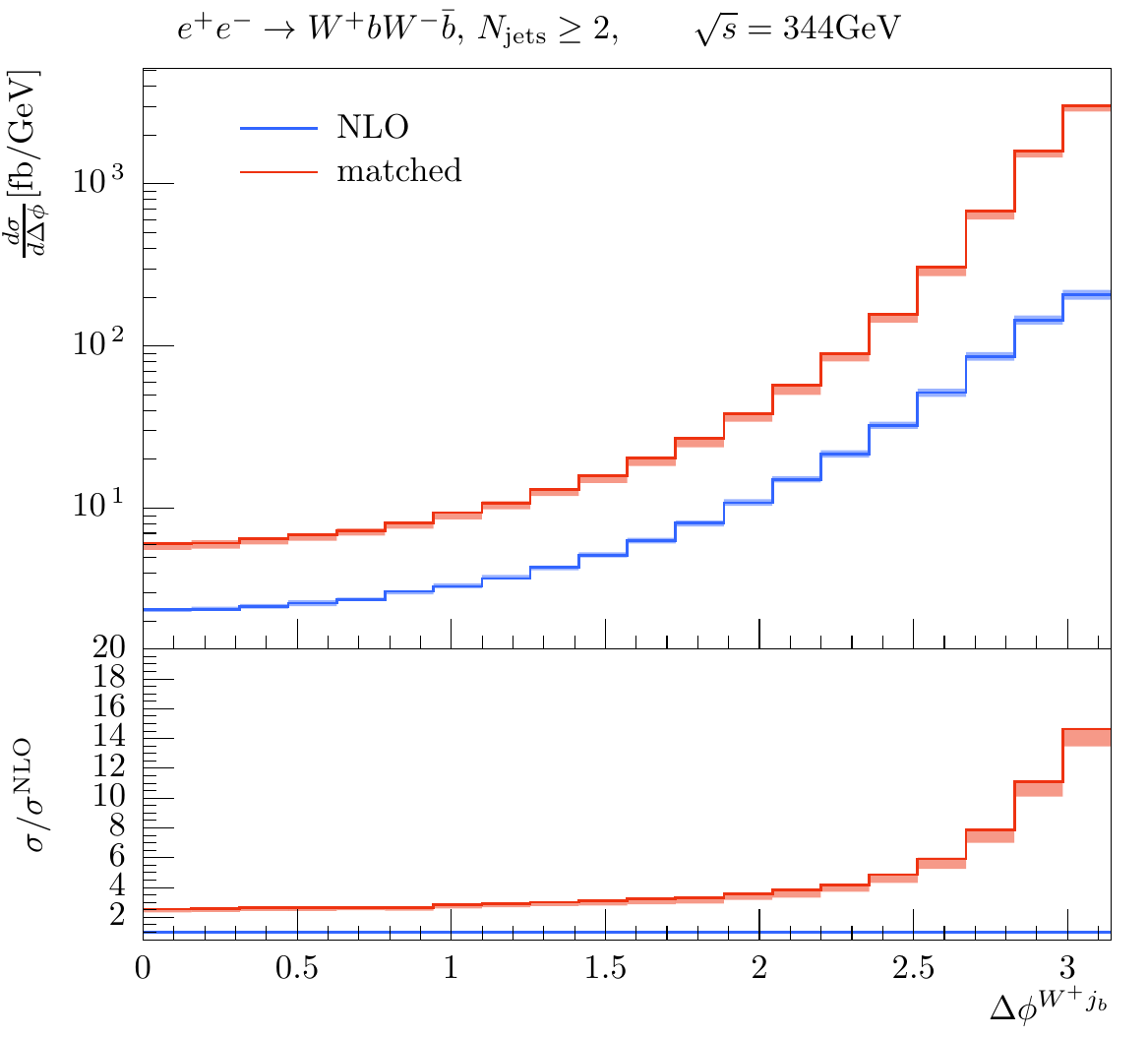}
\caption{Distributions of rapidity (left panel) and azimuthal angle differences (right panel)
  between $b$-jets and $W^+$ bosons.
  Lines, bands and panels as in \cref{fig:top_obs1}.
  }
\label{fig:top_decay_angles}
\end{figure}
In \cref{fig:top_decay_angles}, we show the rapidity and azimuthal angle
differences between $b$-jets and $W^+$ bosons.
These tell us a lot about the kinematics of the top decay and the
underlying background.
In the rapidity difference, we observe already in the QCD-NLO fixed-order 
results a peak around $\Delta R^{W^+ \jb}=3$.
This is quite different from the situation at high energies,  
where for $\sqrt{s}\sim 800$~GeV~\cite{1609.03390}
a rather low rapidity separation of about $1$ is favored.
Obviously, this is related to the boost of the top decay products.
At threshold, the tops have preferably low three momenta
$\absIT{\mathbf p}^{W^+ \jb}$, cf.\ the left plot in
\cref{fig:top_obs3}.
Thus, the back-to-back decay in the top rest frame remains essentially 
unboosted in the lab frame and $W^+$ and \jb move in
opposite directions.
On the other hand, at high energies $W^+$ and \jb will be boosted in the
same direction and thus move preferably in a cone around the original
top momentum, leading to a smaller average $\Delta{R}$.
Going to the matched results, we see that the threshold resummations 
lead to an almost
constant enhancement factor of the $\Delta{R}\geq3$ regime, while barely
enhancing the QCD-NLO results for $\Delta{R}\leq2$.
Thus, we can conclude that the events populating $\Delta{R}\leq2$
represent  dominantly (background) $\WbWb{}$ production not associated
to top production. 
Finally, we note that in the azimuthal angle difference the same physics
is reflected.
Here, we can see a preferred angle separation of $\Delta\Phi=\pi$, as
expected at threshold.
For comparison, for $\sqrt{s}\sim 800$~GeV~\cite{1609.03390}
a value of $\sim\pi/4$ is favored.
Also in this case, the matched (and resummed) results enhance the pure top-decay
topology.
Compared to the R separation, there is no jump in
$\XSmatched/\XSNLO$, though, but a continuous increase
with larger angles.
\begin{figure}[htbp]
\centering
  \includegraphics[width=\halfwidth]
  {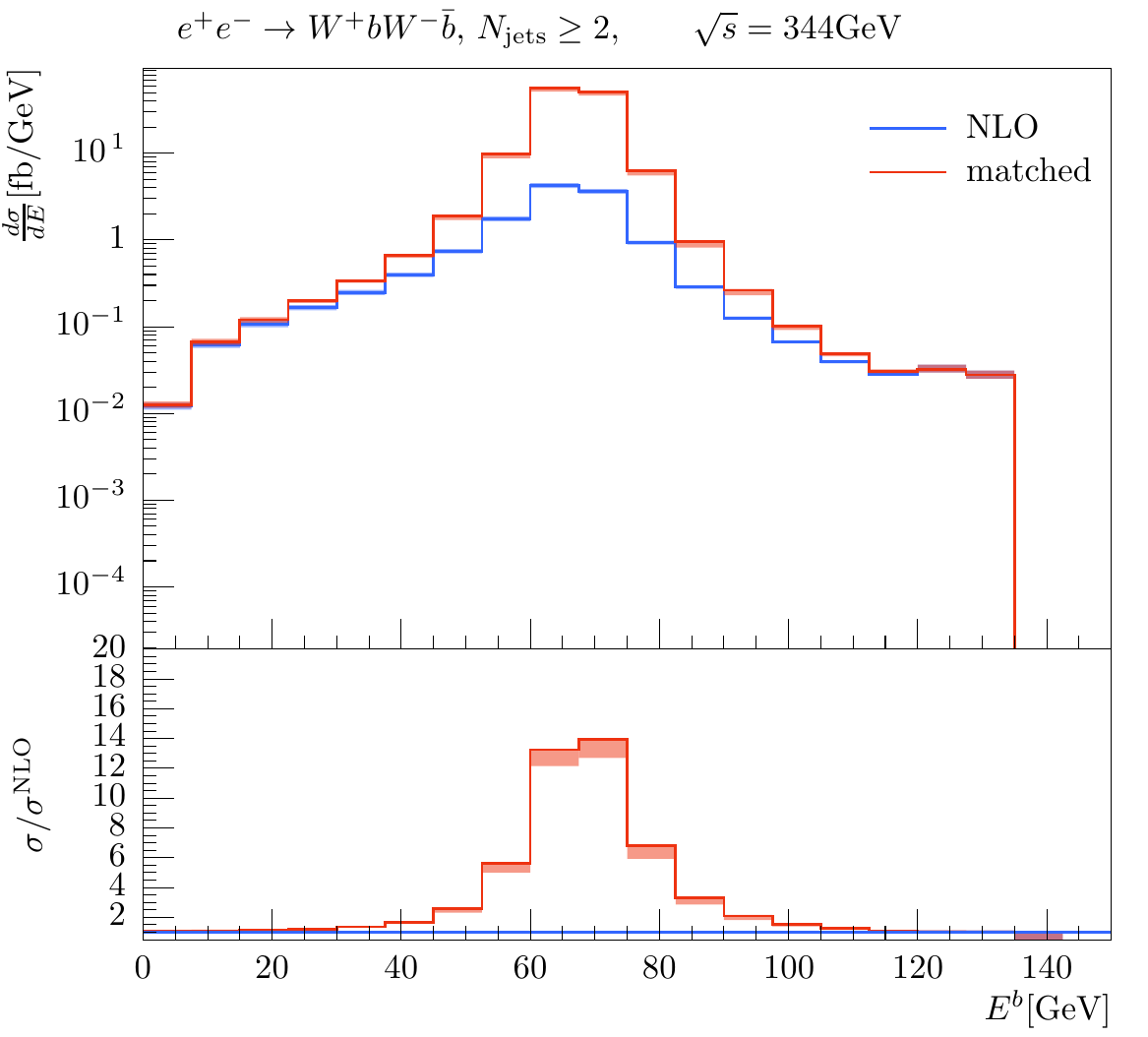}
  \includegraphics[width=\halfwidth]
  {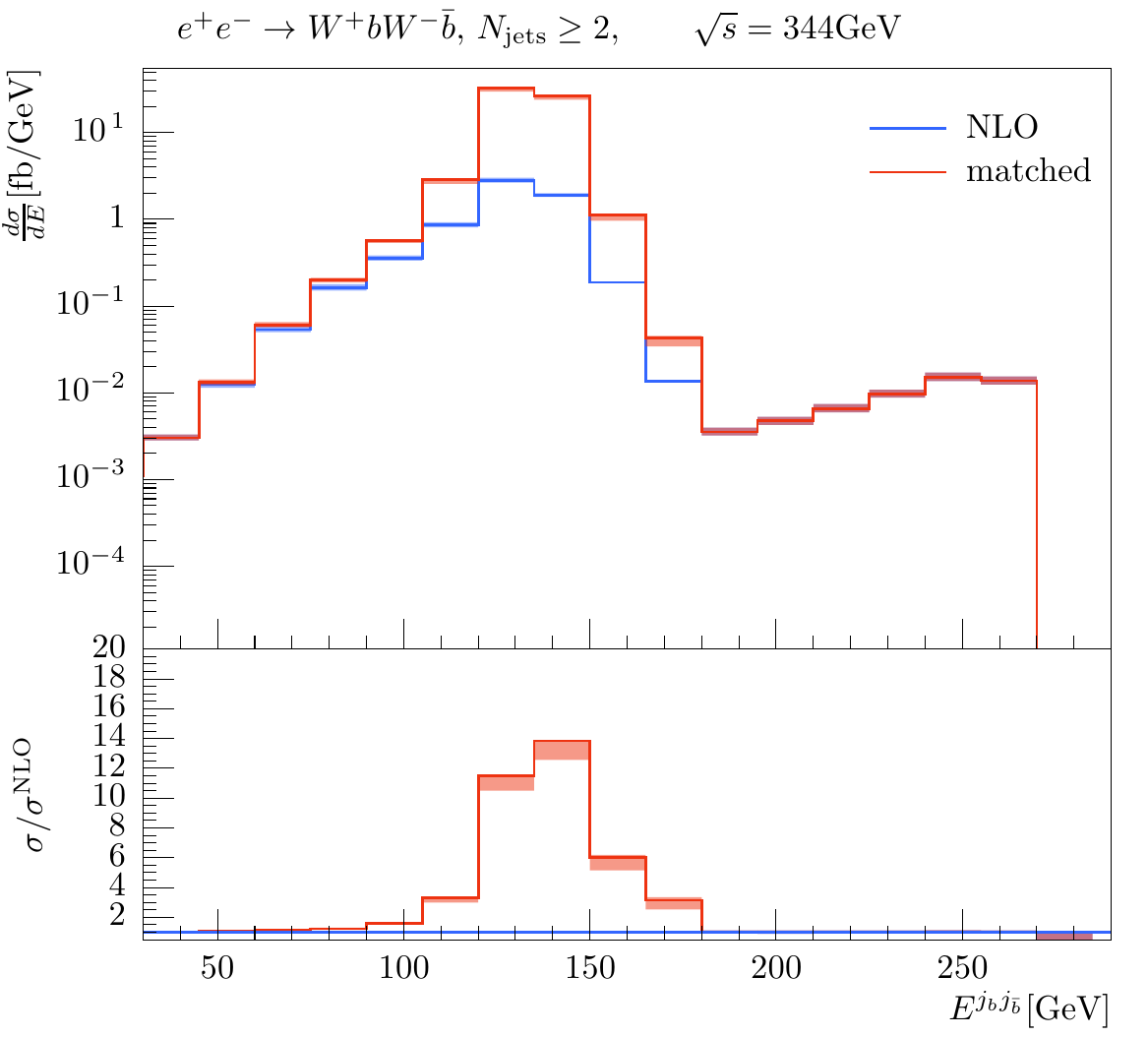}
  \caption{Energy distributions of $b$-jets (left panel) and
  ($b$-jet,$\bar{b}$-jet) pairs (right panel).
  Lines, bands and panels as in \cref{fig:top_obs1}.
  }
\label{fig:b-jet-energies}
\end{figure}
%

While it is fairly obvious that rapidity and azimuthal angle difference
between $W^+$ and $b$-jet will carry information about the top decay, it
is interesting to see whether even single final state distributions of
$b$-jet or $W^+$ carry similar information.
In \cref{fig:b-jet-energies} (left panel), we can clearly confirm this
in the $b$-jet energy distribution.
It peaks around \SI{70}{\GeV} with a major threshold enhancement 
by a factor of $\sim14$.
In fact, the peak position of $b$-jets has been proposed as a
possibility to measure the top quark mass~\cite{1209.0772,1512.02265,1603.03445},
which has been realized by CMS using \SI{8}{\TeV} data~\cite{CMS:2015jwa}.
We note that the peak position is consistent with the (Born-level) rest frame 
energy (cf. \cref{eq:decay_projection} for $m_{b} \ll m_{W}$),
\begin{align}
  E_{b}^* = \frac{\mpole^2 - m_{W}^2}{2\mpole} \approx 
\SI{67.9}{\GeV}
\end{align}
as it has been shown for unpolarized top decays, massless b-quarks and
generic boost directions in \Rcite{1209.0772}. In our case, this is of
course especially expected as nearly no boost of the top decay is
present. The intriguing aspect of this analysis is that no correct
reconstruction of $b$-jets with $W^+$ has to be performed and even
the charge of the $b$-jets is irrelevant.
Considering pairs of $b$ and $\bar{b}$-jets as shown in the right 
panel of \cref{fig:b-jet-energies}, we observe that the peak in $E^\jb$ 
around \SI{70}{\GeV} is translated to a peak in $E^{\jb\jb}$ around
\SI{140}{\GeV}.
We stress, however, that these results have to be interpreted with some 
caution as we have neglected NLL order final state interactions involving 
$b$ and $\bar{b}$, which could affect particularly this observable.

\begin{figure}[htbp]
\centering
  \includegraphics[width=\halfwidth]
  {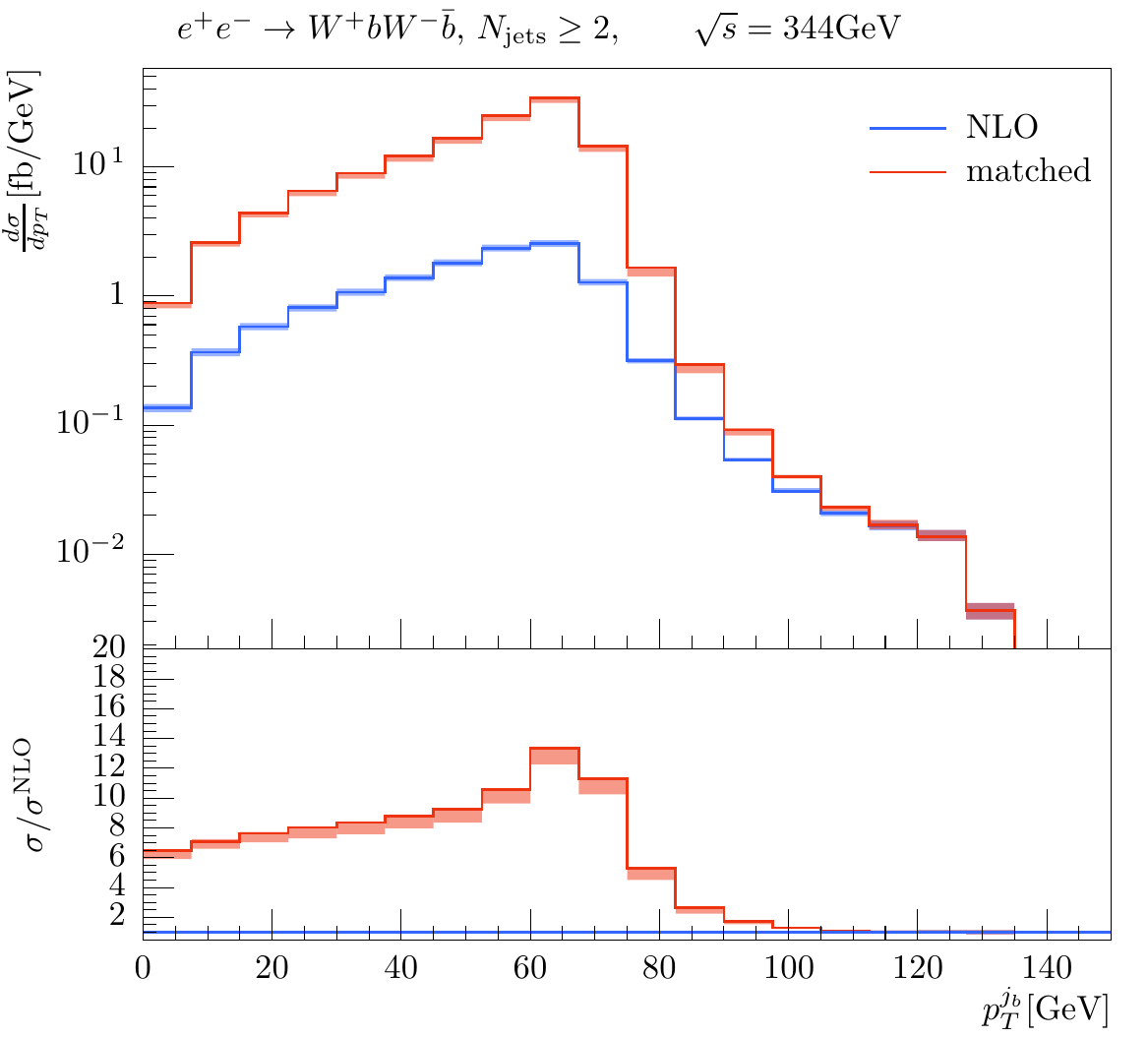}
  \includegraphics[width=\halfwidth]
  {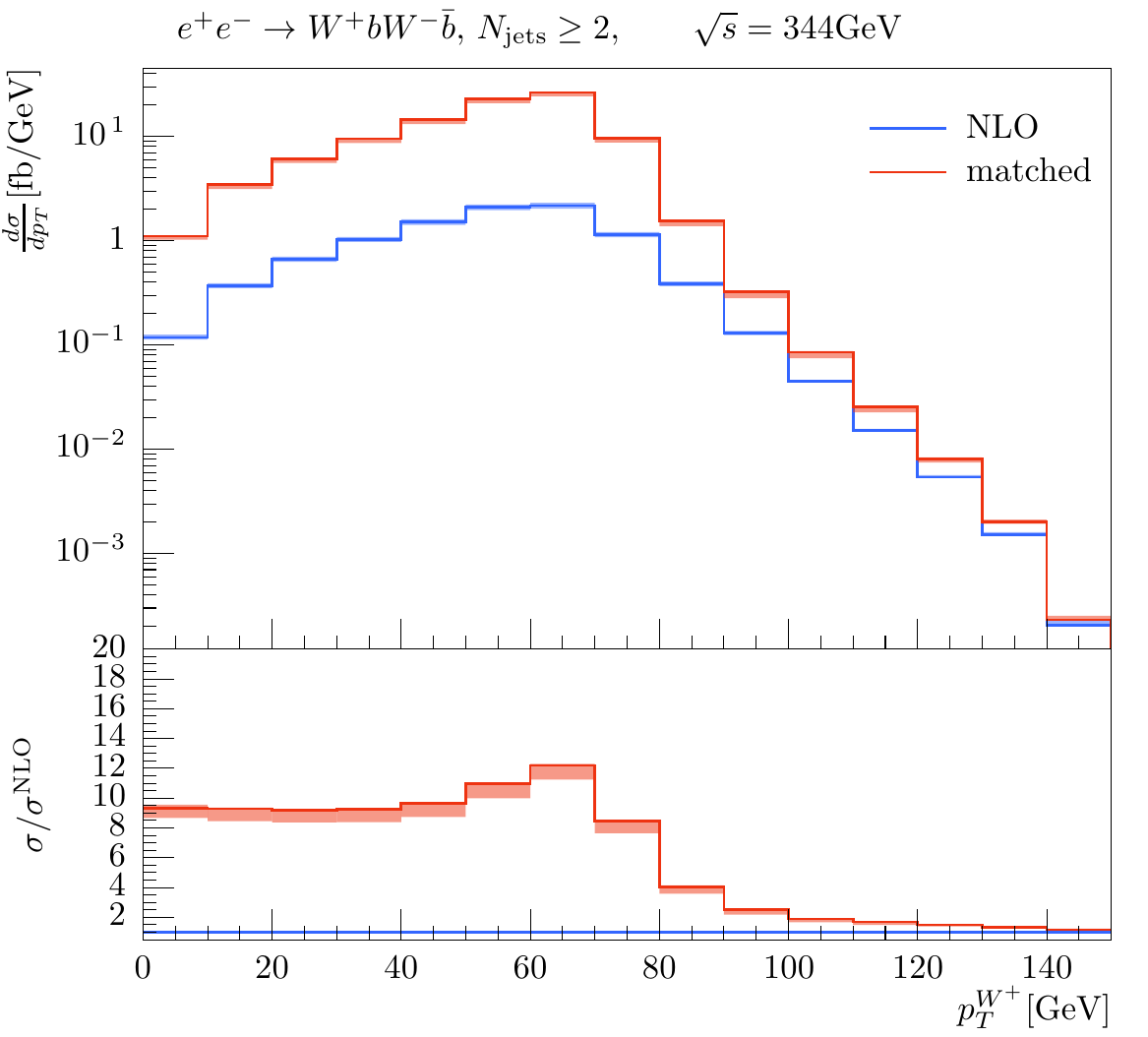}
\caption{Transverse momentum distributions of $b$-jets  (left panel) 
	and $W^+$ bosons  (right panel).
  Lines, bands and panels as in \cref{fig:top_obs1}.
  }
\label{fig:b-jet-W-transverse-mom}
\end{figure}
In \cref{fig:b-jet-W-transverse-mom}, we show the transverse momentum of
$b$-jets  (left panel) and $W^+$ bosons  (right panel).
As we know that the $b$-jet energy peaks around \SI{70}{\GeV}, we
can expect $p_\T{T}^{\jb}$ to have its maximum slightly below this value
due to the small bottom mass $m_{b}=\SI{4.2}{\GeV}$.
As a result of momentum conservation, $p_\T{T}^{W^+}$ has to follow a
similar distribution.
This is exactly what we observe in \cref{fig:b-jet-W-transverse-mom}.
Compared to the $b$-jet energy, the peak is not as pronounced and
more smeared to smaller values, which can still correspond to the
peak jet energy due to the projection to the transverse plane.
Accordingly, $\XSmatched/\XSNLO$ is large, a factor of 6--13, for momenta
below $70$~GeV and quickly drops to 1 above \SI{90}{\GeV}, where
the contributions largely stem from the  $\WbWb{}$ (background) production
not associated to top quark decays.

\begin{figure}[htbp]
\centering
  \includegraphics[width=\standardwidth]
  {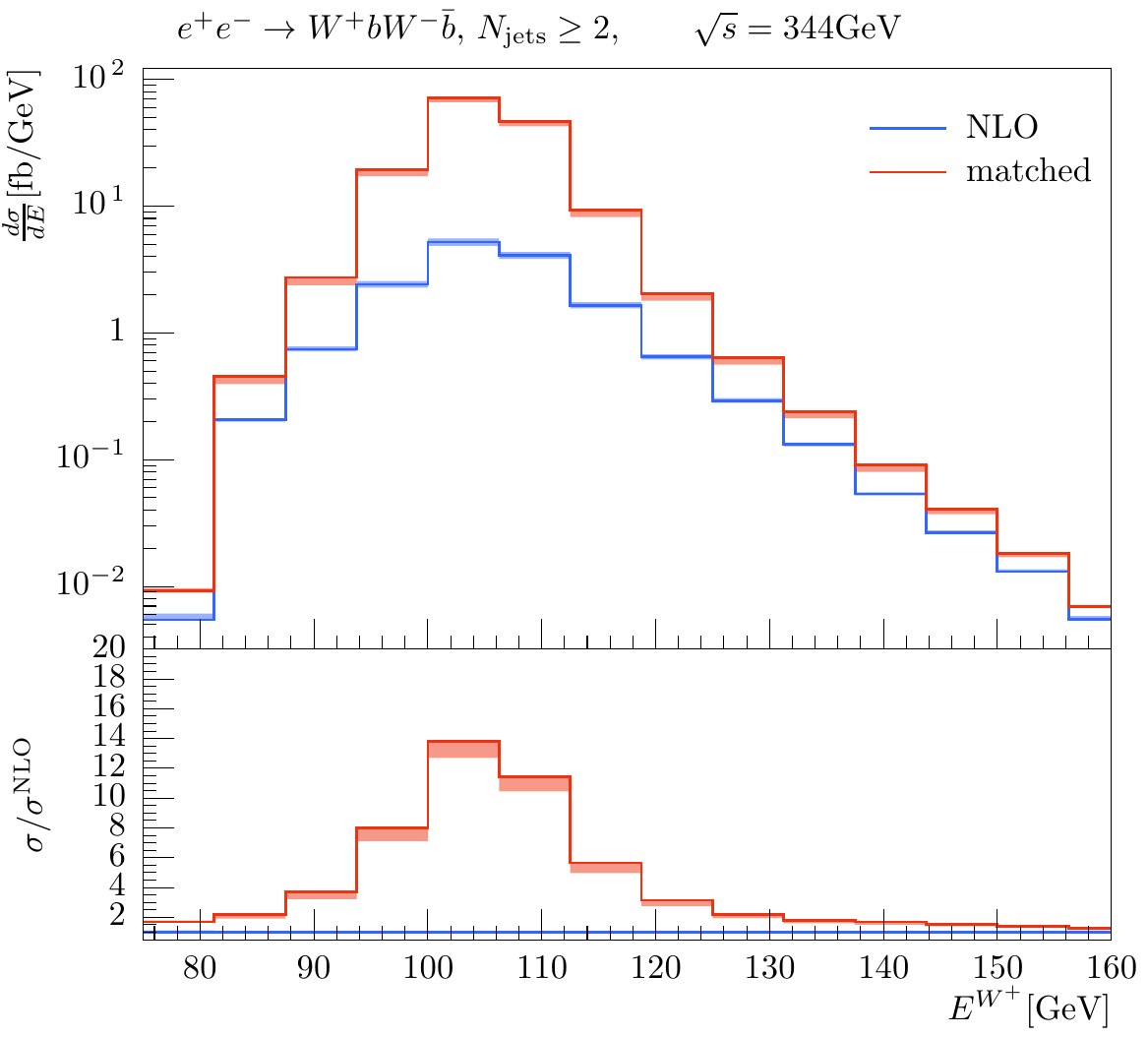}
  \caption{Energy distribution of $W^+$ bosons.
  Lines, bands and panels as in \cref{fig:top_obs1}.
  }
\label{fig:W-energies}
\end{figure}
Finally, we show in \cref{fig:W-energies} the energy of $W^+$ bosons.
Also in this distribution, we can identify the footprint of the top
decay in the peak and large threshold enhancement around the
\SIrange{100}{106}{\GeV} bin.
Compared to \cref{fig:b-jet-energies} though, we observe even for large
$W^+$ boson energies still sizable threshold enhancements of a factor of
$\sim2$.
Thus, the top quark contributions are not as localized as in the
$E^{\jb}$ case.

%
\section{Summary \& conclusions}
\label{s:summary}

The center-of-mass energy scan over the lineshape of the onset of
top-pair production at a future lepton collider represents the most
precise known experimental method to determine the top quark mass. Due
to the high level of theoretical understanding this method allows to
directly access a short-distance top-quark mass that is theoretically
much better defined than the so-called pole mass and yields theoretical as well as experimental
uncertainties well below $0.1$~GeV. Thus, from the point of view of
theoretical control as well as precision, the top-pair threshold scan
is superior to top mass determinations from kinematic reconstruction. In
order to precisely assess the experimental systematics of the threshold scan as well
as to study kinematical distributions in the threshold region e.g. as an
alternative means to measure the top mass, an exclusive calculation
fully differential in the final state is indispensable. 

In this paper, we have for the first time set up a fully exclusive framework 
which allows to access all aspects of $\WbWb{}$ production, is
valid in the threshold region as well as away from threshold, and provides
smooth and continuous predictions over the whole energy range. It is 
thus suitable without restrictions 
for any conceivable staging
plan for a future lepton collider. Our approach combines at the amplitude level
NLO fixed-order relativistic continuum QCD calculations with
the NLL resummation of the threshold corrections determined from a
renormalization group improved extension of NRQCD. The approach is constructed
such that in the threshold region the threshold-resummed predictions dominate, 
while far away from the threshold the predictions are given only by the 
fixed-order calculations.
In order to make this feasible and particularly to avoid
double counting in the intermediate cross-over regions, the $\tT{}$ signal diagrams
containing the resummed nonrelativistic S and P wave form factors 
have been evaluated in a factorized and manifestly gauge-invariant 
double-pole approximation. 
The resummed form factors for both have been expanded to first order 
in $\alpha_s$ and subtracted at the amplitude level from the relativistic 
fixed-order matrix elements to remove double-counting
contributions. Our approach accounts in particular for the
interference contributions that arise from threshold-resummed and
fully relativistic $\tT{}$ resonant as well as non-resonant matrix
elements. For the factorized parts of our approach the top quark 
decays have been evaluated at NLO in QCD. 

On the technical side, the NLL threshold resummations 
for the form factors were realized by incorporating
the numerical \prog{Toppik} code into the MC framework of
\whz{} which already provides automated NLO QCD fixed-order
predictions for multi-leg processes. 
To achieve a gauge-invariant description in the context of the
matrix element evaluations in the factorized contributions, an on-shell
projection regarding intermediate top- and anti-top quark states for
the matrix element evalutation has been performed. This allows to keep
the resummed contributions in the MC framework in the kinematic
regions above threshold, as well as at and below threshold. The
implementation has been thoroughly tested against existing analytical
threshold calculations. Furthermore consistency checks concerning
gauge invariance, the high-energy behavior, and spin correlations have
been performed.

As the factorized contributions containing the form factor with the
threshold-resummed Coulomb-singular and logarithmic terms 
(and their expansions) do not provide a valid description in the
relativistic regime 5-10 GeV away from the toponium peak region, those terms have
to be switched off smoothly in order to transition to the pure NLO
QCD relativistic
description in the continuum. This has been achieved by supplementing 
these contributions with a smooth switch-off function which is unity
in the threshold region and which vanishes away from threshold.
As the form of the switch-off function is not unique, variations of it
have to be performed to determine the theoretical uncertainties in the
intermediate region between threshold and relativistic continuum.
So, to provide a meaningful description of the overall remaining 
theoretical uncertainties, common scale variations in the
relativistic calculation, a scan over soft and hard scaling parameters 
of the threshold-resummed form factors, as well as
a variation over the parameters of the switch-off function
have to be carried out.
As the scale variations from the nonrelativistic scaling parameters are
asymmetric with respect to the default values at the current level of our approximations, 
we in addition suggest to 
symmetrize the corresponding error bands adopting the 
largest deviation from the default as the uncertainty to arrive at a 
conservative
error estimate.

In our current implementation there are a number of NLL effects in the 
threshold region which are not yet implemented and which shall be addressed in
future releases. These include contributions such as the Coulomb potential
due to photon exchange, the resummation of phase space logarithms, a more
systematic implementation of top quark short-distance mass schemes that allows
to simultaneously use the 1S mass scheme in the threshold region and the 
$\overline{\mathrm{MS}}$ scheme at high energies, and final state interactions 
due to
dynamical soft gluon exchange which are known to cancel in the total inclusive 
cross section and cannot be described by form factors. 
The latter effects are essential for differential observables
affected by gluon exchange involving top and anti-top and their decay
products, and such differential results therefore only have LL
precision in the threshold regime in our current implementation.
We also mention that our current implementation lacks a coherent full description of the 
top spin correlations due to current limitations of \openloops. 
An elaborate (but possible) next step is also the inclusion and proper matching 
of purely electroweak NLO corrections and a coherent treatment of initial-state 
photon
radiation which at this level does no longer simply factorize with the top production 
process. It is also straightforward to incorporate the available higher-order
total inclusive cross section results in the normalization via a K-factor approach. 

The obtained differential results at the $\tT{}$ threshold 
have demonstrated that it is possible to clearly separate and identify  areas 
of the $\WbWb{}$ phase space which are $\tT{}$-resonant, 
non-$\tT{}$-resonant or intermediate. These results may provide additional means
to measure top quark parameters such as its
mass, width, and couplings. 

More systematic studies related to these observables are beyond the scope of this
work.
We finally remark that it is straightforward to apply our implementation 
of the combined threshold-resummed and fixed-order relativistic description
of the top pair threshold to other processes where the nonrelativistic 
$\tT{}$ dynamics is essential. Well-known examples of such processes are $\tT 
H$ or $\tT Z$ production in the region close to threshold. As the
Higgs and $Z$ bosons to a good approximation act as colorless
recoilers, these processes can be also described using the approach
detailed in this paper. We also remark that the same techniques  
can also be applied to pair thresholds arising from 
heavy colored resonances beyond the Standard Model
at hadron colliders such as the LHC.

Given the power of the method proposed in this work to 
coherently incorporate and combine available results at threshold and 
in the relativistic region and to access and study arbitrary
observables at the fully differential level, we believe that our 
 MC approach represents a highly suitable
framework to make further progress in top threshold physics.

\acknowledgments{}
We thank T.~Ohl for checking the gauge invariant subsets of
$\eeWbWb{}$. For clarifying technical details we like to thank
T.~Je\v{z}o as well as J.~Lindert, who also provided polarized top
production and decay libraries at NLO. Furthermore, we like to thank
M.E.~Peskin for enlightening discussions on the top threshold
project. BCN, CW and JRR acknowledge funding support by the
Collaborative Research Unit (SFB) 676 of the German Research Council
(DFG), projects B1 and B11. The work of MS has been supported by GFK and by the
PRISMA cluster of excellence at JGU Mainz.
The work of TT has been supported by STFC under the consolidated grant 
ST/L000431/1. AHH acknowledges partial support by the FWF Austrian Science Fund 
under the Project No. P28535-N27.
\clearpage{}
\appendix
\section{NLL current coefficients}
\label{app:NLLc1c3}
At NLL the current coefficient $c_i$ 
with $i=1,3$ reads
\begin{align}
&c_i^\mathrm{NLL}(h,\nu) \equiv 
c_i^\mathrm{NLL}(\AShard,\ASsoft^\mathrm{LL},\ASusoft^\mathrm{LL})  = c_i(h,1) 
\exp \Bigg\{ 
\pi  \AShard \bigg[
A_0^{(i)} \bigg(z-1 -\frac{\log \omega}{\omega} \bigg)
+ A_2^{(i)} (1-z) 
\nonumber\\
&\quad
+ A_3^{(i)} \log z  
+ A_4^{(i)} \bigg(1-z^{1-\frac{13 C_A}{6 \beta_0}}\bigg)
+ A_5^{(i)} \bigg(1-z^{1-\frac{2 C_A}{\beta_0}}\bigg)
+ A_8^{(i)} \bigg(1-z^{1-\frac{C_A}{\beta_0}}\bigg)
\bigg]
\Bigg\} \co
\label{eq:ciNLL}
\end{align}
where
\begin{align}
 z \equiv \frac{\ASsoft^\mathrm{LL}}{\AShard} \co
 \qquad
 \omega \equiv \frac{\ASusoft^\mathrm{LL}}{\ASsoft^\mathrm{LL}} \co
\end{align}
and $c_i(h,1)$ is the one-loop matching condition given in 
\eqs{c1match}{c3match}.
For the $^3S_1$ current coefficient $c_1^\mathrm{NLL}$ we have \cite{Pineda:2001et,hep-ph/0209340}
\begin{align}
A_0^{(1)} &= -\frac{8 C_F (C_A+C_F) (C_A+2 C_F)}{3\beta_0^2}
\co\\
A_2^{(1)} &=\frac{C_F \Big[C_A C_F (9 C_A - 100C_F)
-\beta_0 (26 C_A^2+19 C_A C_F-32 C_F^2)\Big]}{26 \beta_0^2 C_A}
\co\\
A_3^{(1)} &=
\frac{C_F^2 \Big[6 \beta_0^2 (C_A-2 C_F)+2 \beta_0 
C_A (37 C_F-8 C_A)+C_A^2 (9 C_A-100 
C_F)\Big]}{\beta_0^2 (6 \beta_0-13 C_A) (\beta_0-2 C_A)}
,\\
A_4^{(1)} &= \frac{24 C_F^2 (11 C_A-3 \beta_0) (5 C_A+8 C_F)}{13 C_A (6 
\beta_0-13 C_A)^2}
\co\\
A_5^{(1)} &= \frac{C_F^2 (5 \beta_0-13 C_A)}{6 (\beta_0-2 
C_A)^2}
\co\\
A_8^{(1)} &= 0\,.
\co
\end{align}
For the $^3P_1$ current coefficient $c_3^\mathrm{NLL}$ we have \cite{hep-ph/0609151}
\begin{align}
A_0^{(3)} &= -\frac{8 C_A C_F (C_A+4 C_F)}{9 \beta_0^2}
\co \\
A_2^{(3)} &= -\frac{C_F (4 C_A+7 C_F)}{12 \beta_0}
\co \\
A_3^{(3)} &= 0
\co \\
A_4^{(3)} &= 0
\co \\
A_5^{(3)} &= \frac{C_F^2}{12 (\beta_0-2 C_A)}
\co \\
A_8^{(3)} &= -\frac{C_F^2}{3 (\beta_0-C_A)}
\po 
\end{align}
We note that in \Rcite{hep-ph/0609151} the NLL anomalous dimensions of color 
singlet 
heavy quark--anti-quark production currents for all possible $^{2S+1}L_J$ quantum numbers were determined.

\section{Additional validation results}
\label{app:additional_validation_results}

\begin{figure}[h]
\centering
\includegraphics[width=0.45\textwidth]{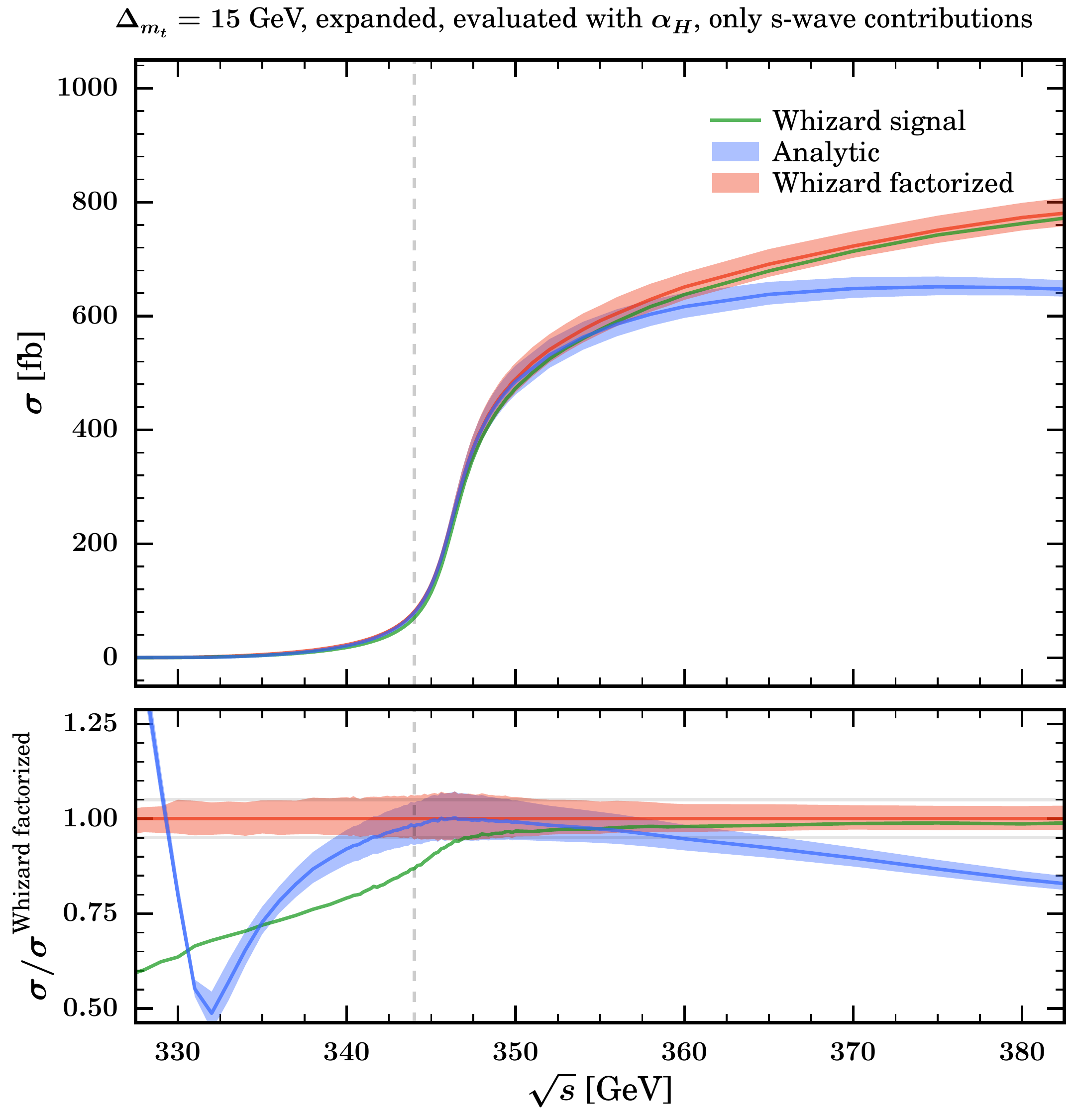}
\includegraphics[width=0.45\textwidth]{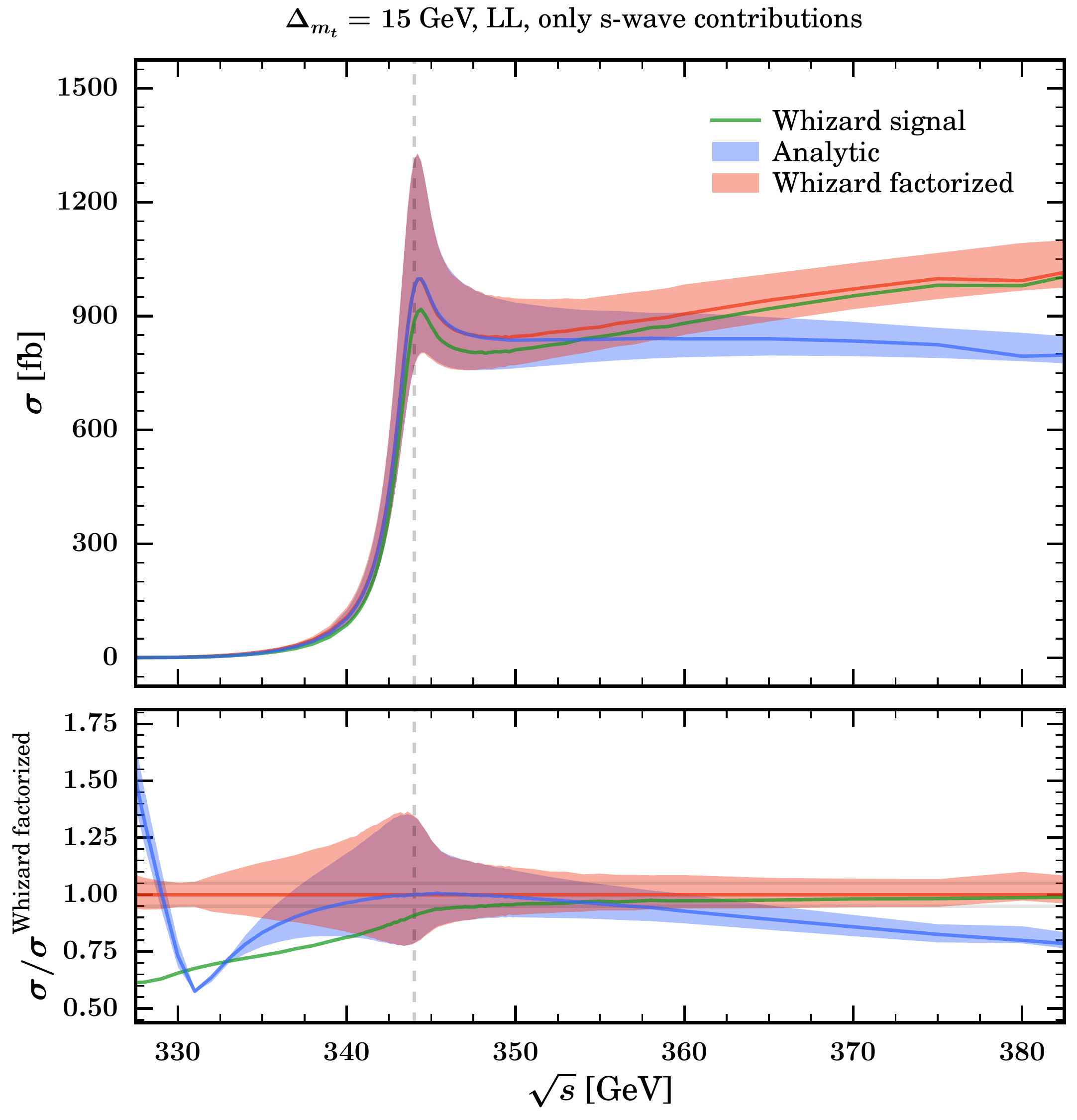} \\
\includegraphics[width=0.45\textwidth]{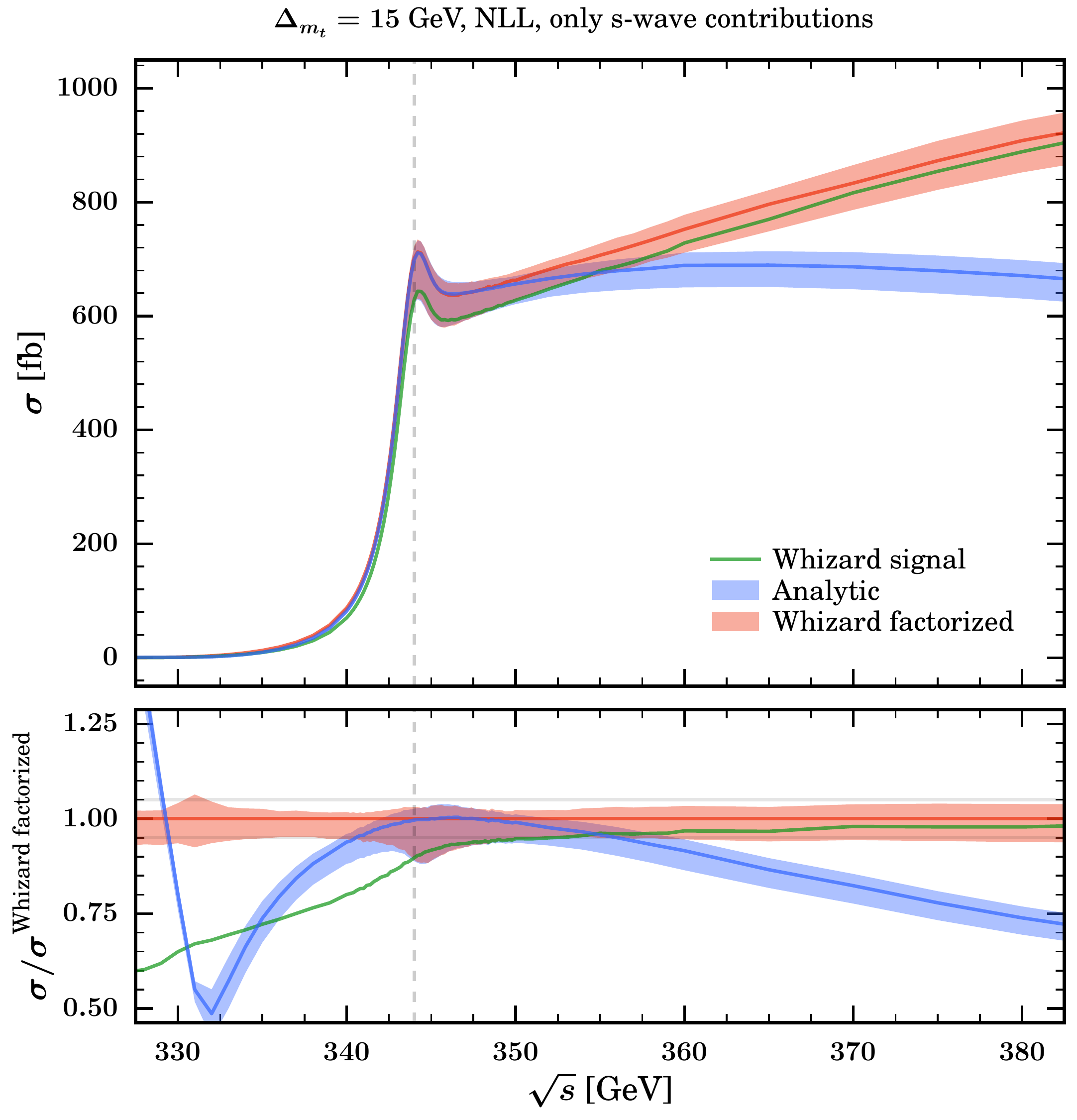}
\caption{Comparison of analytic and  \whz{} results using
  the factorized and the signal-diagram approach with expanded, LL and NLL form factor for the inclusive cross section with a top invariant mass cut of $\DeltaM = \SI{15}{\GeV}$.
  Bands and lines are analogous to the ones in \cref{fig:validation_dm_fixed}.
  }
\label{fig:validation_dm_fixed_extra}
\end{figure}

Completing the validation for the inclusive cross section with cuts on the reconstructed invariant mass of the tops in \cref{ss:validation_sqrts_scans},
we show in \cref{fig:validation_dm_fixed_extra} the corresponding validation plots for a (tight) cut, $\DeltaM = 15$ GeV.

\section{Alternative switch-off functions}
\label{s:alternative_switch_off_functions}
\begin{figure}[htbp]
\centering
\includegraphics[width=0.8\textwidth]{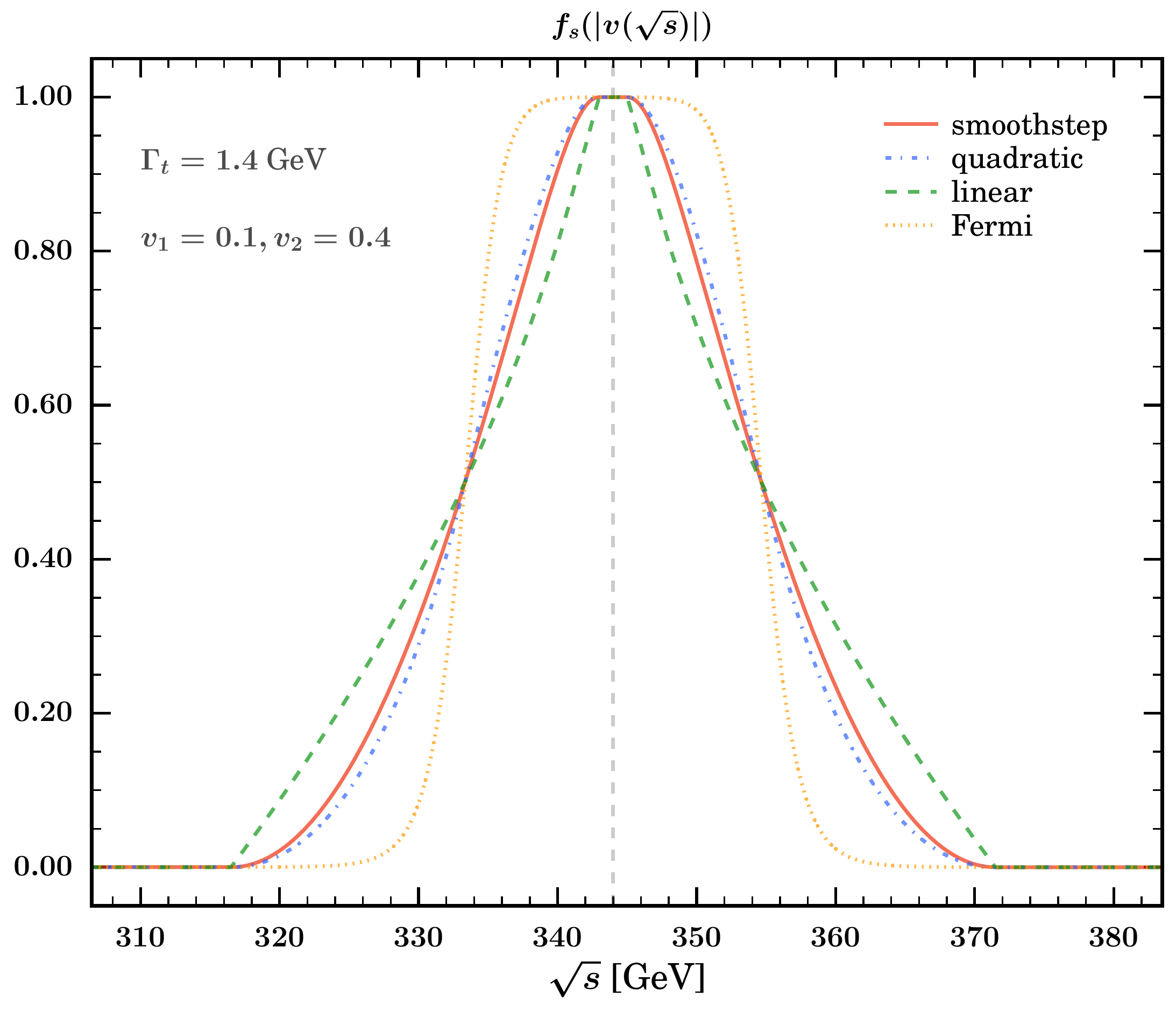}
\caption{Various switch-off functions as explained in the text in a wide
  switch-off window between $v_1=0.1$ and $v_2=0.4$.}
\label{fig:switch_off_functions}
\end{figure}
Here we briefly discuss other possibilities for the switch-off function \switch{} and compare them to the smoothstep function  in \cref{eq:switch_smoothstep}, which we use for the matched predictions in this work.
Specifically, we depict in \cref{fig:switch_off_functions} the smoothstep,
a quadratic, a linear and a so-called Fermi function (Fermi-Dirac distribution).
By trying different versions of \switch{} we have observed that high curvature functions typically lead to
artificial bumps or wiggles in the matched cross section.
Those are absent for the linear switch-off, which on the other hand is not
smoothly differentiable and produces unphysical edges at
$v_1$ and $v_2$.
The quadratic \switch{} we have displayed here actually consists of two quadratic functions:
\begin{align}
  \switch(v) = \begin{cases} 1 & v < v_1 \\
  1 - 2 \frac{(v - v_1)^2}{(v_2 - v_1)^2} & v_1 < v < \frac{v_1 + v_2}2 \\
    2\frac{(v - v_2)^2}{(v_2 - v_1)^2} & \frac{v_1 + v_2}2 < v < v_2 \\
              0 & v > v_2
            \end{cases}\po
  \label{eq:switch}
\end{align}
This is quite close to the smoothstep, but further away from the linear function.
The Fermi-Dirac distribution in \cref{fig:switch_off_functions} has been generated with a mean of $(v_1+v_2)/2$ and a width of $(v_2-v_1)/20$.
Note that while one can get a behavior closer to the linear function
around the mean with a larger width, this leads to \switch{} not being
approximately 1 and 0 at $v_1$ and $v_2$, respectively.
Overall, our smoothstep function appears to be a good compromise between smoothness and little curvature, while
 most other (reasonable) parametrizations give cross section results within the matching uncertainty from varying $v_1$ and $v_2$ as described in \cref{ss:theoretical_uncertainties}.

\bibliography{main}
\end{document}